\documentclass[12pt,a4paper]{article}
\usepackage[utf8]{inputenc}
\usepackage[T1]{fontenc}
\usepackage[font=small]{caption}
\usepackage{subfig}
\usepackage{titling}
\usepackage[subfigure,titles]{tocloft}
\usepackage[english]{babel}
\usepackage{rotating}
\usepackage{amsmath}
\usepackage[backend=biber,style=numeric-comp,sorting=none]{biblatex}
\usepackage[titletoc]{appendix}
\usepackage{xspace}
\usepackage{amssymb}
\usepackage{multirow}
\usepackage{siunitx}
\usepackage[scale=0.8]{geometry}
\usepackage{pdfpages} 
\usepackage{subfig}
\usepackage{blindtext}
\usepackage{hyperref}
\usepackage{lineno}
\usepackage[hang,flushmargin]{footmisc}
\usepackage{authblk} 

\hypersetup{
  colorlinks   = true, urlcolor     = blue, linkcolor    = blue, citecolor    = blue }
\usepackage{cleveref}

\newcommand*{\MEX}{\ensuremath{E_{\text{x}}^{\text{miss}}}\xspace}
\newcommand*{\MEY}{\ensuremath{E_{\text{y}}^{\text{miss}}}\xspace}

\newcommand*{\electronvolt}{\text{e\kern-0.1em V}}

\newcommand*{\TeV}{\ensuremath{\text{T\electronvolt}}}
\newcommand*{\GeV}{\ensuremath{\text{G\electronvolt}}}

\newcommand*{\MGMCatNLOV}[1]{\textsc{MadGraph5}\_aMC@NLO~#1\xspace}

\newcommand*{\PYTHIA}{\textsc{Pythia}}
\newcommand*{\NNPDF}{\textsc{NNPDF}}

\bibliography{paper_draft.bib}
\begin{document}

\title{\textbf{Proposed measurement of longitudinally polarised vector bosons in $WH$ and $ZH$ production at Hadron colliders}}
\author[1]{Rosa Colyer\thanks{r.k.colyer@sms.ed.ac.uk}}
\author[1]{Dominik Duda\thanks{dominik.duda@cern.ch, ORCiD:\href{https://orcid.org/0000-0002-5916-3467}{0000-0002-5916-3467}}}

\affil[1]{School of Physics and Astronomy, The University of Edinburgh\\ Edinburgh EH9 3JZ, Scotland, UK}

\date{\today}

\maketitle

\begin{abstract}
The longitudinal polarisation states of the $W$ and $Z$ bosons arise during electroweak (EW) symmetry breaking as the vector 
bosons absorb the Higgs boson's massless degrees of freedom. Consequently, these states are intrinsically linked to the 
EW symmetry breaking mechanism. Measuring the production of longitudinally polarised vector bosons in association with the 
Higgs boson thus offers a unique probe of EW symmetry breaking and provides potential sensitivity to physics beyond the 
Standard Model. This work studies vector boson polarisation in associated $WH$ and $ZH$ production at the LHC and proposes 
targeted analysis strategies for their measurement. We explore kinematic and angular observables and employ machine learning
techniques to optimise the separation between longitudinal and transverse polarisation states. Furthermore, we perform a 
statistical analysis to project the measurement precision and discovery significance for longitudinally polarised $W^\pm$ 
and $Z$ bosons ($W^\pm_\mathrm{L}H$ and $Z_\mathrm{L}H$) in the $H \rightarrow \gamma\gamma$ and $H \rightarrow b\bar{b}$ 
decay channels. Our studies project that 5.0 standard deviation discovery significances can be achieved at the High-Luminosity 
LHC with integrated luminosities of 210\,fb$^{-1}$ for $Z_\mathrm{L}H$ (in $H \rightarrow b\bar{b}$) and 340\,fb$^{-1}$ 
(425\,fb$^{-1}$) for $W^\pm_\mathrm{L}H$ in the $H\rightarrow b\bar{b}$ ($H\rightarrow\gamma\gamma$) decay modes. With the full High-Luminosity LHC 
dataset of 3000\,fb$^{-1}$, inclusive cross-section measurement precisions are expected to reach up to $13\%$ ($10\%$) 
for $W^\pm_\mathrm{L}H$ and $8\%$ ($35\%$) for $Z_\mathrm{L}H$ in the $H \rightarrow  b\bar{b}$ ($H \rightarrow \gamma\gamma$) 
channels, demonstrating the feasibility of isolating longitudinally polarised vector bosons produced in association with a 
Higgs boson in future LHC analyses.
\end{abstract}
\clearpage
\section{Introduction}

The Standard Model (SM) of particle physics is a well-tested theory that describes
the production, interaction, and decays of elementary particles. One of its most
central principles is the spontaneous breaking of the electroweak (EW) symmetry, achieved through
the Higgs Mechanism~\cite{Higgs_I,Higgs_II}. This mechanism explains how the $W^\pm$ and $Z$ bosons
acquire mass by coupling to a spinless quantum field, the Higgs field, which permeates
the entire Universe. \par 
In addition to gaining mass, the $W^\pm$ and $Z$ bosons also obtain their 
longitudinal polarisation states during EW symmetry breaking as they absorb the massless 
degrees of freedom of the Higgs boson, known as Goldstone bosons. The coupling of the 
Higgs boson to the longitudinally polarised vector bosons is crucial; it prevents the divergence of 
tree-level vector-boson scattering amplitudes at high energies, thereby restoring unitarity at the 
TeV scale~\cite{Lee:1977yc,Alboteanu:2008my}. Consequently, longitudinally polarised $W^\pm$ and $Z$ 
bosons play a crucial role in the SM and are intrinsically linked to the mechanism of EW symmetry 
breaking. Therefore, dedicated measurements of the Higgs boson's coupling to longitudinally polarised 
vector bosons offer a unique opportunity to enhance our understanding of the underlying structure of 
EW symmetry breaking, potentially opening a window to physics beyond the SM (BSM).\par
With the discovery of the Higgs boson~\cite{HIGG-2012-27,CMS-HIG-12-028} in 2012 by the ATLAS and CMS collaborations~\cite{ATLAS,CMS} 
at the Large Hadron Collider (LHC)~\cite{LHC}, we entered a new era of probing the nature of EW symmetry 
breaking. As the LHC experiments collect more data, our sensitivity to previously undetectable effects increases, 
thereby motivating novel analyses of unstudied Higgs boson properties. The unprecedentedly large dataset to be accumulated in 
the proton--proton collisions of the High Luminosity LHC~\cite{ZurbanoFernandez:2020cco} will therefore offer a great 
opportunity to perform dedicated measurements of the Higgs boson couplings to longitudinally and transversely polarised vector bosons.\par
\begin{figure}[!h]
\begin{center}
\subfloat[]{\includegraphics[width=0.195\textwidth]{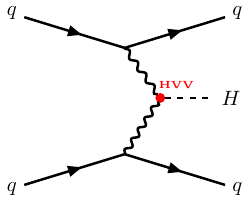}}
\hspace{1cm}
\subfloat[]{\includegraphics[width=0.2425\textwidth]{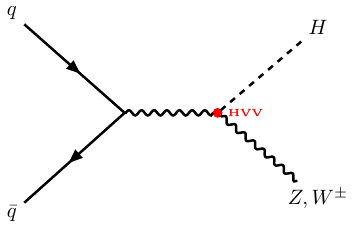}}
\hspace{1cm}
\subfloat[]{\includegraphics[width=0.195\textwidth]{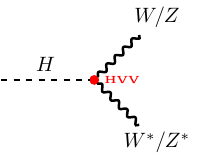}}
\end{center}
\caption{Representative lowest-order diagrams for the production of a Higgs boson via 
(a) vector-boson fusion, (b) associated production with a vector boson, and (c) Higgs boson decays to
vector bosons. The HVV vertex is indicated via a red dot.}
\label{fig:FeynmanDiagrams_HVV}
\end{figure}

The Higgs boson coupling to polarised vector bosons can be studied in several production and decay modes. These include
the Higgs-Strahlung ($VH$, where $V=W^\pm,Z$) and vector-boson-fusion (VBF) production~\cite{Anderson:2013afp,Brehmer:2014pka}, 
as well as Higgs boson decays into a pair of vector bosons ($H\rightarrow VV$)~\cite{Maina:2021xpe,Grossi:2024jae}. Representative 
lowest-order diagrams for these processes are depicted in Fig.~\ref{fig:FeynmanDiagrams_HVV}. Each of these modes provides sensitivity 
to the HVV vertex. However, VBF Higgs production and Higgs boson decays constrain such studies to energy scales consistent with the 
Higgs boson mass. In contrast, focusing on the $VH$ production mode allows us to study the Higgs boson coupling properties up to the 
TeV scale, significantly exceeding the EW symmetry breaking scale. Consequently, the $VH$ production mode is particularly promising 
for observing contributions from BSM physics~\cite{Anderson:2013afp,Liu:2018pkg,SALLY}. Moreover, $VH$ production  
offers an unique opportunity to eventually probe for electroweak restoration\footnote{Electroweak restoration is a consequence 
of the Goldstone boson equivalence theorem~\cite{Butterworth:2002tt}, which predicts that at sufficiently high energies 
(well above the $W^\pm$ and $Z$ boson masses), the longitudinal states of the vector bosons become equivalent to 
their associated Goldstone bosons, while only the transversely polarised states remain.}, i.e. the phenomenon where the 
longitudinal polarisation states of vector bosons revert back to their associated Goldstone bosons at sufficiently high 
energies, as argued in Ref.~\cite{Huang:2020iya}. This makes the $VH$ production mode of broad interest, extending beyond 
BSM considerations.\par

This article proposes a strategy for measuring the production cross-sections of polarised vector bosons
produced in association with a Higgs boson ($W^\pm H$ and $ZH$) using proton--proton collision data at the LHC.
To evaluate the feasibility of such measurements, we explore kinematic and angular observables capable of
distinguishing longitudinally polarised vector bosons ($W^\pm_\mathrm{L}$ and $Z_\mathrm{L}$) from their
transversely polarised counterparts ($W^\pm_\mathrm{T}$ and $Z_\mathrm{T}$). Furthermore, machine learning
techniques are employed to enhance the distinction between these polarisation states, and a statistical
analysis is performed to project the expected measurement precision for $W^\pm_\mathrm{L}H$ and
$Z_\mathrm{L}H$ production. For this, we focus on events where the vector boson decays leptonically 
(i.e. $W^\pm\rightarrow \ell^\pm\nu$ or $Z\rightarrow \ell^+\ell^-$) and the Higgs boson decays either to two 
photons ($H\rightarrow \gamma\gamma$) or to two bottom quarks ($H\rightarrow b\bar{b}$). We therefore study the 
$\ell^\pm\nu \gamma\gamma$, $\ell^+\ell^-\gamma\gamma$, $\ell^\pm\nu b\bar{b}$, and $\ell^+\ell^- b\bar{b}$ final states. 
However, our strategies can be adapted with relative ease to other Higgs boson decay modes. \par
The proposed measurements would complement existing studies by the ATLAS and CMS collaborations of longitudinally and transversely
polarised vector boson production cross-sections in di-boson processes such as $ZZ$~\cite{ATLAS:2023zrv,ATLAS:2026fhf}, $WZ$~\cite{ATLAS:2024qbd,CMS-SMP-20-014}, 
and $W^{\pm}W^{\pm}$~\cite{ATLAS:2025wuw,CMS-SMP-20-006}. They would also extend initial studies of Higgs boson coupling properties
to longitudinally and transversely polarised vector bosons in VBF Higgs production~\cite{HIGG-2017-13,ATLAS:2026ovz}. \par
Deviations from the SM predictions in the relative contributions of $V_\mathrm{L}H$ and $V_\mathrm{T}H$ production could
indicate that the Higgs boson is a composite state~\cite{Kaplan:1983fs}. Alternatively, an altered composition of these
polarisation states could be induced by on-shell or off-shell contributions from new heavy gauge bosons ($W'$ and $Z'$) via
$W'\rightarrow W_\mathrm{L}H$ or $Z'\rightarrow Z_\mathrm{L}H$ decays~\cite{Pappadopulo:2014qza}.
Hence, explicit measurements of the $W^{\pm}_\mathrm{L}H$ and $Z_\mathrm{L}H$ production cross-sections serve as a natural
extension of direct searches for $W^\pm H$ and $ZH$ resonances at the LHC~\cite{ATLAS:2020qiz,cmsVHqqbb2017,CMS:2021lyi,ATLAS:2022enb,ATLAS:2024rcu,CMS:2024hbn}.
More broadly, such polarisation measurements provide a distinct and model-independent probe of electroweak symmetry breaking,
offering a complementary strategy to previous studies focused on non-standard $HVV$ tensor structures through effective field
theory fits and angular asymmetries~\cite{Franceschini:2017xkh, Banerjee:2018bio, Banerjee:2019pks, Hernandez-Juarez:2023dor, Hernandez-Juarez:2025nzh}.\par
The structure of this article is organised as follows: Sec.~\ref{sec:Setup} describes the event simulation, object definitions, 
selection criteria, and the machine learning strategy used to separate longitudinal and transverse polarisation states. Sec.~\ref{sec:StatAna} 
outlines the statistical framework used to evaluate measurement sensitivities at the LHC. Sec.~\ref{sec:Results} presents the projected results, 
and Sec.~\ref{sec:Conclusion} provides our concluding remarks.
 \section{Proposed analysis techniques}\label{sec:Setup}

\subsection{Signal and background processes}\label{sec:SignalAndBackgrounds}

\begin{figure}[!h]
\centering
\subfloat[]{\includegraphics[width=0.4225\textwidth]{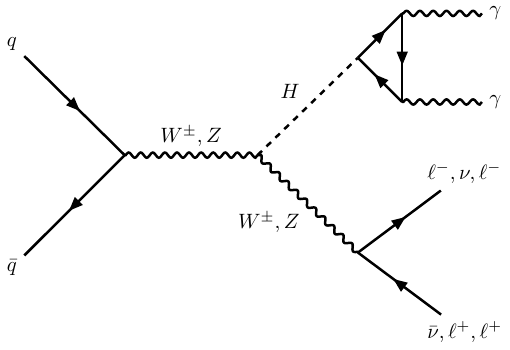}}
\hspace{1cm}
\subfloat[]{\includegraphics[width=0.4225\textwidth]{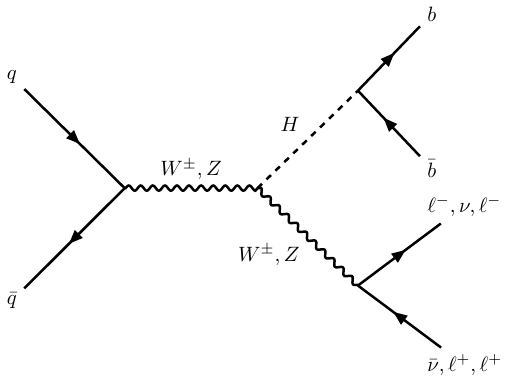}}
\caption{Representative lowest-order Feynman diagrams for the associated production of a Higgs boson and a 
         vector boson ($W^\pm$ or $Z$). The subsequent decays into the $\ell^\pm \nu\gamma\gamma$, 
         $\ell^\pm\ell^\mp\gamma\gamma$, $\ell^\pm\nu b\bar{b}$, and $\ell^+\ell^- b\bar{b}$
         final states are also depicted, where $\ell =$ $e$, $\mu$, and $\tau$.}
\label{fig:FeynmanDiagrams_signal}
\end{figure}

This analysis targets the associated $W^\pm H$ and $ZH$ production modes in the $W^\pm H\rightarrow\ell^\pm\nu \gamma\gamma$, 
$ZH\rightarrow\ell^+\ell^-\gamma\gamma$, $W^\pm H\rightarrow \ell^\pm\nu b\bar{b}$, and $ZH\rightarrow \ell^+\ell^- b\bar{b}$ 
final states (where $\ell = e, \mu$). Representative lowest-order Feynman diagrams for these signal processes are shown in 
Fig.~\ref{fig:FeynmanDiagrams_signal}. \par
These specific decay channels offer complementary strengths. Higgs boson decays to photon pairs ($H\rightarrow\gamma\gamma$) 
provide a very clean experimental signature with an excellent mass resolution and manageable, well-constrained backgrounds. 
Despite its small branching ratio of $(0.227 \pm 0.007)\%$~\cite{LHCHiggsCrossSectionWorkingGroup}, the 
$H\rightarrow\gamma\gamma$ channel has yields some of the most precise measurements of Higgs boson properties to 
date~\cite{HIGG-2020-16,HIGG-2019-16,HIGG-2019-13,CMS-HIG-19-015,CMS:2025ihj}. In contrast, the $H\rightarrow b\bar{b}$ decay 
mode benefits from the largest branching fraction of ($58.09 \pm 0.73)\%$~\cite{LHCHiggsCrossSectionWorkingGroup}, providing high 
statistics. However, it suffers from a complex background composition, requiring sophisticated background reduction 
techniques~\cite{ATLAS:2024yzu,CMS:2023vzh}.\par
Targeting the leptonic decays of the associated $W^\pm$ and $Z$ bosons ensures high triggering efficiencies and enables the 
reconstruction of observables sensitive to vector boson polarisation, as detailed in Sec.~\ref{sec:PolarisationObservables}. \par
\begin{figure}[!h]
\centering
\subfloat[]{\includegraphics[width=0.4225\textwidth]{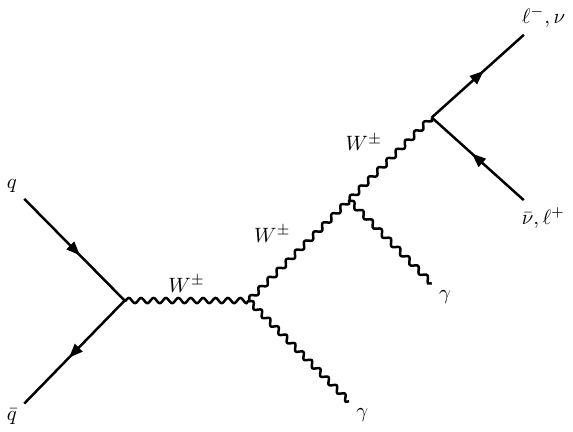}}
\hspace{1cm}
\subfloat[]{\includegraphics[width=0.2825\textwidth]{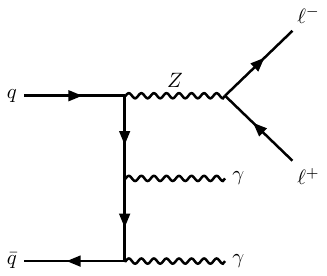}}
\caption{Representative lowest-order Feynman diagrams for the (a) $W\gamma\gamma$ and (b) $Z\gamma\gamma$ production processes.
The subsequent decays of the vector bosons via $W^\pm\rightarrow \ell^\pm\nu$ and $Z\rightarrow \ell^\pm\ell^\mp$  are also depicted, 
where $\ell =$ $e$, $\mu$, and $\tau$.}
\label{fig:FeynmanDiagrams_bkg_yy}
\end{figure}
\begin{figure}[!h]
\centering
\subfloat[]{\includegraphics[width=0.325\textwidth]{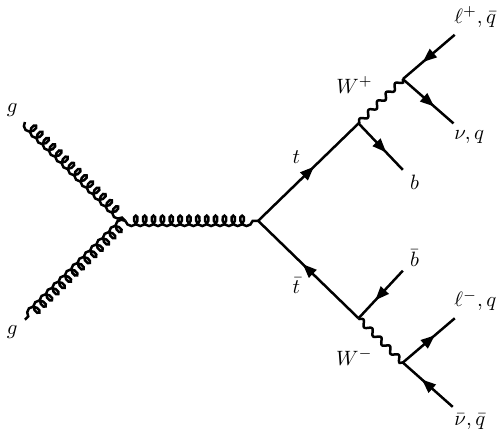}}
\hspace{0.5cm}
\subfloat[]{\includegraphics[width=0.325\textwidth]{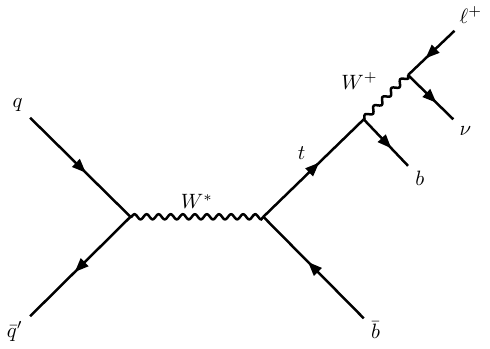}}
\hspace{0.5cm}
\subfloat[]{\includegraphics[width=0.225\textwidth]{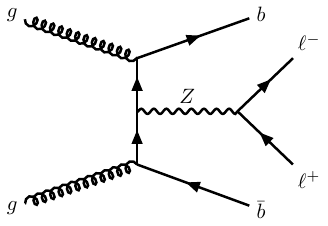}}\\
\subfloat[]{\includegraphics[width=0.225\textwidth]{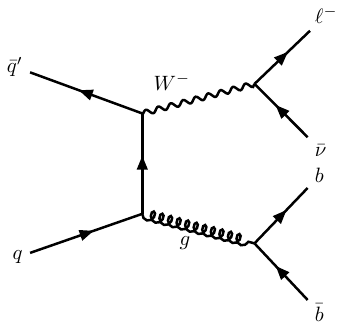}}
\hspace{0.5cm}
\subfloat[]{\includegraphics[width=0.225\textwidth]{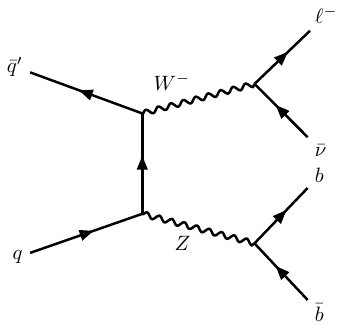}}
\hspace{0.5cm}
\subfloat[]{\includegraphics[width=0.225\textwidth]{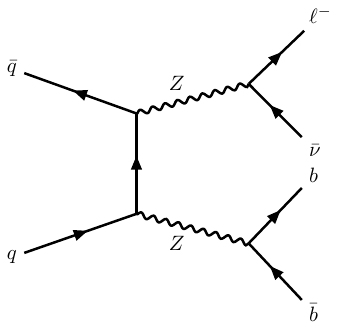}}
\caption{Representative lowest-order Feynman diagrams for (a) top-quark pair, (b) $s$-channel single-top-quark, 
(c) $Z b\bar{b}$, (d) $W b\bar{b}$, (e) $WZ$, and (f) $ZZ$ production. The subsequent decays of all unstable 
particles into the $\ell^\pm\nu b\bar{b}$ or $\ell^+\ell^- b\bar{b}$ final states are also depicted, 
where $\ell = e, \mu, \tau$.}
\label{fig:FeynmanDiagrams_bkg_bb}
\end{figure}

The main backgrounds for the $H\rightarrow \gamma\gamma$ decay channel are triboson $W^\pm \gamma\gamma$ and $Z\gamma\gamma$ 
production, as these processes share the same final state as the signal. Representative lowest-order Feynman diagrams for these 
backgrounds are shown in Fig.~\ref{fig:FeynmanDiagrams_bkg_yy}. For the $H\rightarrow b\bar{b}$ decay channel, the dominant 
backgrounds consist of $t\bar{t}$ production, $s$-channel single-top-quark production, diboson $W^\pm Z\rightarrow \ell^\pm \nu b\bar{b}$ 
and $ZZ\rightarrow \ell^+ \ell^- b\bar{b}$ production, as well as vector-boson production in association with $b$-jets ($W^\pm/Z + b\bar{b}$). 
Representative lowest-order Feynman diagrams for these processes are shown in Fig.~\ref{fig:FeynmanDiagrams_bkg_bb}. 
Backgrounds originating from non-prompt and misidentified photons or leptons are not considered in either analysis channel.

\clearpage

\subsection{Event generation}\label{sec:EventGeneration}

Our signal samples were simulated using the \MGMCatNLOV3.5.1 generator with the SM UFO model~\cite{Degrande:2011ua,Darme:2023jdn} and the 
\NNPDF2.3nlo~\cite{NNPDF} set of parton distribution functions (PDFs) from the \textsc{LHAPDF}~6.3.0 package~\cite{LHAPDF}, 
interfaced with \PYTHIA8.3.0.6~\cite{Sjostrand:2014zea} to model the parton shower (PS), hadronisation, and underlying event. The 
matrix element (ME) was calculated at leading order (LO) accuracy in quantum chromodynamics (QCD) for diagrams with up to two additional 
parton emissions, assuming a center-of-mass energy of $\sqrt{s}=14\,\TeV$. The MLM jet merging algorithm~\cite{Hoeche:2005vzu} was used to 
remove redundancies between identical partonic final states generated at the ME and the PS stage, with the merging scale 
parameter \textit{xqcut} set to $30\,\GeV$. Furthermore, the renormalisation and factorisation scales were set to the 
transverse mass of the $VH$ system.\par 
Separate samples of $VH$ events were generated for transversely and longitudinally polarised vector bosons. The following 
syntax was used in \MGMCatNLOV{} for generating $V_\mathrm{L}H$ events: 

\begin{verbatim}
    import model loop_sm

    define p = g u c d s u~ c~ d~ s~
    define j = g u c d s u~ c~ d~ s~
    define v = z w+ w-
    define lep = e- mu- ta- ve vm vt
    define antilep = e+ mu+ ta+ ve~ vm~ vt~

    generate p p > v{0} h, v > lep antilep
    add process p p > v{0} h j, v > lep antilep
    add process p p > v{0} h j j, v > lep antilep
\end{verbatim}

\noindent For $V_\mathrm{T}H$ events, the syntax was as follows:

\begin{verbatim}
    import model loop_sm

    define p = g u c d s u~ c~ d~ s~
    define j = g u c d s u~ c~ d~ s~
    define v = z w+ w-
    define lep = e- mu- ta- ve vm vt
    define antilep = e+ mu+ ta+ ve~ vm~ vt~

    generate p p > v{T} h, v > lep antilep
    add process p p > v{T} h j, v > lep antilep
    add process p p > v{T} h j j, v > lep antilep
\end{verbatim}

\noindent The calculations were performed using the following experimental values as input parameters: $m_{H}=125.000\,$\GeV, 
$m_{W}=80.419\,$\GeV, $m_{Z}=91.188\,$\GeV, $m_{t}=173.000\,$\GeV, $m_{b}=4.700\,$\GeV, $m_{\tau}=1.777\,$\GeV, 
and $G_{F}=1.16639\cdot 10^{-5}\,$\GeV$^{-2}$, while other particles were considered massless. The decay widths of all 
relevant massive particles were set to their LO values, and the \textsc{Madspin}~\cite{MADSPIN} package was used to decay 
Higgs bosons into photon and bottom-quark pairs. During event generation, the vector boson polarisations were calculated in 
the partonic rest frame. We generated approximately $4.4\cdot 10^{6}$ $W^\pm_\mathrm{L}H$ events, $5.4\cdot 10^{6}$ $W^\pm_\mathrm{T}H$ 
events, $6.1\cdot 10^{6}$ $Z_\mathrm{L}H$ events, and $8.2\cdot 10^{6}$ $Z_\mathrm{T}H$ events for each Higgs boson decay channel.\par
The simulated $W^\pm H$ and $ZH$ samples were normalised to match state-of-the-art calculations of their production cross-sections as well as 
the $H\rightarrow \gamma\gamma$ and $H\rightarrow b\bar{b}$ branching ratios~\cite{LHCHiggsCrossSectionWorkingGroup}.
The production cross-sections were calculated at next-to-next-to-leading order (NNLO) in QCD, including 
next-to-leading order (NLO) electroweak (EW) corrections, using the programmes \textsc{VH@NNLO}~\cite{Brein:2012ne} 
and \textsc{HAWK}~\cite{Denner:2014cla}. The branching ratios were calculated with the programmes \textsc{HDECAY}~\cite{Djouadi:1997yw} 
and \textsc{Prophecy4f}~\cite{Bredenstein:2006nk}. The generated event samples were normalised such that the sum of the $V_\mathrm{L}H$ 
and $V_\mathrm{T}H$ components matched the total cross-section prediction. These cross-sections are listed in Table~\ref{tab:XSecs}. \par
The event samples for the background processes were generated at LO accuracy in QCD using the \MGMCatNLOV{} SM UFO model,
employing the same software packages as used for the signal generation. Dedicated event samples were produced for
$W^\pm\gamma\gamma$, $Z\gamma\gamma$, top-quark pair, $s$-channel single-top-quark, diboson, and ($W^\pm/Z + b\bar{b}$) production.\par

\begin{table}[h!]
\centering
\caption{Production cross-sections for the signal and background processes used in this analysis. The cross-sections for 
the $W^\pm H$ and $ZH$ processes are evaluated at NNLO QCD and NLO EW accuracy using the \textsc{VH@NNLO} and \textsc{HAWK} 
programmes. The $W^\pm \gamma\gamma$ and $Z\gamma\gamma$ cross-sections are calculated at LO in QCD with \textsc{MadGraph5\_aMC@NLO}, 
requiring an invariant diphoton mass of $100\,\mathrm{GeV} < m_{\gamma\gamma} \leq 150\,\mathrm{GeV}$ at the matrix-element 
level. A similar requirement of $m_{b\bar{b}} > 60\,\mathrm{GeV}$ on the bottom-quark pair invariant mass is applied for 
$W^\pm/Z + b\bar{b}$ processes. The $t\bar{t}$ samples, sliced in intervals of the transverse momentum of the leptonically decaying 
$W$ boson $p_{\mathrm{T},W}$, are normalised such that 
their summed cross-section matches the total NNLO QCD prediction including next-to-next-to-leading logarithmic (NNLL) 
soft-gluon resummation calculated with \textsc{Top++2.0}~\cite{Beneke:2011mq,Czakon:2012zr}.}
\renewcommand{\arraystretch}{1.25}
\begin{tabular}{c | c}
\hline
\hline
Process         & Production cross section $[\mathrm{fb}]$   \\
\hline
$pp \rightarrow W^{\pm}_\mathrm{L}H\rightarrow \ell^\pm\nu\gamma\gamma$ & 0.588  \\
$pp \rightarrow W^{\pm}_\mathrm{T}H\rightarrow \ell^\pm\nu\gamma\gamma$ & 0.526  \\
$pp \rightarrow Z_\mathrm{L}H\rightarrow \ell^+\ell^-\gamma\gamma$  & 0.114  \\
$pp \rightarrow Z_\mathrm{T}H\rightarrow \ell^+\ell^-\gamma\gamma$  & 0.112  \\
$pp \rightarrow W^\pm(\rightarrow \ell^\pm\nu)\gamma\gamma$ ($m_{\gamma\gamma}$ filtered) & 3.516 \\
$pp \rightarrow Z(\rightarrow \ell^{+}\ell^{-})\gamma\gamma$ ($m_{\gamma\gamma}$ filtered) & 1.163 \\
\hline
$pp \rightarrow W^{\pm}_\mathrm{L}H\rightarrow \ell^\pm\nu b\bar{b}$ & 152.5 \\
$pp \rightarrow W^{\pm}_\mathrm{T}H\rightarrow \ell^\pm\nu b\bar{b}$ & 135.6 \\
$pp \rightarrow Z_\mathrm{L}H\rightarrow \ell^+\ell^- b\bar{b}$  & 29.11 \\
$pp \rightarrow Z_\mathrm{T}H\rightarrow \ell^+\ell^- b\bar{b}$  & 28.84 \\
$pp \rightarrow W^\pm(\rightarrow \ell^\pm\nu) b\bar{b}$  ($m_{b\bar{b}}$ filtered) & 5006 \\
$pp \rightarrow Z(\rightarrow \ell^+\ell^-) b\bar{b}$ ($m_{b\bar{b}}$ filtered) & 8407 \\
$pp \rightarrow WZ^\pm \rightarrow \ell^\pm\nu b\bar{b}$   &  1312 \\
$pp \rightarrow ZZ \rightarrow \ell^+\ell^- b\bar{b}$  &  304 \\
$pp \rightarrow t\bar{t}\rightarrow \ell^\pm\nu q \bar{q} b\bar{b}  \,\,\mathrm{or}\,\, \ell^+\nu \ell^{-}\nu b\bar{b}$                 & 537200 \\
$pp \rightarrow t\bar{b}/b\bar{t} \rightarrow \ell^\pm\nu b\bar{b}$ & 2202 \\
\hline
\hline
\end{tabular}
\label{tab:XSecs}
\end{table}

To ensure sufficient statistics in the phase-space regions probed by this analysis, a parton-level invariant diphoton mass
selection of $100\,\GeV < m_{\gamma\gamma} \le 150\,\GeV$ was applied during the matrix-element calculation for the
$W^\pm\gamma\gamma$ and $Z\gamma\gamma$ samples. Similarly, a requirement of $m_{b\bar{b}} > 60\,\GeV$ on the invariant mass
of the bottom-quark pair was applied when producing the $W^\pm/Z + b\bar{b}$ samples. Furthermore, the $t\bar{t}$ samples were
filtered to require at least one leptonically decaying $W$ boson. These samples were also sliced into intervals of the $W$ boson transverse momentum
($p_{\mathrm{T},W}$): $p_{\mathrm{T},W} < 50\,\GeV$, $50\,\GeV \le p_{\mathrm{T},W} < 140\,\GeV$, and $p_{\mathrm{T},W} \ge 140\,\GeV$.\par
Approximately $3.6 \times 10^{6}$ $W^\pm(\rightarrow \ell^\pm\nu) \gamma\gamma$ and $6.0 \times 10^{6}$ $Z(\rightarrow \ell^+\ell^-)\gamma\gamma$
events were generated. Around $15 \times 10^{6}$ events were produced for each of the $W^\pm(\rightarrow \ell^\pm\nu) b\bar{b}$,
$Z(\rightarrow \ell^+\ell^-) b\bar{b}$,  $s$-channel single-top-quark, $W^\pm Z \rightarrow \ell^\pm\nu b\bar{b}$, and $ZZ \rightarrow \ell^+\ell^- b\bar{b}$
channels. For $t\bar{t}$ production, $30 \times 10^{6}$ events were generated for each of the two lower $p_{\mathrm{T},W}$ intervals,
and $40 \times 10^{6}$ events were produced for the highest $p_{\mathrm{T},W}$ interval. \par
The showered signal and background events were then passed through \textsc{DELPHES}~3.5.0~\cite{DELPHES} (using the default ATLAS card) to simulate 
detector effects, including energy and momentum smearing, as well as object reconstruction and identification. 

\subsection{Event selection}\label{sec:EventSelection}
For the following studies, we select events in which all final-state objects are fully reconstructed and identified. Furthermore, 
selection requirements are applied to reconstruction-level observables to mimic typical experimental strategies adopted by the LHC 
experiments in similar studies~\cite{HIGG-2020-16,CMS:2022wpo,ATLAS:2024yzu,CMS:2023vzh}. A comprehensive overview of the object 
definitions is provided in Tab.~\ref{tab:ObjectRequirements}, with further details available in Ref.~\cite{DELPHES}. The event-level 
selection criteria are summarised in Tab.~\ref{tab:SelectionRequirements} and Tab.~\ref{tab:SelectionRequirements_VHbb}.\par

The feasibility studies are performed across four distinct analysis channels. The first channel targets the $H\rightarrow\gamma\gamma$ decay mode in the 
$\ell^\pm\nu\gamma\gamma$ final state, requiring exactly one charged lepton and transverse missing momentum ($E^\mathrm{miss}_\mathrm{T}$) exceeding 
$10\,\mathrm{GeV}$. The second channel targets the $\ell^+\ell^-\gamma\gamma$ final state, selecting events with exactly two same-flavor, 
opposite-charge (SFOC) charged leptons whose invariant mass ($m_{\ell\ell}$) satisfies $81\,\mathrm{GeV} < m_{\ell\ell} \le 101\,\mathrm{GeV}$. 
Both $H\rightarrow\gamma\gamma$ channels require at least two photons, with the two leading photons forming the Higgs boson candidate. The invariant 
mass of the diphoton system ($m_{\gamma\gamma}$) is restricted to $115\,\mathrm{GeV} < m_{\gamma\gamma} \le 135\,\mathrm{GeV}$. Additionally, the 
leading and subleading photons are required to satisfy $p_\mathrm{T}/m_{\gamma\gamma} > 0.35$ and $0.25$, respectively, to suppress continuum 
$W^\pm\gamma\gamma$ and $Z\gamma\gamma$ background contributions.\par

The remaining two channels target $H\rightarrow b\bar{b}$ decays in the 1-lepton ($\ell^\pm\nu b\bar{b}$) and 2-lepton ($\ell^+\ell^- b\bar{b}$) 
final states. These channels retain the same lepton multiplicity requirements as their $H\rightarrow\gamma\gamma$ counterparts, but impose higher 
lepton $p_\mathrm{T}$ thresholds to reflect real experimental trigger limits~\cite{TRIG-2022-01}. While the $H\rightarrow\gamma\gamma$ analyses can use 
highly efficient photon triggers to keep lepton thresholds low, the $H\rightarrow b\bar{b}$ analyses must rely directly on single- or dilepton triggers.\footnote{As discussed 
in Sec.~\ref{sec:PolarisationObservables}, lower lepton $p_\mathrm{T}$ thresholds in the diphoton channels are particularly advantageous for studying 
transversely polarised vector bosons, which tend to produce softer leptons.} To further discriminate signal from background in the $H\rightarrow b\bar{b}$ 
channels, $E^\mathrm{miss}_\mathrm{T} > 30\,\mathrm{GeV}$ is required in the 1-lepton channel, whereas $E^\mathrm{miss}_\mathrm{T} < 40\,\mathrm{GeV}$ 
is required in the 2-lepton channel. Both channels require at least two jets with $p_\mathrm{T} > 20\,\mathrm{GeV}$ for $\vert{}\eta\vert{} < 2.5$ 
and $p_\mathrm{T} > 30\,\mathrm{GeV}$ for $2.5 < \vert{}\eta\vert{} < 4.5$. In the 1-lepton channel, events with more than three jets are vetoed to 
suppress contributions from the $t\bar{t}$ background. Among the selected jets, exactly two are required to be $b$-tagged, forming the Higgs boson 
candidate. Flavour tagging is emulated using $p_\mathrm{T}$-dependent efficiency parameterisations derived from a recent ATLAS flavor-tagging algorithm 
operating at a nominal $77\%$ $b$-jet efficiency working point~\cite{ATLAS:2025dkv}. Jets are classified as true $b$-, $c$-, or light-flavor jets based on 
spatial matching with truth-level hadrons, and tagging decisions are assigned via random sampling according to the corresponding efficiency and 
mis-tagging rates. Finally, both $H\rightarrow b\bar{b}$ channels require the leading $b$-tagged jet to have $p_\mathrm{T} > 45\,\mathrm{GeV}$, 
the di-$b$-jet invariant mass to fall within $80\,\mathrm{GeV} < m_{b\bar{b}} \le 145\,\mathrm{GeV}$, and the angular separation of the two $b$-tagged 
jets to satisfy $\Delta R_{b\bar{b}} < 180/p_{\mathrm{T},b\bar{b}} + 0.7$, with $p_{\mathrm{T},b\bar{b}}$ being the transverse momentum of the
reconstructed Higgs boson candidate. \par

To enhance signal separation in the $H\rightarrow b\bar{b}$ channels, events passing all selection criteria are divided into four mutually exclusive 
regions based on the transverse momentum of the reconstructed vector boson candidate ($p_{\mathrm{T},V}$): $p_{\mathrm{T},V} < 75\,\mathrm{GeV}$, 
$75\,\mathrm{GeV} \le p_{\mathrm{T},V} < 150\,\mathrm{GeV}$, $150\,\mathrm{GeV} \le p_{\mathrm{T},V} < 250\,\mathrm{GeV}$, and 
$p_{\mathrm{T},V} \ge 250\,\mathrm{GeV}$. Here, $p_{\mathrm{T},V}$ is reconstructed as the four-vector sum of the two charged leptons in the 2-lepton 
channel, and of the charged lepton and the reconstructed neutrino in the 1-lepton channel. Because the neutrino four-vector is not directly detectable 
experimentally, the technique detailed in Ref.~\cite{TOPQ-2012-18} is used. Specifically, the neutrino transverse momentum components ($p_{x}^{\nu}$ 
and $p_{y}^{\nu}$) are set equal to the missing transverse momentum components (\MEX and \MEY), while the longitudinal component ($p_{z}^{\nu}$) is 
calculated by applying an on-shell $W$-boson mass constraint to the charged lepton--neutrino system. This approach yields a quadratic equation with zero,
one, or two real solutions. In cases with no real solution, the missing transverse momentum vector $\vec{p}^{\,\mathrm{miss}}_\mathrm{T}$ is rotated by 
the minimal amount required to obtain a real solution. When two real solutions are present, the solution with the smallest $\vert{}p^\nu_{z}\vert{}$ 
is chosen.\par

Furthermore, a multivariate strategy is employed to maximise the overall experimental sensitivity. Separate boosted decision trees ($\mathrm{BDT}_{VH}$) 
are trained in each $p_{\mathrm{T},V}$ interval for both the 1-lepton and 2-lepton final states to further discriminate signal from background processes. 
The choice of input features, training procedure, and $p_{\mathrm{T},V}$ binning scheme follow the methodology outlined in Ref.~\cite{ATLAS:2024yzu}, 
with minor adaptations tailored specifically to optimise sensitivity to longitudinally and transversely polarised vector bosons produced in 
association with a Higgs boson. Additional details regarding the training and performance of the $\mathrm{BDT}_{VH}$ classifiers are provided in 
App.~\ref{App:BDT_Training}. \par

Tables~\ref{tab:SelectionRequirements} and~\ref{tab:SelectionRequirements_VHbb} also present the signal and 
background selection efficiencies after each selection requirement. Following the full selection criteria of the 
$VH(\rightarrow \gamma\gamma)$ analyses, approximately $18.3\%$ ($13.9\%$) of the generated $W_\mathrm{L}H$ 
($W^\pm_\mathrm{T}H$) events and $9.6\%$ ($8.0\%$) of the $Z_\mathrm{L}H$ ($Z_\mathrm{T}H$) events are retained. 
Meanwhile, the $W^\pm\gamma\gamma$ and $Z\gamma\gamma$ continuum backgrounds are reduced to approximately 
$3.7\%$ and $2.3\%$ of their initial yields, respectively. For the $VH(\rightarrow b\bar{b})$ channels, 
approximately $5.3\%$ ($2.8\%$) of the generated $W_\mathrm{L}H$ ($W^\pm_\mathrm{T}H$) events and $5.0\%$ ($4.5\%$) 
of the $Z_\mathrm{L}H$ ($Z_\mathrm{T}H$) events survive the selection requirements. In the 1-lepton channel, 
the $W^\pm b\bar{b}$, $Z b\bar{b}$, $W^\pm Z$, $ZZ$, $t\bar{t}$, and $tb$ backgrounds are reduced to $3.1\%$, 
$0.4\%$, $2.0\%$, $0.6\%$, $0.8\%$, and $2.4\%$ of their original yields, respectively. In the 2-lepton channel, 
the $Z b\bar{b}$, $ZZ$, and $t\bar{t}$ backgrounds are suppressed to $2.2\%$, $2.6\%$, and $0.01\%$, 
respectively.\par

\begin{table}[h!]
\centering
\caption{Selection requirements applied to electrons, muons, photons, and jets in the $H \rightarrow \gamma\gamma$ and $H \rightarrow b\bar{b}$ analyses.}
\renewcommand{\arraystretch}{1.25}
\begin{tabular}{c | c}
\hline
\hline
Observable                                        & Selection requirement \\
\hline
\multirow{2}{*}{Transverse momentum of electrons} & $p_{\mathrm{T},e} > 10\,\GeV$ (in $H\rightarrow \gamma\gamma$ events) \\
                                                  & $p_{\mathrm{T},e} > 27\,\GeV$ (in $H\rightarrow b\bar{b}$ events) \\
Pseudorapidity of electrons                       & $|\eta_{e}| \leq 2.47$ (excluding $1.37 <|\eta_{e}| \leq 1.52$) \\
Transverse impact parameter of electrons          & $|d_{0,e}| < 5\,$mm \\
Longitudinal impact parameter of electrons        & $|z_{0,e}| \cdot \sin\theta_e < 5\,$mm \\
Isolation variable of electrons                   & $I(e) < 0.45$ \\
\hline
\multirow{2}{*}{Transverse momentum of muons}     & $p_{\mathrm{T},\mu} > 10\,\GeV$ (in $H\rightarrow \gamma\gamma$ events) \\
                                                  & $p_{\mathrm{T},\mu} > 25\,\GeV$ (in $H\rightarrow b\bar{b}$ events) \\
Pseudorapidity of muons                           & $|\eta_{\mu}| \leq 2.7$ \\
Transverse impact parameter of muons              & $|d_{0,\mu}| < 5\,$mm \\
Longitudinal impact parameter of muons            & $|z_{0,\mu}| \cdot \sin\theta_\mu < 5\,$mm \\
Isolation variable of muons                       & $I(\mu) < 0.45$ \\
\hline
Transverse momentum of photons                    & $p_{\mathrm{T},\gamma} > 20\,\GeV$ \\
Pseudorapidity of photons                         & $|\eta_{\gamma}| \leq 2.37$ (excluding $1.37 <|\eta_{\gamma}| \leq 1.52$) \\
\hline
\multirow{2}{*}{Transverse momentum of jets}      & $p_{\mathrm{T},\mathrm{jet}} > 20\,\GeV$ (for $|\eta_\mathrm{jet}| \leq 2.5$) \\
                                                  & $p_{\mathrm{T},\mathrm{jet}} > 30\,\GeV$ (for $2.5 <|\eta_\mathrm{jet}| \leq 4.5$)\\
Pseudorapidity of jets                            & $|\eta_\mathrm{jet}| \leq 4.5$ \\
\hline
\hline
\end{tabular}
\label{tab:ObjectRequirements}
\end{table}

\begin{table}[h!]
\centering
\caption{Selection requirements applied to event-level observables in the $H \rightarrow \gamma\gamma$ analysis, alongside their corresponding 
selection efficiencies.}
\resizebox{1.0\textwidth}{!}{
\begin{tabular}{c | c c}
\hline
\hline 
Observable                                          & \multicolumn{2}{c}{Selection requirement} \\
\hline
                                                    & $W^\pm H$ Analysis   &  $ZH$ Analysis      \\
                                                    & {\scriptsize($W^\pm_\mathrm{L}H$, $W^\pm_\mathrm{T}H$, $W^\pm\gamma\gamma$)}  & {\scriptsize($Z_\mathrm{L}H$, $Z_\mathrm{T}H$, $Z\gamma\gamma$)}  \\
\hline
\multirow{2}{*}{Number of selected charged leptons} &   $ N_{\ell}= 1$                                     &        $N_{\ell}=2$  (SFOC)         \\
                                                    &   {\scriptsize($46.7\%$, $39.2\%$, $39.1\%$)}        &  {\scriptsize($24.7\%$, $21.6\%$, $27.3\%$)}   \\
\hline
\multirow{2}{*}{Invariant di-lepton mass}           &   \multirow{2}{*}{\textemdash}                       &  $81\,\GeV < m_{\ell\ell} \leq 101\,\GeV $   \\
                                                    &                                                      &   {\scriptsize($21.8\%$, $19.1\%$, $24.2\%$)}   \\
\hline
\multirow{2}{*}{Missing transverse momentum}        &   $E^\mathrm{miss}_\mathrm{T} > 10\,\GeV$            &    \multirow{2}{*}{\textemdash} \\
                                                    &   {\scriptsize($45.9\%$, $37.6\%$, $37.2\%$)}        &                                              \\
\hline
\multirow{2}{*}{Number of selected photons}         &   \multicolumn{2}{c}{$N_{\gamma}\geq 2$}                                                            \\
                                                    &   {\scriptsize($24.1\%$, $18.3\%$, $15.1\%$)}        &  {\scriptsize($12.5\%$, $10.4\%$, $9.9\%$)}     \\
\hline
\multirow{2}{*}{Invariant di-photon mass}           &   \multicolumn{2}{c}{$115\,\GeV < m_{\gamma\gamma} \leq 135\,\GeV $}            \\
                                                    &   {\scriptsize($21.7\%$, $16.3\%$, $5.8\%$)}        &  {\scriptsize($11.4\%$, $9.5\%$, $3.8\%$)}     \\
\hline
Transverse momentum of the leading photon           &   \multicolumn{2}{c}{$p_{\mathrm{T},\gamma_{1}}/m_{\gamma\gamma} > 0.35$}       \\
relative to the invariant di-photon mass            &   {\scriptsize($21.5\%$, $16.0\%$, $5.4\%$)}        &  {\scriptsize($11.3\%$, $9.3\%$, $3.4\%$)}     \\
\hline
Transverse momentum of the subleading photon        &   \multicolumn{2}{c}{$p_{\mathrm{T},\gamma_{2}}/m_{\gamma\gamma} > 0.25$}       \\
relative to the invariant di-photon mass            &   {\scriptsize($18.3\%$, $13.9\%$, $3.7\%$)}        &  {\scriptsize($9.6\%$, $8.0\%$, $2.3\%$)}     \\
\hline
\hline
\end{tabular}}
\label{tab:SelectionRequirements}
\end{table}
\clearpage

\begin{sidewaystable}[h!]
\caption{Selection requirements applied to event-level observables in the $H \rightarrow b\bar{b}$ analysis, alongside their corresponding
selection efficiencies.}
\centering
\resizebox{1.0\textwidth}{!}{
\begin{tabular}{c | c c }
\hline
\hline
Observable                         & \multicolumn{2}{c}{Selection requirement} \\
\hline
                                   & $W^\pm H$ Analysis   &  $ZH$ Analysis \\
                                   & {\scriptsize($W^\pm_\mathrm{L}H$, $W^\pm_\mathrm{T}H$, $W^\pm b \bar{b}$, $Z b \bar{b}$, $W^\pm Z$, $ZZ$, $t\bar{t}$, tb)}  & {\scriptsize($Z_\mathrm{L}H$, $Z_\mathrm{T}H$, $Z b\bar{b}$, $ZZ$, $t\bar{t}$)}  \\
\hline
\multirow{2}{*}{Number of selected charged leptons} &   $ N_{\ell}= 1$                                                                                       &        $N_{\ell}=2$  (SFOC)    \\
                                                    &   {\scriptsize($38.7\%$, $29.9\%$, $35.9\%$, $28.5\%$, $32.6\%$, $30.8\%$, $32.7\%$, $31.0\%$)}        &  {\scriptsize($24.2\%$, $20.9\%$, $27.5\%$ ,$25.0\%$, $2.9\%$)}   \\
\hline
\multirow{2}{*}{Invariant di-lepton mass}           &   \multirow{2}{*}{\textemdash}                       &  $81\,\GeV < m_{\ell\ell} \leq 101\,\GeV $               \\
                                                    &                                                      &   {\scriptsize($22.2\%$, $19.3\%$, $25.5\%$, $23.1\%$, $0.4\%$)}  \\
\hline
\multirow{2}{*}{Missing transverse momentum}        &   $E^\mathrm{miss}_\mathrm{T} > 30\,\GeV$ & $E^\mathrm{miss}_\mathrm{T} < 40\,\GeV$           \\
                                                    &   {\scriptsize($31.0\%$, $19.0\%$, $23.9\%$, $6.3\%$, $21.7\%$, $8.3\%$, $26.8\%$, $23.8\%$)}        &  {\scriptsize($18.7\%$, $17.5\%$, $24.2\%$, $21.8\%$, $0.08\%$}   \\
\hline
\multirow{2}{*}{Number of selected jets}            &    $2 \leq N^\mathrm{jets} \leq 3$                          & $N^\mathrm{jets}\geq 2$                      \\
                                                    &   {\scriptsize($17.6\%$, $10.8\%$, $16.7\%$, $3.6\%$, $12.7\%$, $4.8\%$, $8.1\%$, $17.2\%$)}        &  {\scriptsize($16.6\%$, $15.4\%$, $16.7\%$, $14.5\%$, $0.07\%$)} \\
\hline
\multirow{2}{*}{Number of selected $b$-tagged jets} &   \multicolumn{2}{c}{$N^{b\mathrm{-tags}} = 2$}                                                           \\
                                                    &   {\scriptsize($7.1\%$, $3.9\%$, $8.1\%$, $1.3\%$, $4.0\%$, $1.3\%$, $3.0\%$, $7.8\%$)}        &  {\scriptsize($7.1\%$, $6.3\%$, $7.0\%$, $5.1\%$, $0.04\%$)} \\
\hline
\multirow{2}{*}{Invariant di-jet mass}              &   \multicolumn{2}{c}{$80\,\GeV < m_{b\bar{b}} \leq 145\,\GeV $}            \\
                                                    &   {\scriptsize($6.1\%$, $3.4\%$, $3.7\%$, $0.5\%$, $2.4\%$, $0.7\%$, $1.1\%$, $2.8\%$)}        &  {\scriptsize($6.0\%$, $5.4\%$, $3.0\%$, $3.1\%$, $0.01\%$)} \\
\hline
\multirow{2}{*}{Transverse momentum of the leading $b$-jet}          &   \multicolumn{2}{c}{$p_{\mathrm{T},b_1}$ > 45\,\GeV}                                                               \\
                                                    &   {\scriptsize($5.9\%$, $3.2\%$, $3.3\%$, $0.4\%$, $2.1\%$, $0.6\%$, $1.0\%$, $2.6\%$)}        &  {\scriptsize($5.9\%$, $5.1\%$, $2.3\%$, $2.7\%$, $0.01\%$)} \\
\hline
\multirow{2}{*}{Azimuthal angular difference between the two $b$-jets} &   \multicolumn{2}{c}{$\Delta R_{b\bar{b}}(p_{\mathrm{T},b\bar{b}}) < 180/p_{\mathrm{T},b\bar{b}}+ 0.7$} \\
                                                      &   {\scriptsize($5.3\%$, $2.8\%$, $3.1\%$, $0.4\%$, $2.0\%$, $0.6\%$, $0.8\%$, $2.4\%$)}        &  {\scriptsize($5.0\%$, $4.5\%$, $2.1\%$, $2.6\%$, $0.01\%$)} \\
\hline
\multirow{8}{*}{Transverse momentum of vector boson}  &   \multicolumn{2}{c}{$p_{\mathrm{T},V} \leq 75\,\GeV$}  \\
                                                      &   {\scriptsize($1.5\%$, $1.9\%$, $2.1\%$, $0.2\%$, $1.1\%$, $0.3\%$, $0.4\%$, $1.5\%$)}                                & {\scriptsize($1.7\%$, $2.9\%$, $1.4\%$, $1.6\%$, $0.007\%$)}               \\
                                                      &   \multicolumn{2}{c}{$75 \,\GeV < p_{\mathrm{T},V} \leq 150\,\GeV$}  \\
                                                      &   {\scriptsize($2.1\%$, $0.8\%$, $0.7\%$, $0.1\%$, $0.6\%$, $0.2\%$, $0.4\%$, $0.9\%$)}                                & {\scriptsize($2.0\%$, $1.2\%$, $0.6\%$, $0.7\%$, $0.004\%$)}               \\
                                                      &   \multicolumn{2}{c}{$150 \,\GeV < p_{\mathrm{T},V} \leq 250\,\GeV$}  \\
                                                      &   {\scriptsize($1.1\%$, $0.2\%$, $0.2\%$, $0.02\%$, $0.2\%$, $0.06\%$, $0.05\%$, $0.06\%$)}                                & {\scriptsize($1.0\%$, $0.2\%$, $0.1\%$, $0.2\%$, $0.0001\%$)}                \\
                                                      &   \multicolumn{2}{c}{$p_{\mathrm{T},V} > 250\,\GeV$}  \\
                                                      &   {\scriptsize($0.5\%$, $0.03\%$, $0.05\%$, $0.002\%$, $0.06\%$, $0.02\%$, $0.004\%$, $0.003\%$)}                                 & {\scriptsize($0.4\%$, $0.04\%$, $0.008\%$,  $0.05\%$, $0\%$)}                \\
\hline
\hline
\end{tabular}}
\label{tab:SelectionRequirements_VHbb}
\end{sidewaystable}
\clearpage

\subsection{Polarisation-dependent observables}\label{sec:PolarisationObservables}

The precise measurement of the cross sections for the $W^\pm_{\text{L}}H$, $W^\pm_{\text{T}}H$, $Z_{\text{L}}H$, and $Z_{\text{T}}H$ 
processes depends on the ability to effectively separate longitudinally and transversely polarised vector bosons. This separation is
achieved by exploiting significant differences in the angular and kinematic distributions of the final-state particles.\par

\begin{figure}[!h]
\begin{center}
\includegraphics[width=0.425\textwidth]{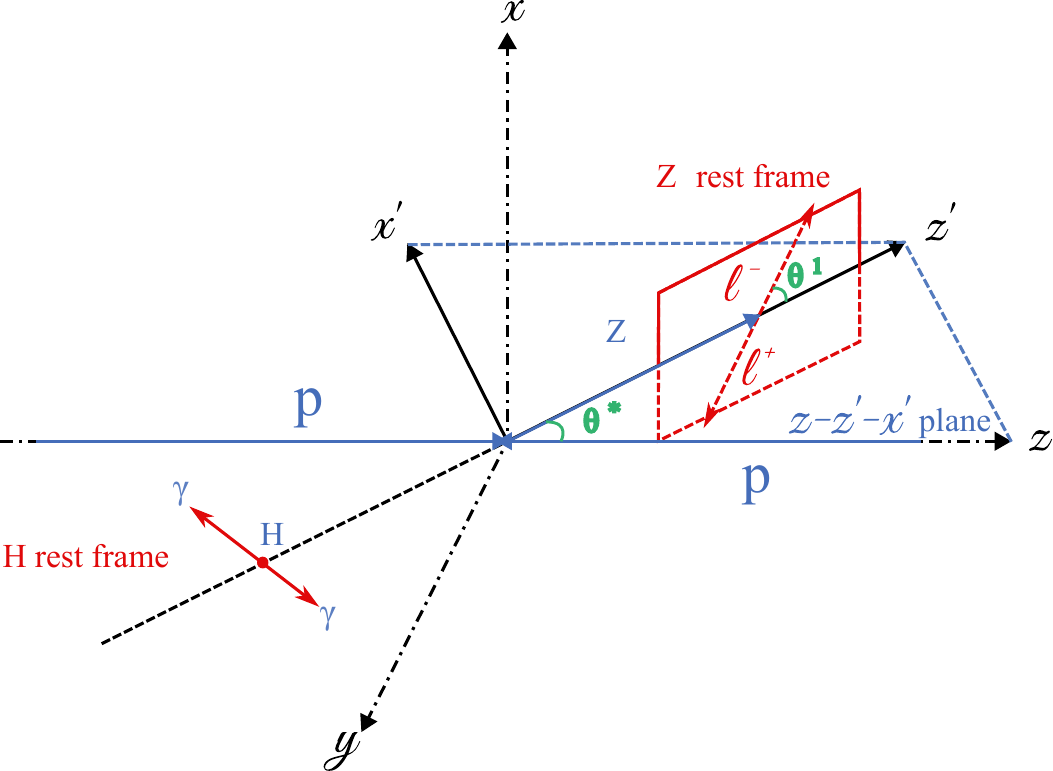}
\end{center}
\caption{Schematic representation of angular observables for separating longitudinally and transversely polarised
vector bosons, and the corresponding reference frames. The $xyz$ frame represents the laboratory
frame, with the $z$-axis aligned with the beam direction. The $z'$-axis is defined as the direction
of motion of the $Z$ boson in the rest frame of the four final state particles. The $x'$-axis defines the reaction 
plane containing the laboratory $z$-axis and the $z'$-axis. Due to the Higgs boson's Spin-0 nature, angles derived from the 
Higgs boson or its decay products do not provide information for distinguishing between longitudinally and transversely 
polarised vector bosons. The same relations apply for $W^\pm$ boson decays.}
\label{fig:VectorBosonDecayPlanes}
\end{figure}

\begin{figure}[!h]
\centering
\subfloat[]{\includegraphics[width=0.285\textwidth]{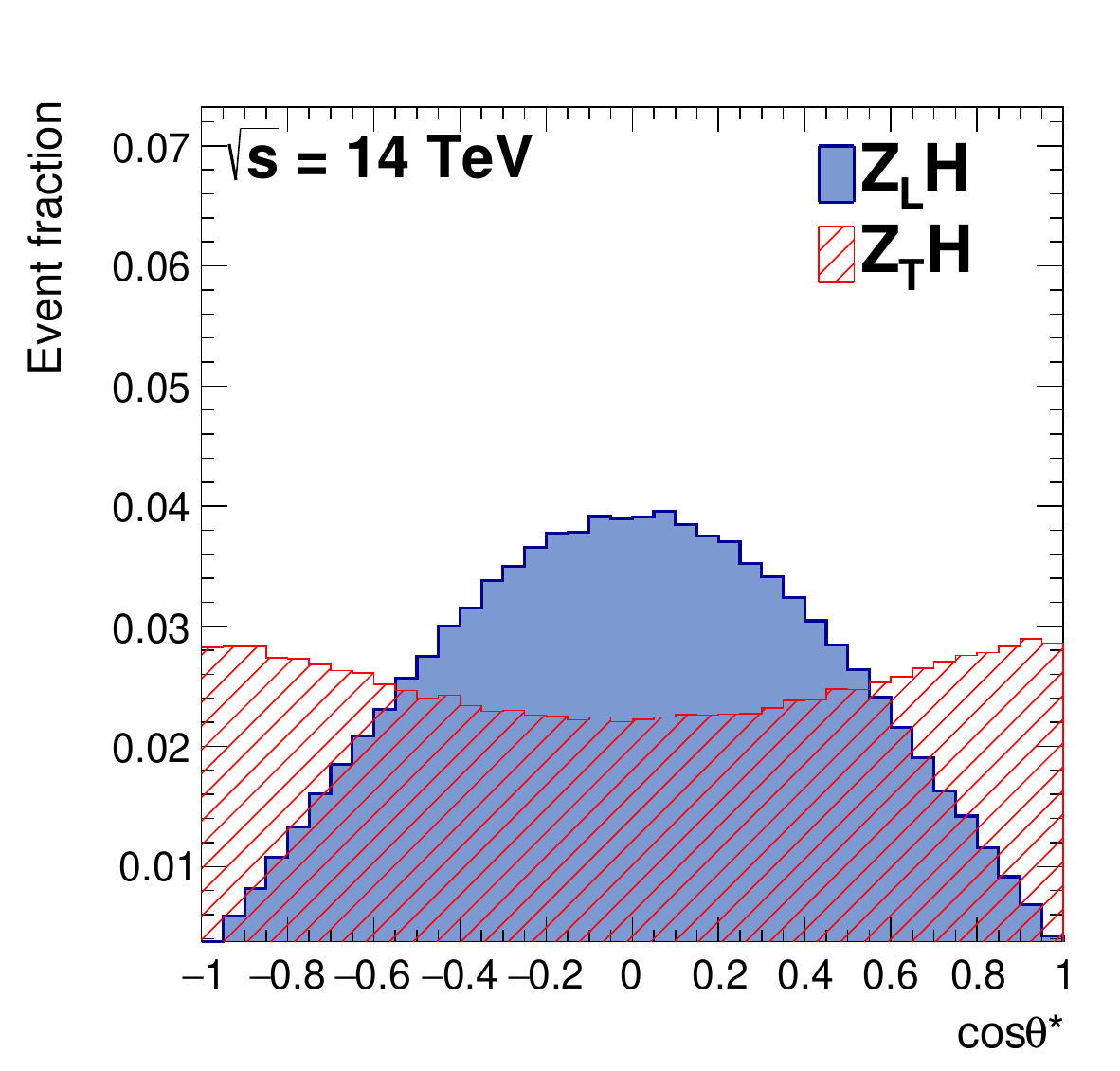}}
\subfloat[]{\includegraphics[width=0.285\textwidth]{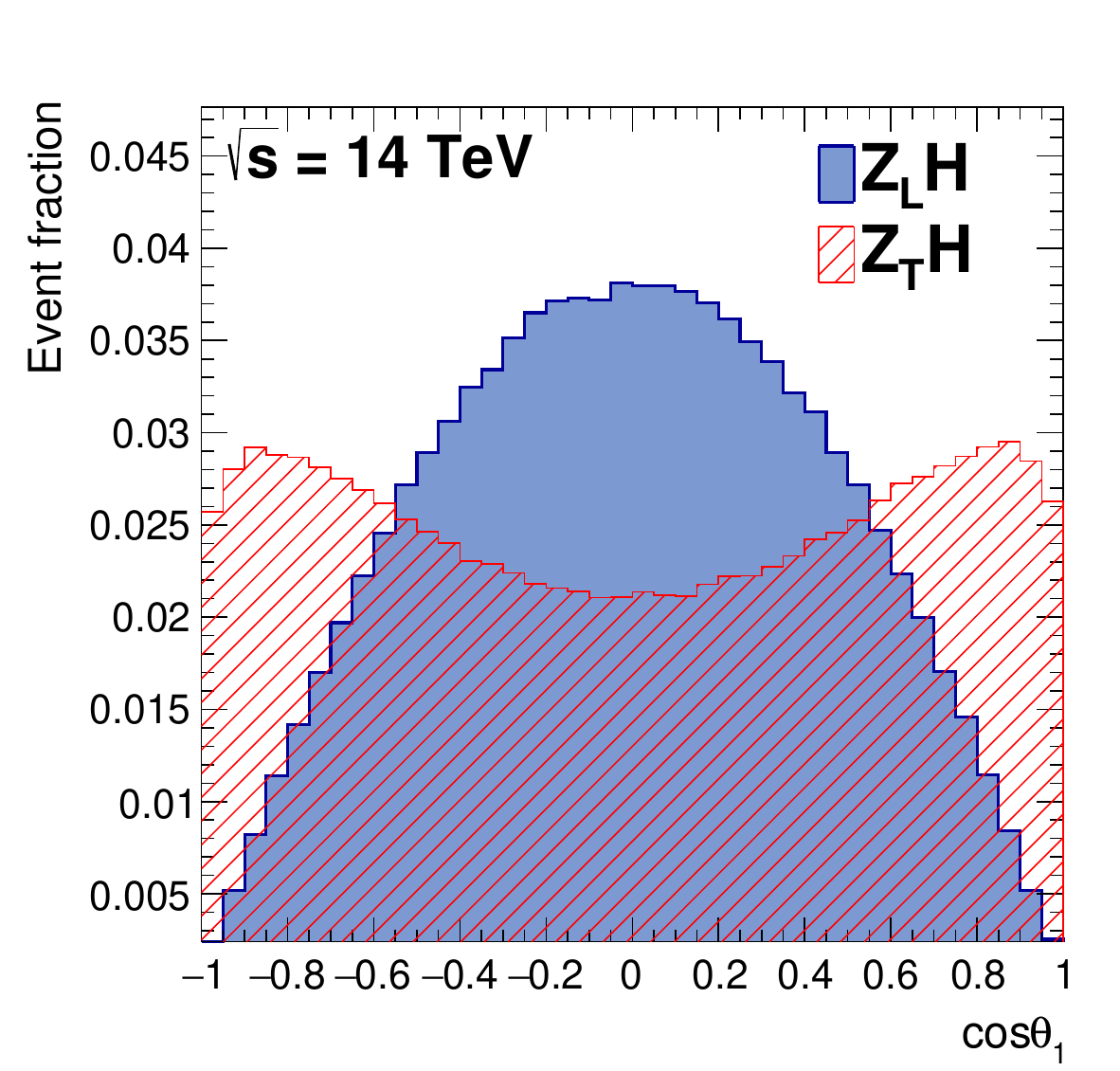}}\\
\subfloat[]{\includegraphics[width=0.285\textwidth]{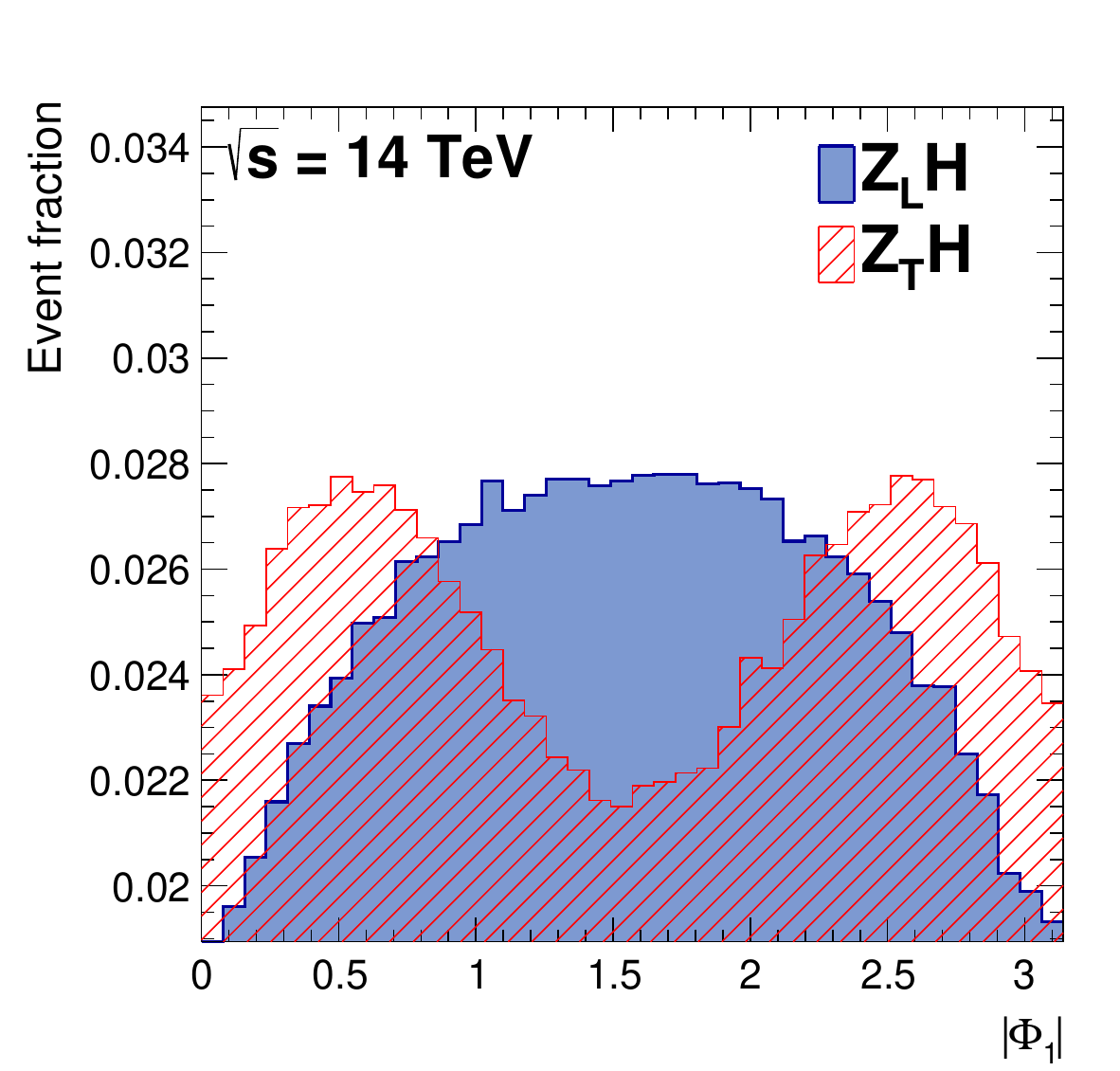}}
\subfloat[]{\includegraphics[width=0.285\textwidth]{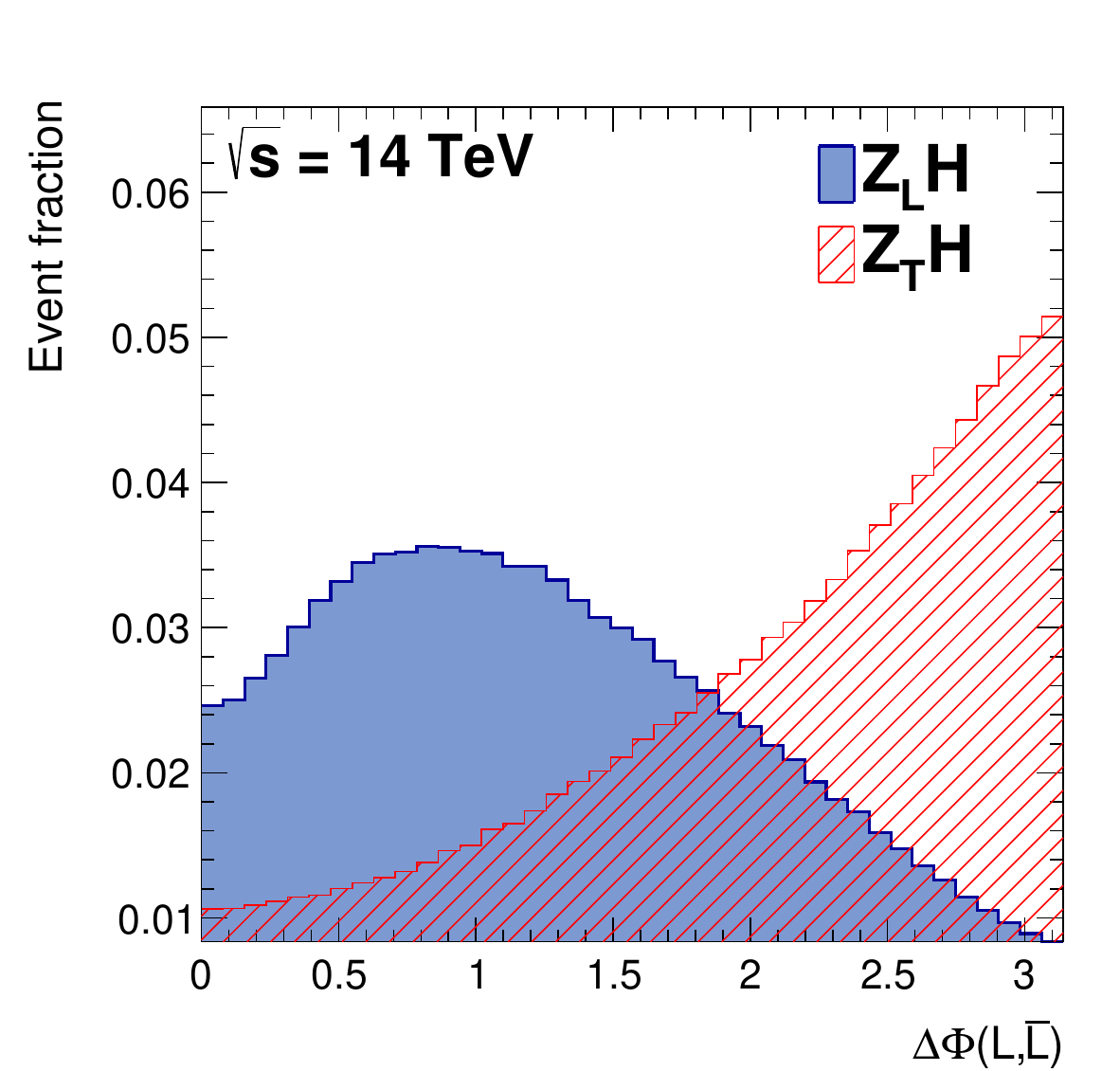}}
\caption{Distributions of the (a) $\cos\theta^{*}$, (b) $\cos\theta_{1}$, (c) $\Phi_{1}$, and
(d) $\Delta\Phi(L,\bar{L})$  observables for simulated $ZH\rightarrow \ell^+\ell^- \gamma\gamma$ 
events. The distributions are presented separately for events with longitudinally ($Z_\mathrm{L}H$) and transversely 
($Z_\mathrm{T}H$) polarised $Z$ bosons. All distributions are normalised to unit area.}
\label{fig:PolarisationObservables_I}
\end{figure}

Following previous studies~\cite{ATLAS:2023zrv,ATLAS:2013xga,Bolognesi:2012mm,Nakamura:2017ihk,Rao:2018abz,Banerjee:2018bio, Banerjee:2019pks}, we 
identified four angular observables that 
characterise the production and decay processes in $pp\rightarrow W^\pm H$ and $pp\rightarrow Z H$ events across all four covered final states. These 
observables comprise the production angle $\theta^{*}$ and three decay angles: $\theta_{1}$, $\Phi_{1}$, and $\Delta\Phi(L,\bar{L})$. The production 
angle $\theta^{*}$ is defined as the polar angle of the vector boson in the center-of-mass frame of the $VH$ system with respect to the beam axis.
The angle $\theta_{1}$ is defined as
\begin{equation}
\theta_{1} = \cos^{-1}\left(-\frac{\boldsymbol{p_{H}} \cdot \boldsymbol{p_{L}}}{|\boldsymbol{p_{H}}| |\boldsymbol{p_{L}}|}\right)\,,
\end{equation}
with all three-momenta evaluated in the vector boson rest frame. Thus, $\theta_{1}$ describes the angle between
the final-state lepton and the Higgs boson's direction in the vector boson rest frame. The azimuthal angle $\Phi_{1}$
between the vector boson's decay plane and the partonic scattering plane in the partonic rest frame is defined as
\begin{equation}
\Phi_{1} = \frac{\boldsymbol{p_{V}}\cdot \left(\boldsymbol{\hat{n}_{V}} \times \boldsymbol{\hat{n}_{sc}} \right)}{|\boldsymbol{p_{V}} \cdot \left(\boldsymbol{\hat{n}_{V}} \times \boldsymbol{\hat{n}_{sc}} \right) |} \times \cos^{-1}\left(\boldsymbol{\hat{n}_{V}} \cdot \boldsymbol{\hat{n}_{sc}}\right)\,,
\end{equation}
where the normal vectors to these two planes are given by
\begin{equation}
\boldsymbol{\hat{n}_{V}} = \frac{\boldsymbol{p_{L}} \times \boldsymbol{p_{\bar{L}}}}{|\boldsymbol{p_{L}} \times \boldsymbol{p_{\bar{L}}}|} \,\,\, \mathrm{and} \,\,\, \boldsymbol{\hat{n}_{sc}} = \frac{\boldsymbol{\hat{n}_{z}} \times \boldsymbol{p_{V}}}{|\boldsymbol{\hat{n}_{z}} \times \boldsymbol{p_{V}}|}\,. \nonumber
\end{equation}
Finally, the angle $\Delta\Phi(L,\bar{L})$ is defined as the azimuthal separation between the lepton and anti-lepton 
in the center-of-mass frame of the four final state particles. Fig.~\ref{fig:VectorBosonDecayPlanes} illustrates these 
production and decay angles, while Fig.~\ref{fig:PolarisationObservables_I} depicts representative distributions for 
these observables, shown separately for $Z_\mathrm{L}H$ and $Z_\mathrm{T}H$ events.\par

\begin{figure}[!h]
\centering
\subfloat[]{\includegraphics[width=0.305\textwidth]{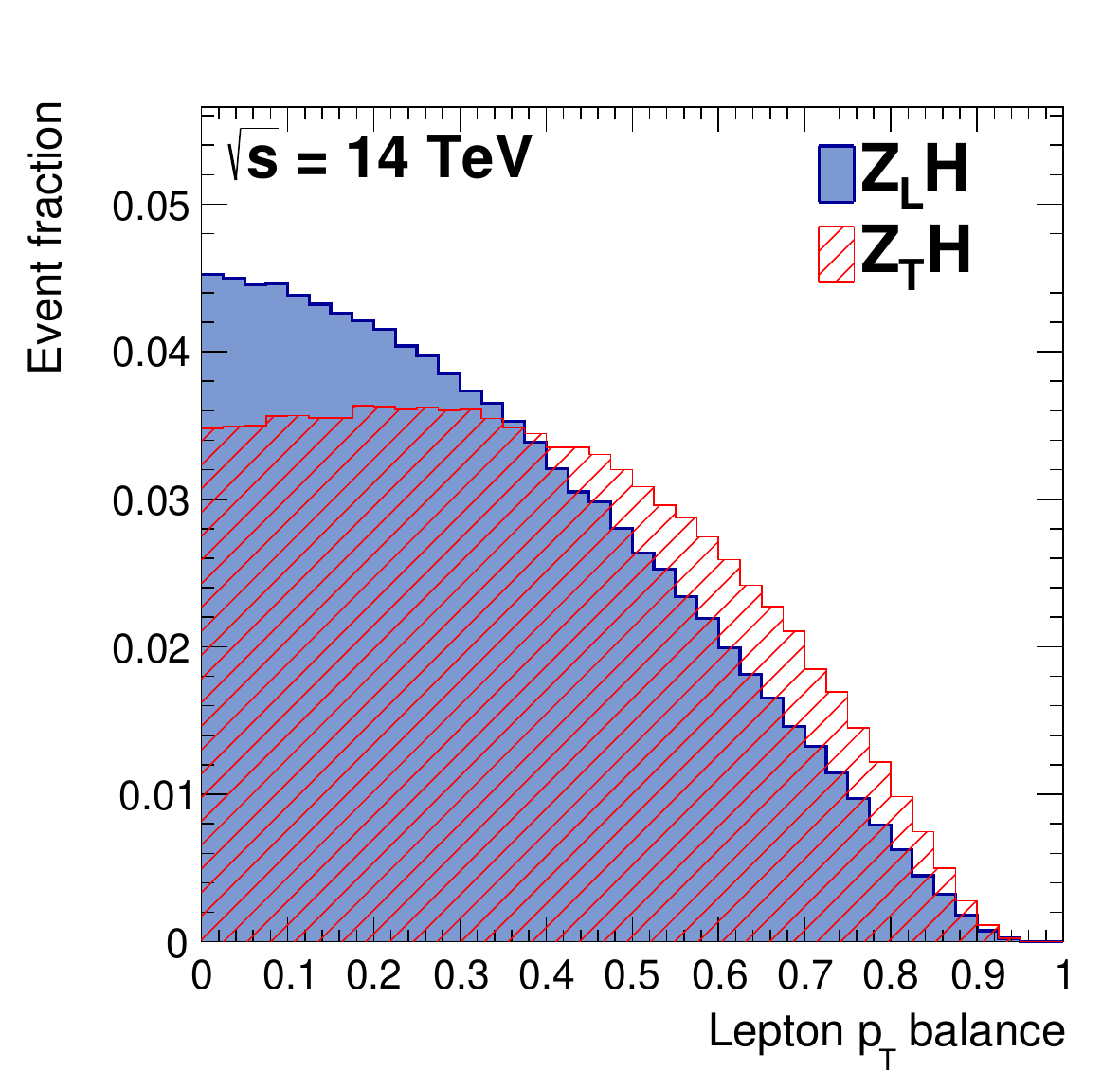}}
\subfloat[]{\includegraphics[width=0.305\textwidth]{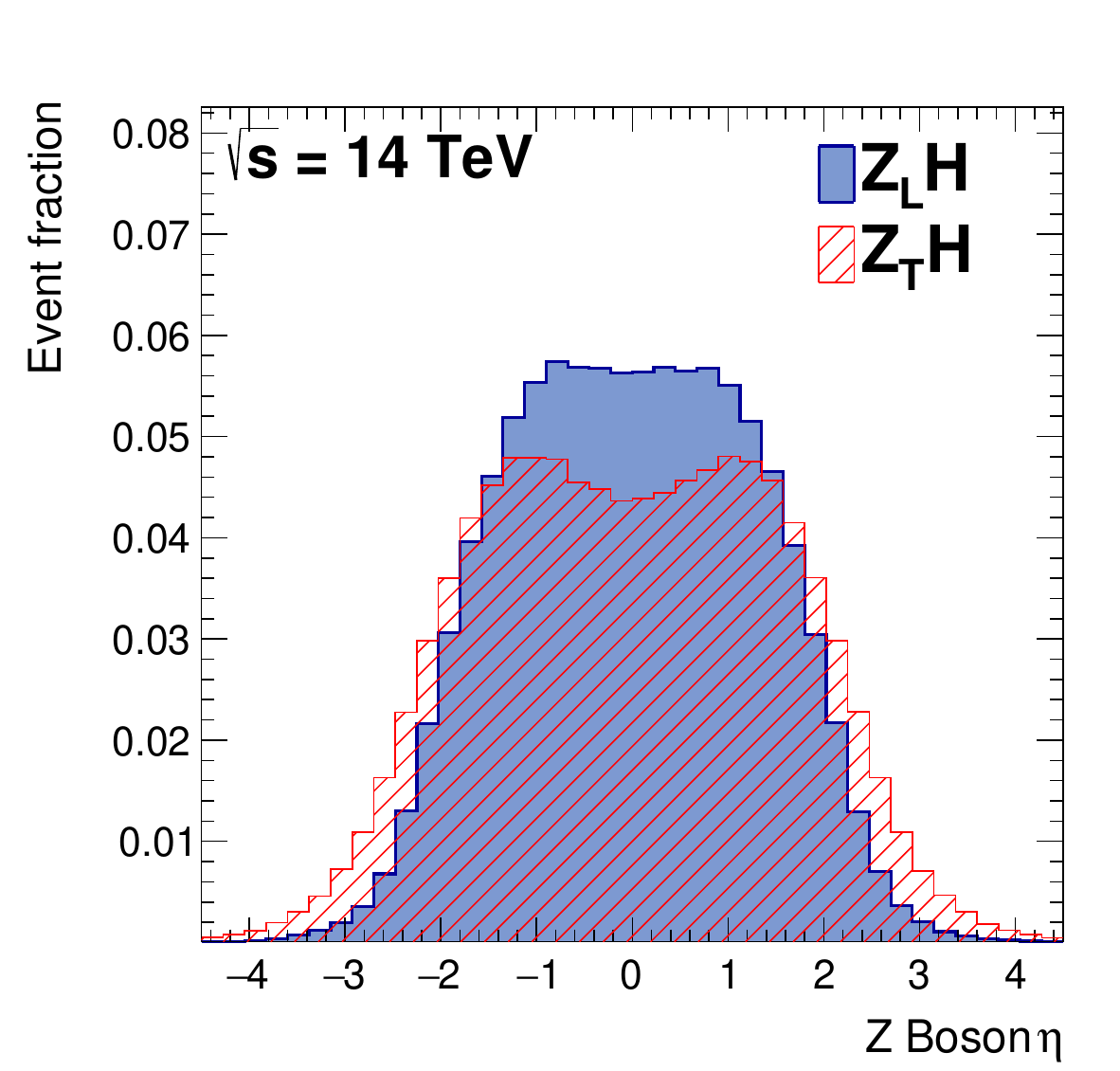}}
\subfloat[]{\includegraphics[width=0.305\textwidth]{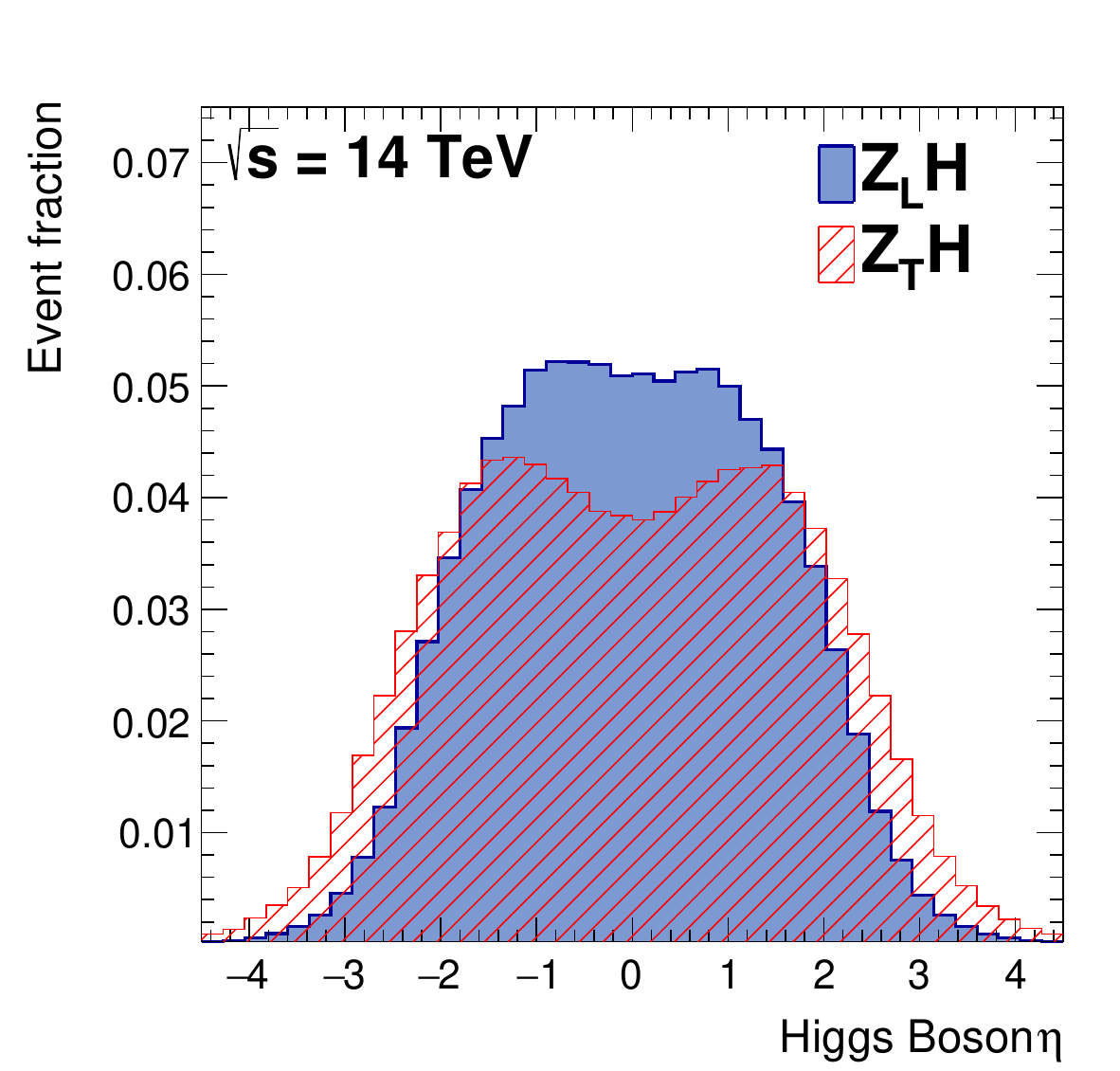}}\\
\subfloat[]{\includegraphics[width=0.305\textwidth]{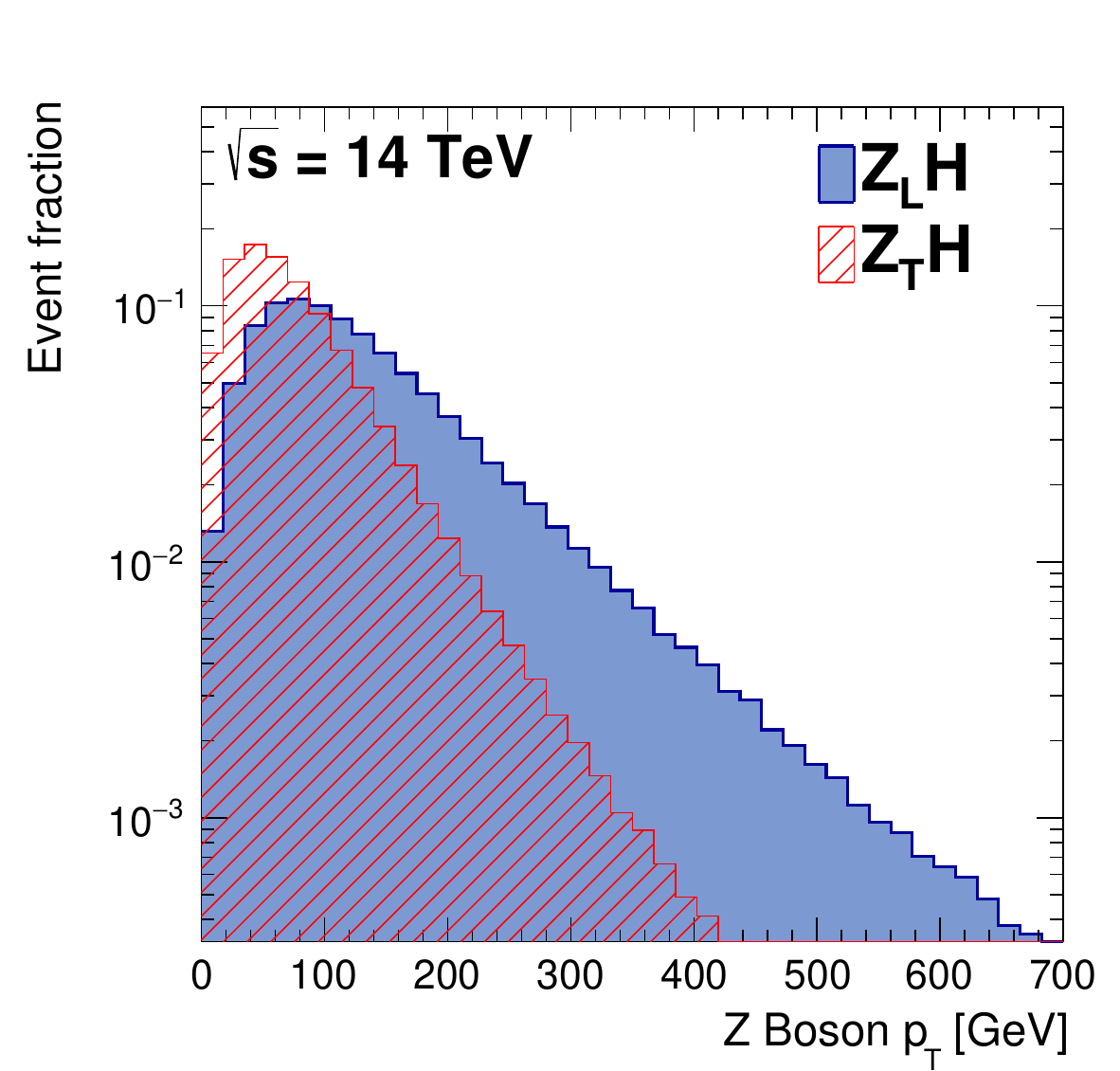}}
\subfloat[]{\includegraphics[width=0.305\textwidth]{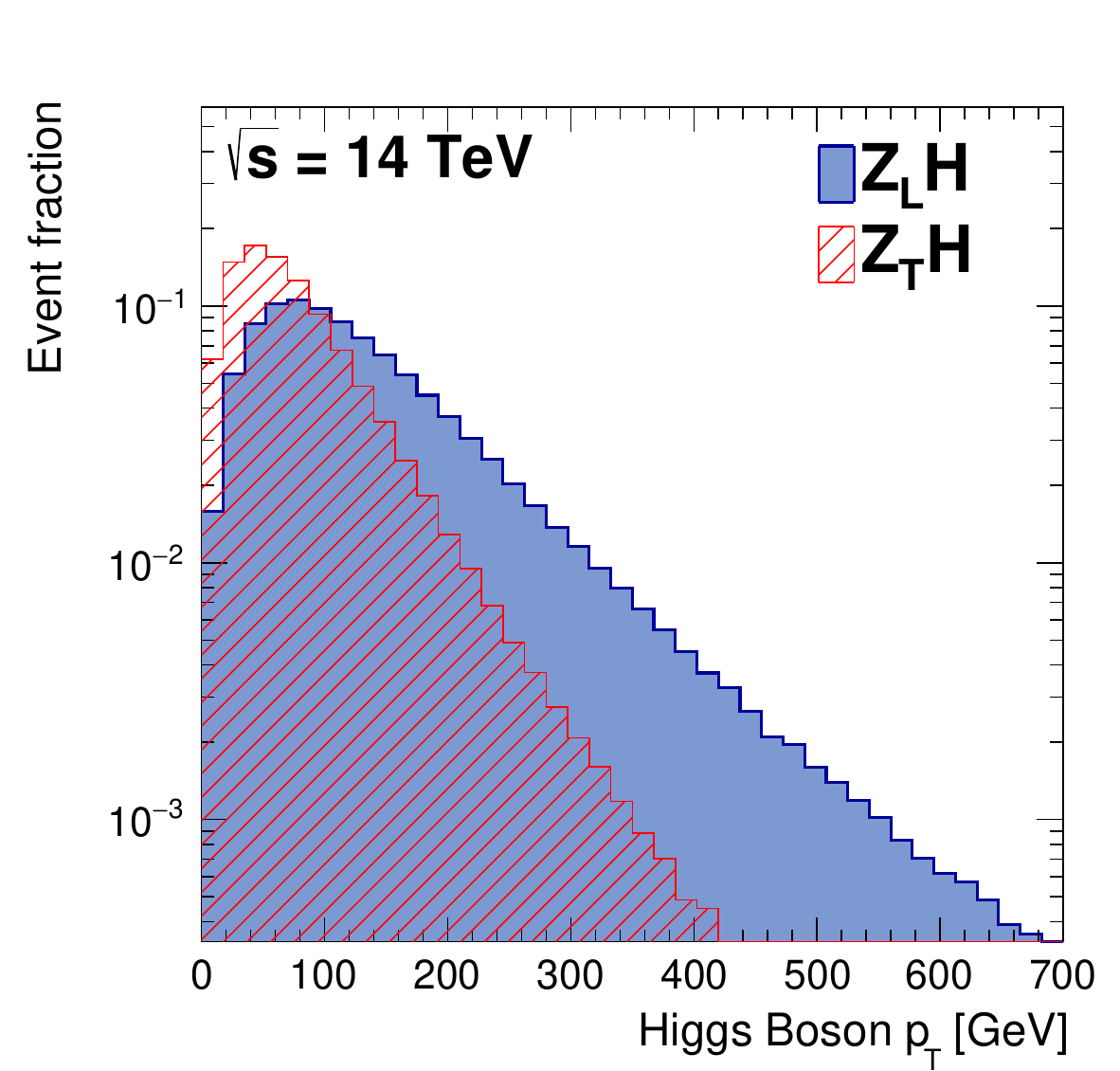}}
\subfloat[]{\includegraphics[width=0.305\textwidth]{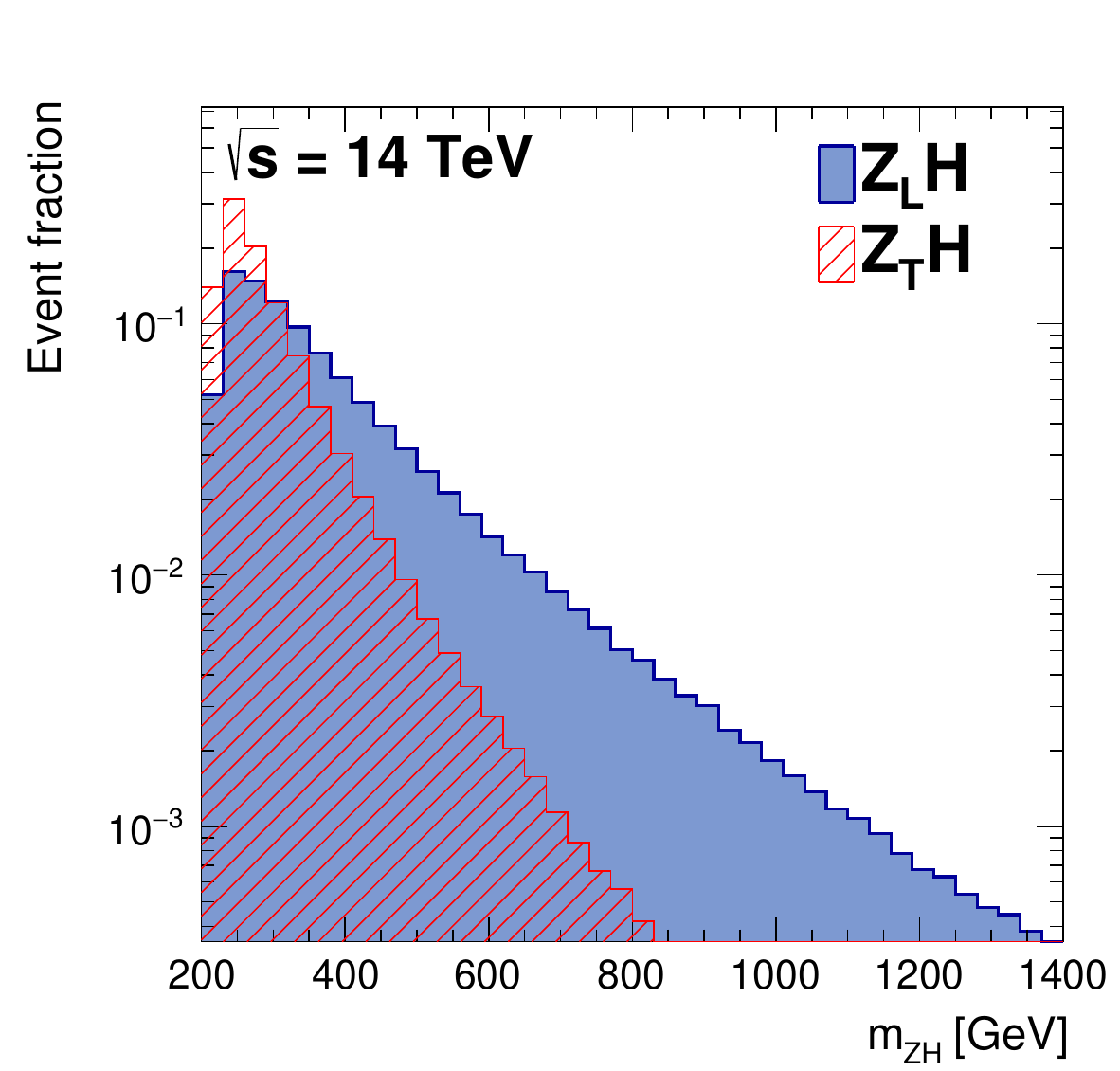}}
\caption{Distributions of the (a) lepton $p_\mathrm{T}$ balance, (b) $Z$ boson $\eta$, (c) Higgs boson $\eta$,
(d) $Z$ boson $p_\mathrm{T}$, (e) Higgs boson $p_\mathrm{T}$, and (f) invariant mass $m_{ZH}$ of the $ZH$ system
observables for simulated $ZH\rightarrow \ell^+\ell^- \gamma\gamma$ events. The distributions are presented 
separately for events with longitudinally ($Z_\mathrm{L}H$) and transversely ($Z_\mathrm{T}H$) polarised
$Z$ bosons. All distributions are normalised to unit area.}
\label{fig:PolarisationObservables_II}
\end{figure}

In addition, various kinematic observables provide substantial separation power between events with longitudinally and transversely 
polarised vector bosons. These include the transverse momentum ($p_\mathrm{T}$) and rapidity ($\eta$) of both the vector boson and 
the Higgs boson, as well as the invariant mass of the $VH$ system ($m_{VH}$). The lepton $p_\mathrm{T}$ balance, calculated as 
$\frac{p_\mathrm{T,L_{1}} - p_\mathrm{T,L_{2}}}{p_\mathrm{T,L_{1}} + p_\mathrm{T,L_{2}}}$, where $p_\mathrm{T,L_{1}}$ and 
$p_\mathrm{T,L_{2}}$ are the transverse momenta of the leading and subleading leptons, also contributes significantly. 
The separation power derived from the Higgs boson's kinematics originate from its recoil against the vector boson. Representative
distributions for these observables are depicted in Fig.~\ref{fig:PolarisationObservables_II} separately for 
$Z_\mathrm{L}H$ and $Z_\mathrm{T}H$ events. \par

These observables obtain their separation power from distinct kinematic behaviors. Longitudinally polarised vector bosons are preferentially 
produced centrally in the detector, unlike transversely polarised bosons which exhibit more forward or backward production. Furthermore, 
the decay products of longitudinally polarised vector bosons are characterised by a spin axis perpendicular to the vector boson's direction 
of motion, resulting in decay products predominantly emitted perpendicular to this direction. This contrasts with transversely polarised vector 
bosons, whose decay products are more frequently emitted parallel or antiparallel to the boson's direction.\par

\clearpage

Most of the polarisation-sensitive observables discussed above require the full reconstruction of the vector boson's four-momentum. In the 2-lepton
channels, this is calculated as the four-momentum sum of the two charged leptons. In the 1-lepton channels, the $W^\pm$ boson kinematics are
reconstructed by combining the charged lepton and the neutrino, where the neutrino four-momentum is determined using the on-shell $W$-boson mass
constraint described in Section~\ref{sec:EventSelection}. \par

\subsection{Training of the boosted decision trees}\label{sec:BDT_Training}

Dedicated sets of boosted decision trees ($\mathrm{BDT}_\mathrm{Pol}$) are trained on simulated $W^\pm H$ and $ZH$ events to 
separate longitudinally polarised vector bosons ($W^\pm_\mathrm{L}H$ and $Z_\mathrm{L}H$) from their transversely polarised 
counterparts ($W^\pm_\mathrm{T}H$ and $Z_\mathrm{T}H$). The training relies exclusively on the kinematic properties of the 
final-state particles and the reconstructed vector and Higgs bosons, without including background processes. \par
The $\mathrm{BDT}_\mathrm{Pol}$ classifiers are implemented into the analysis using the TMVA package~\cite{TMVA}. Each ensemble 
comprises 600 decision trees trained via the Gradient Boosting algorithm~\cite{GradientBoost} with a learning rate of 0.1 and a 
maximum tree depth of six. Ten input features are used, as listed in Table~\ref{tab:MVAFeatures} and detailed in 
Section~\ref{sec:PolarisationObservables}. Among these, the angles $\Delta\Phi(L, \bar{L})$, $\cos\theta^{*}$ and $\Phi_{1}$ 
exhibit the highest feature importance in discriminating between longitudinal and transverse polarisation states.\par
To prevent overtraining biases, a two-fold cross-validation scheme is adopted. Events are randomly split into two equal subsamples, 
$A$ and $B$. Independent BDT ensembles trained on sample $A$ are evaluated on sample $B$, and vice versa, with the resulting response 
distributions merged for the subsequent analysis.\par
This scheme produces four distinct $\mathrm{BDT}_\mathrm{Pol}$ classifiers for the $H\rightarrow\gamma\gamma$ decay channels (two each for 
the $W^\pm H$ and $ZH$ sub-channels). For the $H\rightarrow b\bar{b}$ decay channels, separate classifiers are trained across specific 
$p_{\mathrm{T},V}$ intervals. Specifically, four BDTs are trained for $W^\pm H(\rightarrow b\bar{b})$ across two bins 
($p_{\mathrm{T},V} < 75\,\GeV$ and $p_{\mathrm{T},V} \geq 75\,\GeV$), while six BDTs are trained for $ZH(\rightarrow b\bar{b})$ 
across three bins ($p_{\mathrm{T},V} < 75\,\GeV$, $75\,\GeV \leq p_{\mathrm{T},V} < 150\,\GeV$, and 
$p_{\mathrm{T},V} \geq 150\,\GeV$).\par
\begin{figure}[h!]
\centering
\subfloat[]{\includegraphics[width=0.5\textwidth]{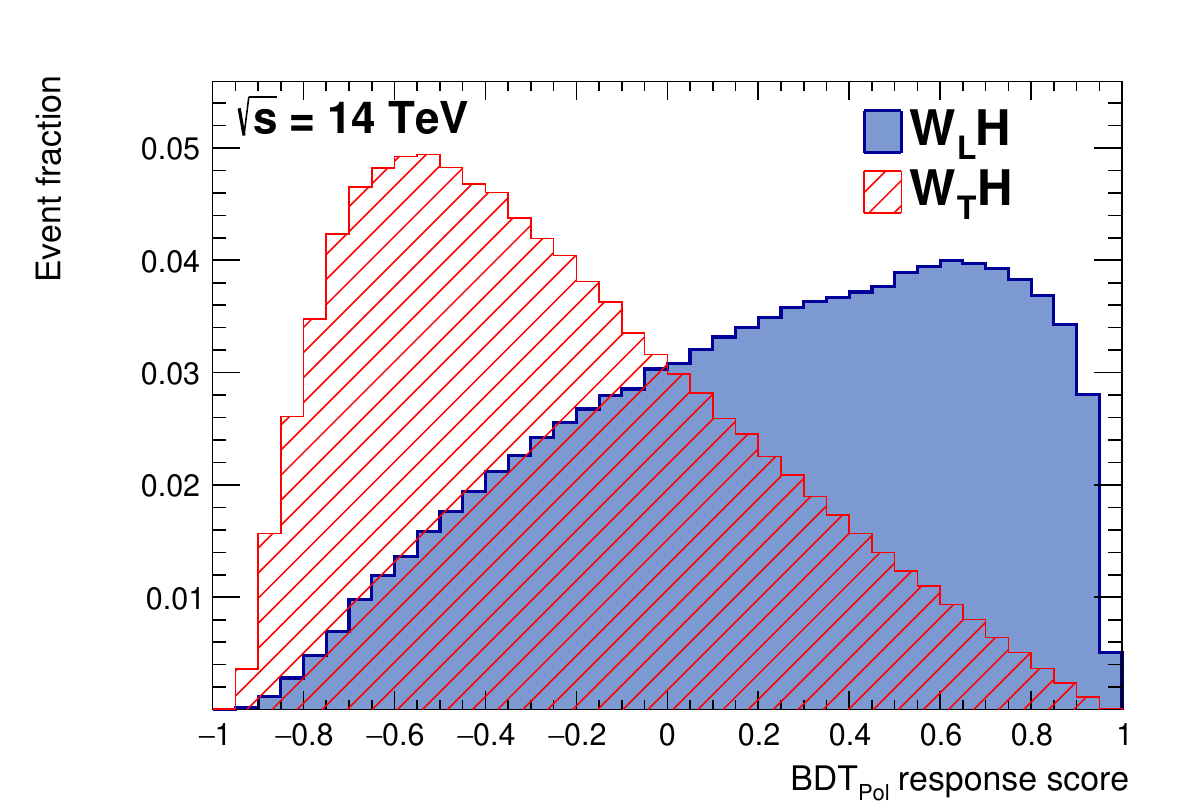}}
\subfloat[]{\includegraphics[width=0.5\textwidth]{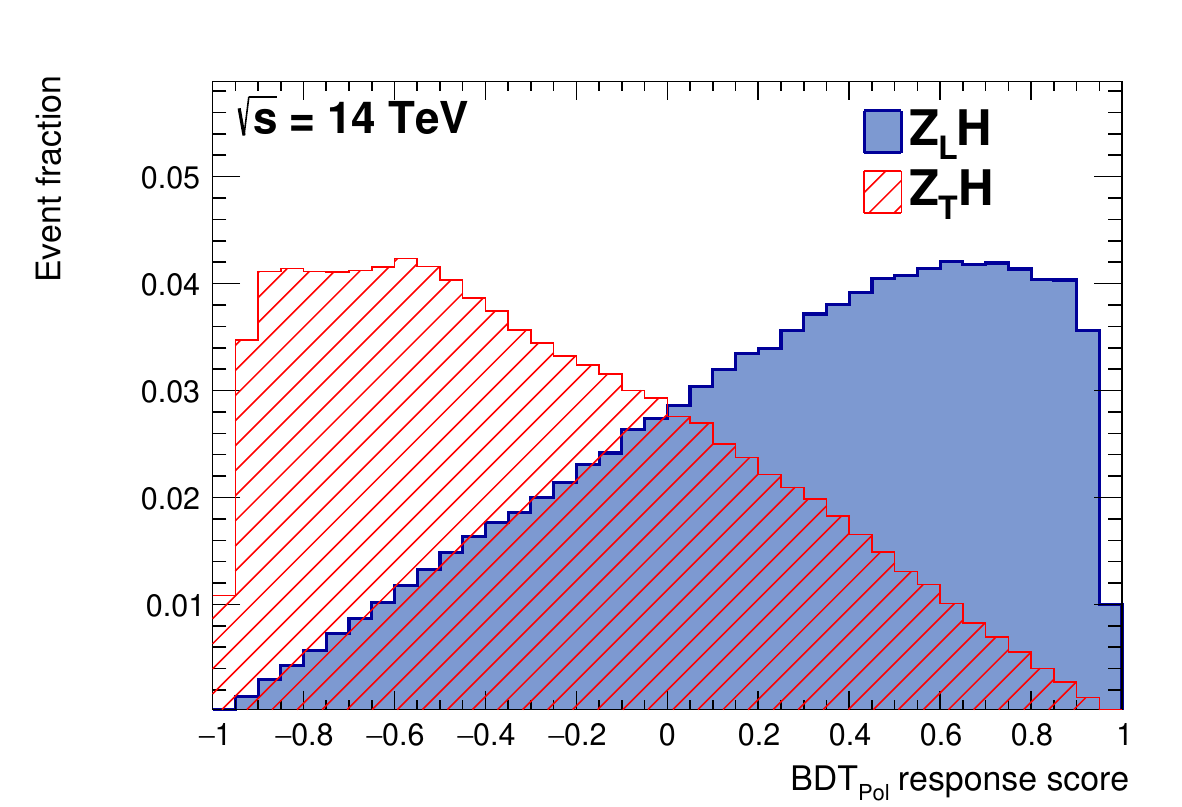}}
\vspace{0.125cm}
\caption{Distributions of $\mathrm{BDT}_\mathrm{Pol}$ response score observables for (a) simulated
$W^\pm H\rightarrow \ell^\pm\nu \gamma\gamma$ events, and (b) simulated $ZH\rightarrow \ell^\pm\ell^\mp \gamma \gamma$ events. 
The distributions are presented separately for events with longitudinally ($W^\pm_\mathrm{L}H$ and $Z_\mathrm{L}H$) 
and transversely ($W^\pm_\mathrm{T}H$ and $Z_\mathrm{T}H$) polarised vector bosons. All distributions are 
normalised to unit area.}
\label{fig:PolBDTScores}
\end{figure}

The normalised score distributions for the $H\rightarrow\gamma\gamma$ classifiers are shown in Figure~\ref{fig:PolBDTScores}. The response 
distributions for $H\rightarrow b\bar{b}$ events exhibit very similar separation characteristics.\par
The BDTs achieve strong separation between longitudinally and transversely polarised vector bosons in $H\rightarrow\gamma\gamma$ events, 
yielding Receiver Operating Characteristic (ROC) integrals of 0.78 for $W^\pm H$ and 0.80 for $ZH$. This minor difference stems from ambiguities 
in the reconstruction of the prompt neutrino in $W^\pm H$ events, which slightly degrades the resolution of the polarisation-sensitive observables. 
The separation power in the $H\rightarrow b\bar{b}$ channels tends to be lower, with ROC integrals ranging from 0.73 down to 0.70 for $ZH$ events and 
from 0.70 down to 0.65 for $W^\pm H$ events, decreasing monotonically with increasing $p_{\mathrm{T},V}$ values. This degradation in final states with 
$H\rightarrow b\bar{b}$ decays stems from unreconstructed neutrinos originating from semileptonic $b$-hadron decays, which blur the Higgs boson kinematic 
reconstruction while also degrading the $E_{\mathrm{T}}^{\mathrm{miss}}$ resolution required to accurately reconstruct the neutrino four-momentum from $W$ 
boson decays.

\begin{table}[h!]
\centering
\caption{Definitions of the observables employed in the boosted decision tree training to distinguish between 
longitudinally and transversely polarised vector bosons in $W^\pm H$ and $ZH$ events.}
\renewcommand{\arraystretch}{1.25}
\resizebox{1.0\textwidth}{!}{
\begin{tabular}{| c | c | c |}
\hline
\hline
                                      &  Symbol                                   & Definition \\
\hline
\multirow{8}{4cm}{Angular observables}  & \multirow{2}{*}{$\Delta\Phi(L,\bar{L})$}  &  Azimuthal separation between the lepton and anti-lepton \\
                                      &                                             &  in the center-of-mass frame of the reconstructed $VH$ system \\
\cline{2-3}
                                     & \multirow{2}{*}{$\cos\theta^{*}$}         &  Direction of flight of the $Z$ ($W^\pm$) boson  \\
                                     &                                           &  in the center-of-mass frame of the reconstructed $VH$ system \\
\cline{2-3}
                                     & \multirow{2}{*}{$\cos\theta_{1}$}         &  Angle between the lepton ($\ell$ or $\nu$) and the vector boson's \\
                                     &                                           &  direction in the vector boson rest frame \\
\cline{2-3}
                                     & \multirow{2}{*}{$\Phi_{1}$}               & Azimuthal angle between the vector boson decay plane and the $VH$ \\
                                     &                                           & production plane in the $VH$ center-of-mass frame \\
\hline
\multirow{7}{4cm}{Energy-dependent observables} & \multirow{2}{*}{$p_\mathrm{T}$ balance}   &  \multirow{2}{*}{$\displaystyle\frac{p_\mathrm{T,L_{1}} - p_\mathrm{T,L_{2}}}{p_\mathrm{T,L_{1}} + p_\mathrm{T,L_{2}}}$} \\
                                                &                                           &                                          \\
\cline{2-3}
                                     &  $p_{\mathrm{T},V}$                       &  Transverse momentum of the vector boson\\
\cline{2-3}
                                     & $\eta_V$                                  &  Pseudorapidity of the vector boson \\
\cline{2-3}
                                     & $p_{\mathrm{T},H}$                        &  Transverse momentum of the Higgs boson \\
\cline{2-3}
                                     & $\eta_H$                                  &  Pseudorapidity of the Higgs boson \\
\cline{2-3}
                                     & $m_{VH}$                                  &  Invariant mass of the reconstructed $VH$ system \\
\hline
\hline
\end{tabular}}
\label{tab:MVAFeatures}
\end{table}
\clearpage 

 \section{Statistical analysis}\label{sec:StatAna}

We derive our results by performing separate binned maximum-likelihood fits~\cite{Cowan:2010js} to the response score 
distributions of the $\mathrm{BDT}_{\mathrm{Pol}}$ classifier in the $W^\pm H(\rightarrow \gamma\gamma)$ and 
$ZH(\rightarrow \gamma\gamma)$ analysis channels. This approach follows the strategy adopted in measurements of 
$Z_{\mathrm{L}}Z_{\mathrm{L}}$ production in four-lepton final states~\cite{ATLAS:2026fhf,ATLAS:2023zrv}. \par
In contrast, for the $W^\pm H(\rightarrow b\bar{b})$ and $ZH(\rightarrow b\bar{b})$ channels, the observables input
to the likelihood fit are the $\mathrm{BDT}_{\mathrm{VH}}$ response score distributions across the four $p_{\mathrm{T},V}$
intervals, chosen to maximise sensitivity given the lower signal-to-background ratio and more complex background composition in final
states with $H\rightarrow b\bar{b}$ decays.
In the $ZH$ channel, a uniform threshold of $\mathrm{BDT}_{\mathrm{Pol}} > 0.0$ ($< 0.0$) defines the $V_{\mathrm{L}}$-like 
($V_{\mathrm{T}}$-like) categories across these intervals. In the $W^\pm H$ channel, $p_{\mathrm{T},V}$-dependent thresholds of 
$0.0$, $-0.1$, and $0.2$ are applied in the $p_{\mathrm{T},V} < 75\,\mathrm{GeV}$, $75\,\mathrm{GeV} \leq p_{\mathrm{T},V} < 150\,\mathrm{GeV}$, 
and $150\,\mathrm{GeV} \leq p_{\mathrm{T},V} < 250\,\mathrm{GeV}$ intervals, respectively. In the highest interval 
($p_{\mathrm{T},V} \geq 250\,\mathrm{GeV}$), no $V_{\mathrm{L}}$/$V_{\mathrm{T}}$ categorisation is performed in either channel 
due to the negligible contribution of $V_{\mathrm{T}}H$ events (as detailed in App.~\ref{sec:AppI}). This strategy loosely resembles 
the approach adopted in recent searches for $CP$-violating effects in $W^\pm H(\rightarrow b\bar{b})$ events~\cite{ATLAS:2026urg,ATLAS:2024yzu}. 
The fits in the $H\rightarrow b\bar{b}$ analysis channels are carried out separately for $W^\pm H$ and $ZH$ events, but simultaneously 
across all $p_{\mathrm{T},V}$ intervals.\par
Our statistical model is implemented using the \textsc{HistFactory} workspace schema~\cite{HistFactory}, with the \textit{pyhf} 
framework~\cite{PYHF} employed to perform the likelihood minimisation. The key aspects of this procedure are summarised below.
\clearpage

\subsection{Likelihood function and test statistic}

The likelihood function, $\mathcal{L}(\mathbf{N} | \mu_{i,\mathrm{L}}, \mu_{i,\mathrm{T}}, \mathbf{\Theta})$, depends on the 
parameters of interest (PoIs), $\mu_{i,\mathrm{L}}$ and $\mu_{i,\mathrm{T}}$, corresponding to the signal strengths of the 
longitudinally and transversely polarised vector boson production modes ($V_{\mathrm{L}}H$ and $V_{\mathrm{T}}H$) with 
$i \in \{W^\pm H, ZH\}$, as well as a set of nuisance parameters $\mathbf{\Theta} = \{\Theta_{1}, \Theta_{2}, \dots\}$~\cite{Conway:2011in} 
accounting for experimental and theoretical uncertainties. Given the observed or expected bin yields $\mathbf{N} = \{N_{1}, N_{2}, \dots\}$, 
the likelihood function is constructed as a product of three probability distribution functions: 

\begin{equation*}
\begin{split}
\mathcal{L}\left(\mu_{i,\mathrm{L}}, \mu_{i,\mathrm{T}}, \mathbf{\Theta}\right) = {} & \prod_{j}^{N^\mathrm{bins}} P\left( N_{j} \left| \sum_{s \in \{\mathrm{L}, \mathrm{T}\}} \mu_{i,s} S_{j,s}(\mathbf{\Theta}) + \sum_{k}^{N^\mathrm{bkgs}} B_{jk}(\mathbf{\Theta}) \right.\right) \\
& \cdot \prod_{l}^{N^\mathrm{syst}} G\left(\upsilon_{l} \mid \Theta_{l}\right) \cdot \prod_{m}^{N^\mathrm{bins}} P\left(\xi_{m} \mid \zeta_m\right) \,. 
\end{split}
\end{equation*}

The first term is the product of Poisson probabilities for observing $N_{j}$ events in the $j$-th bin of the fit input distribution, 
given the sum of longitudinally and transversely polarised contributions $\sum_{s \in \{\mathrm{L}, \mathrm{T}\}} \mu_{i,s} S_{j,s}(\mathbf{\Theta})$ 
and the expected sum of background yields $\sum_{k}^{N^\mathrm{bkgs}} B_{jk}(\mathbf{\Theta})$. \par
The second term is the product of Gaussian functions, each constraining a nuisance parameter that accounts for an 
experimental or theoretical uncertainty. Each function takes the form $G\left(\upsilon_l | \Theta_l\right)=e^{-(\upsilon_l-\Theta_l)^2/2}/\sqrt{2\pi}$, 
where $\upsilon_l$ represents the central value of a measurement and $\Theta_l$ is the associated systematic variation from that 
central value. \par
Finally, the third term accounts for uncertainties arising from the finite size of the simulated signal and background 
Monte Carlo samples~\cite{Barlow:1993dm}. These statistical uncertainties are modeled via additional Poisson constraint terms, 
where $\xi_m$ and $\zeta_m$ represent the central value of a yield estimate and its corresponding statistical variation, respectively.\par
Given that we do not have collision data for our studies, we build Asimov datasets~\cite{Cowan:2010js} for each analysis channel
as the sum of the signal and the relevant background templates. \par 
The profiled likelihood-ratio test statistic~\cite{Cowan:2010js} is used to evaluate the compatibility of the background-only and
background-and-signal hypotheses. It is defined as

\begin{equation}\label{eq:NLL}
q(\mu_{i,\mathrm{L}}, \mu_{i,\mathrm{T}}) = -2 \ln\left( \frac{\mathcal{L}(\mu_{i,\mathrm{L}}, \mu_{i,\mathrm{T}}, \mathbf{\hat{\hat{\Theta}}}_{\mu_{i,\mathrm{L}}, \mu_{i,\mathrm{T}}})}{\mathcal{L}(\hat{\mu}_{i,\mathrm{L}}, \hat{\mu}_{i,\mathrm{T}}, \mathbf{\hat{\Theta}})} \right)\,.
\end{equation}

The denominator of Eq.~\ref{eq:NLL} represents the unconditional maximum of the likelihood function, maximised simultaneously over all possible
values for the signal strengths and the nuisance parameters $\mathbf{\Theta}$, yielding best-fit values $\hat{\mu}_{i,\mathrm{L}}$,
$\hat{\mu}_{i,\mathrm{T}}$, and $\mathbf{\hat{\Theta}}$. The numerator is maximised over $\mathbf{\Theta}$ for fixed, conditional values of
$\mu_{i,\mathrm{L}}$ and $\mu_{i,\mathrm{T}}$, where the conditional maximum-likelihood estimates of the nuisance parameters are denoted as
$\mathbf{\hat{\hat{\Theta}}}_{\mu_{i,\mathrm{L}}, \mu_{i,\mathrm{T}}}$.\par

The $p_0$ value is calculated independently for each polarisation state using the test statistic $q(\mu_{i,\mathrm{L}}=0)$ or $q(\mu_{i,\mathrm{T}}=0)$, 
evaluated at zero signal strength for the state under study. It represents the probability of obtaining a test statistic value at least as large as 
the observed value under the background-only hypothesis for that polarisation state. All $p_0$ values are calculated using the asymptotic 
approximation~\cite{Cowan:2010js}, which assumes that the negative log-likelihood (NLL) behaves like a $\chi^{2}$ distribution. Finally, the local 
significance ($z_{0}$) is determined as the one-sided tail of a Gaussian distribution using $z_0 = \sqrt{2} \, \mathrm{erf}^{-1}(1-2p_0)$. \par

Different fit configurations are performed to extract the signal-strength parameters for longitudinally and transversely polarised 
vector bosons. One-dimensional fits are carried out for each signal strength ($\mu_{i,\mathrm{L}}$ and $\mu_{i,\mathrm{T}}$) 
independently, keeping the unprobed component fixed to its Standard Model expectation of unity. In addition, two-dimensional fits 
are performed where both parameters float freely. The maximum-likelihood procedure is applied individually to these configurations, 
generating negative log-likelihood (NLL) curves as a function of the relevant signal-strength parameter. Best-fit values are determined 
from the NLL minima, and central confidence level (CL) intervals are extracted at $2\Delta\mathrm{NLL} = 1$ ($1\sigma$) and 
$2\Delta\mathrm{NLL} = 3.84$ ($2\sigma$).\par

\subsection{Systematic and statistical uncertainties}
For our analysis, we use simplified fit models that incorporate a small set of experimental and theory uncertainties impacting the 
yields of the signal and background processes. These uncertainties enter the ML fits as nuisance parameters.\par
The uncertainties in the state-of-the-art cross-section predictions for the $W^\pm H$ and $ZH$ processes are 
taken from Ref.~\cite{LHCHiggsCrossSectionWorkingGroup}. They account for modelling effects on the PDF sets
and $\alpha_{S}$, as well as variations of the renormalisation and factorisation scales ($\mu_{R}$ and $\mu_{F}$) 
to cover uncertainties from missing higher orders. For the $W^\pm H$ production mode, these uncertainties are $^{+0.4\%}_{-0.7\%}$ (QCD scales) 
estimated by independently scaling $\mu_{R}$ and $\mu_{F}$ up and down by a factor of three, and $\pm1.8\%$ (PDF + $\alpha_{S}$) 
calculated following the \textit{PDF4LHC} recommendations~\cite{Butterworth:2015oua}. For the $ZH$ production mode, these 
uncertainties are $^{+3.8\%}_{-3.3\%}$ (QCD scales) and $\pm1.6\%$ (PDF + $\alpha_{S}$), respectively.\par
Theory uncertainties on the $H\rightarrow \gamma\gamma$ and $H\rightarrow b\bar{b}$ branching ratios are based on calculations from the 
\textsc{HDECAY} and \textsc{Prophecy4f} programmes~\cite{LHCHiggsCrossSectionWorkingGroup}. For the $H\rightarrow \gamma\gamma$ decay channel, 
uncertainties of $^{+1.73\%}_{-1.72\%}$ (missing higher-order corrections), $^{+0.93\%}_{-0.99\%}$ (quark mass effects), and 
$^{+0.61\%}_{-0.62\%}$ ($\alpha_{S}$ dependence) are assigned. For the $H\rightarrow b\bar{b}$ decay channel, the corresponding uncertainties are 
$\pm 0.65\%$, $^{+0.72\%}_{-0.74\%}$, and $^{+0.78\%}_{-0.80\%}$, respectively. These uncertainties on the production cross sections and branching ratio 
are treated as correlated between events with longitudinally and transversely polarised vector bosons. \par
In the $H\rightarrow\gamma\gamma$ analysis channels, an uncertainty of $\pm 25\%$ is assigned to the normalisations of the $W^\pm\gamma\gamma$ and 
$Z\gamma\gamma$ background processes. This uncertainty is comparable to the precision of the latest ATLAS production cross-section measurements for 
these processes~\cite{ATLAS:2022wmu,ATLAS:2023avk}.\par
In the $H\rightarrow b\bar{b}$ analysis channels, background normalisation uncertainties are derived from recent ATLAS $VH(\rightarrow b\bar{b})$ 
studies~\cite{ATLAS:2024yzu}. For the 1-lepton channel, uncertainties of $\pm 7\%$ are assigned to the $W^\pm/Z b\bar{b}$ background, 
$^{+13\%}_{-11\%}$ to the diboson backgrounds, and $\pm 4\%$ to backgrounds involving top quarks. For the 2-lepton channel, an uncertainty of 
$\pm 6\%$ is assigned to the $Z b\bar{b}$ background, while the diboson and top-quark background normalisations have uncertainties 
of $^{+13\%}_{-11\%}$ and $\pm 4\%$, respectively. \par
Relevant experimental uncertainties for the probed final states include those associated with trigger efficiencies, photon and charged-lepton reconstruction, 
identification, and isolation criteria, as well as the lepton and photon energy scale and resolution. Additionally, uncertainties in the missing transverse 
momentum reconstruction, the jet energy scale, jet energy resolution, and flavour-tagging efficiencies are relevant for the targeted channels. Given the high 
precision with which these uncertainties are constrained by the ATLAS and CMS experiments~\cite{EGAM-2018-01,MUON-2018-03,MUON-2022-01,EGAM-2021-02,CMS-MUO-21-001,CMS-EGM-17-001,JETM-2018-05,ATLAS:2026bvi}, 
their overall impact on the sensitivity is expected to be subdominant compared to the modelling uncertainties in the signal and background processes and 
the expected statistical uncertainty. Consequently, we neglect these uncertainties in our studies. The only experimental uncertainty we consider is 
a global luminosity uncertainty of $1\%$, which aligns with recent luminosity calibration by the ATLAS collaboration~\cite{DAPR-2021-01}. 
This uncertainty is applied to the normalisation of the $W^\pm_\mathrm{L}H$, $W^\pm_\mathrm{T}H$, $Z_\mathrm{L}H$, $Z_\mathrm{T}H$, 
and the background templates, treated as correlated across all processes. \par
The uncertainties due to the finite statistics of the Monte Carlo samples range from $0.2\%$ to $1.0\%$ per bin of the $\mathrm{BDT}_\mathrm{Pol}$ score 
distribution for the $W^\pm H(\rightarrow \gamma\gamma)$ and $ZH(\rightarrow \gamma\gamma)$ processes, and from $0.6\%$ to $4.0\%$ for the 
$W^\pm\gamma\gamma$ and $Z\gamma\gamma$ processes. For the $H \rightarrow b\bar{b}$ channels, where the background composition and sample statistics 
vary significantly across the analysis regions, the Monte Carlo statistical uncertainties span a slightly wider range. For the $ZH(\rightarrow b\bar{b})$ 
channel, the statistical uncertainties on the signal process distributions range from $0.7\%$ to $4.5\%$ for $Z_\mathrm{L}H$ and from $0.8\%$ to $5.0\%$ 
for $Z_\mathrm{T}H$, while the statistical uncertainties on the background templates range from $0.8\%$ to $3.5\%$ for $Z+b\bar{b}$, $0.7\%$ to $6.2\%$ 
for diboson, and $3.7\%$ to $12.0\%$ for top-quark processes. Similarly, for the $W^\pm H(\to b\bar{b})$ channel, the statistical uncertainties on the 
signal processes range from $1.1\%$ to $4.9\%$ for $W_\mathrm{L}H$ and from $1.0\%$ to $8.0\%$ for $W_\mathrm{T}H$ production, with background 
uncertainties ranging from $0.6\%$ to $6.0\%$ for $V+b\bar{b}$, $1.0\%$ to $7.8\%$ for diboson, and $0.4\%$ to $10.2\%$ for top-quark events.

\section{Results}\label{sec:Results}

\begin{figure}[!h]
\centering
\subfloat[]{\includegraphics[width=0.5\textwidth]{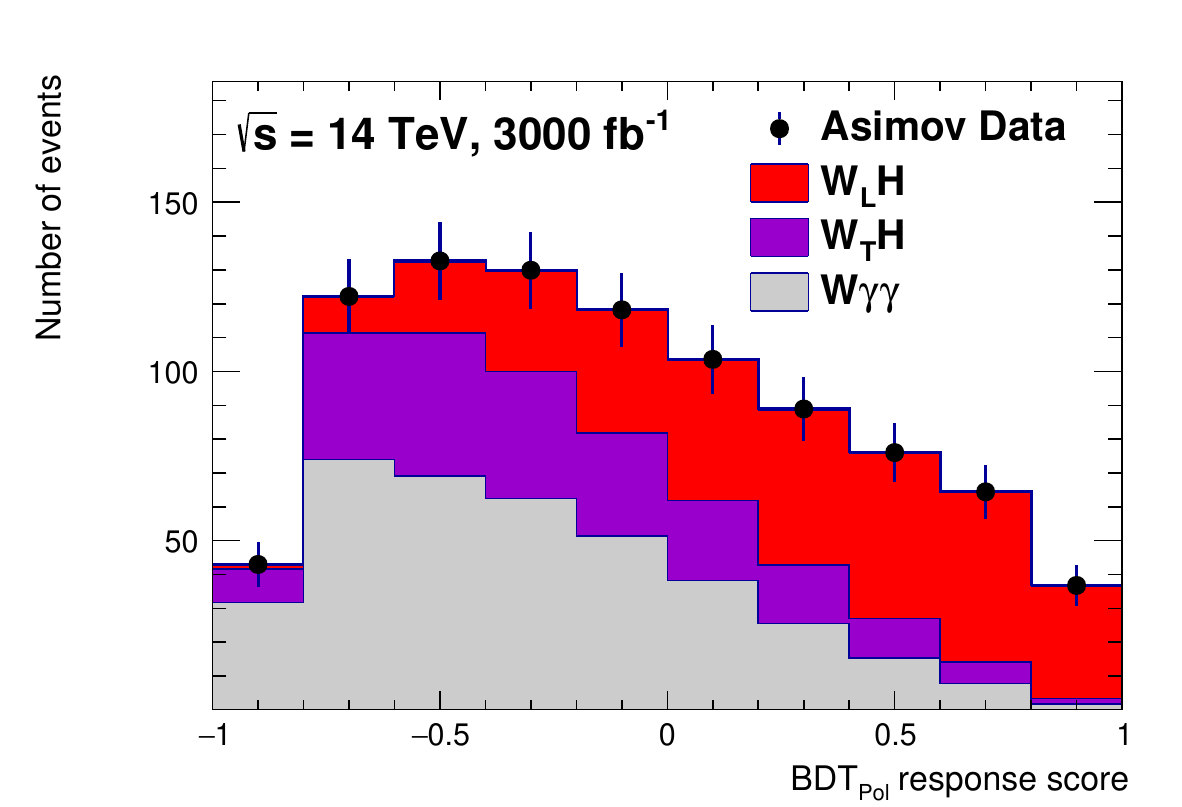}}
\subfloat[]{\includegraphics[width=0.5\textwidth]{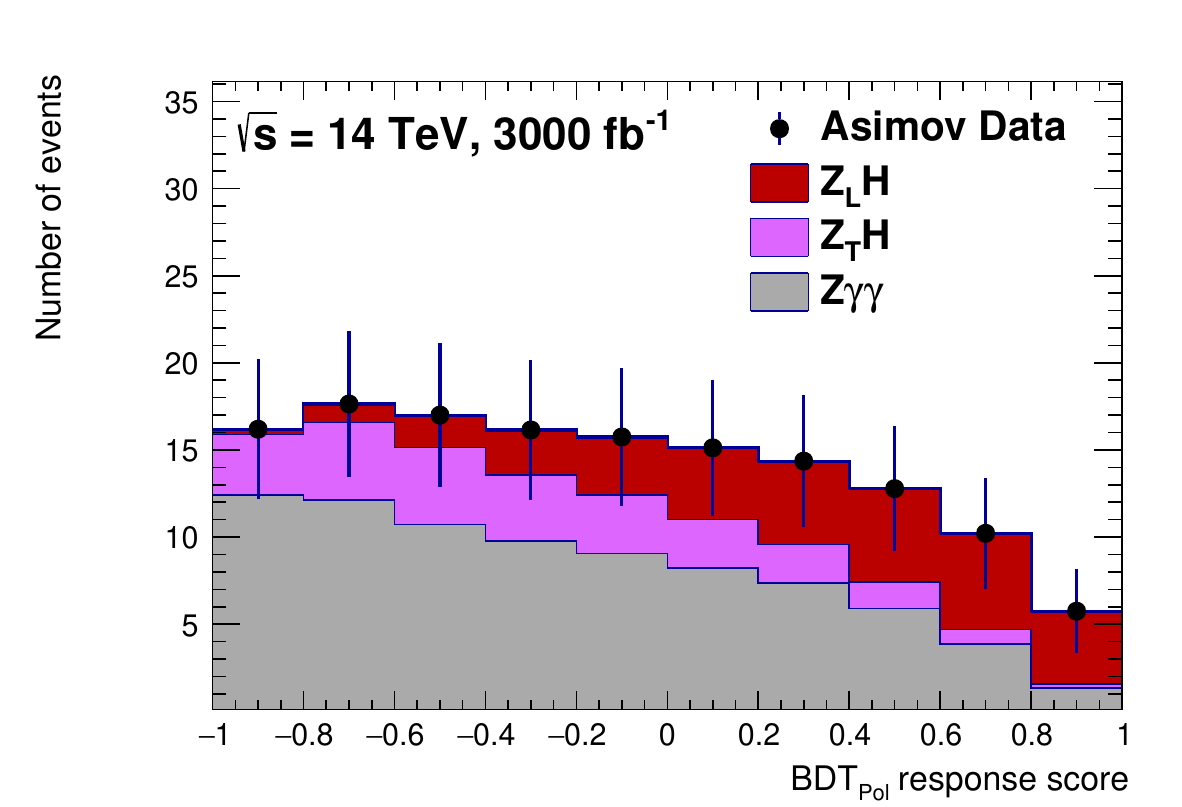}}\\
\caption{Distributions of the $\mathrm{BDT}_{\mathrm{Pol}}$ response scores in the (a) $W^\pm H(\rightarrow\gamma\gamma)$
analysis channel and in the (b) $ZH(\rightarrow\gamma\gamma)$ analysis channel. The predicted signal and background
contributions are shown assuming an integrated luminosity of 3000\,fb$^{-1}$.}
\label{fig:FitInputDistributions}
\end{figure}

\begin{figure}[!h]
\centering
\subfloat[]{\includegraphics[width=0.405\textwidth]{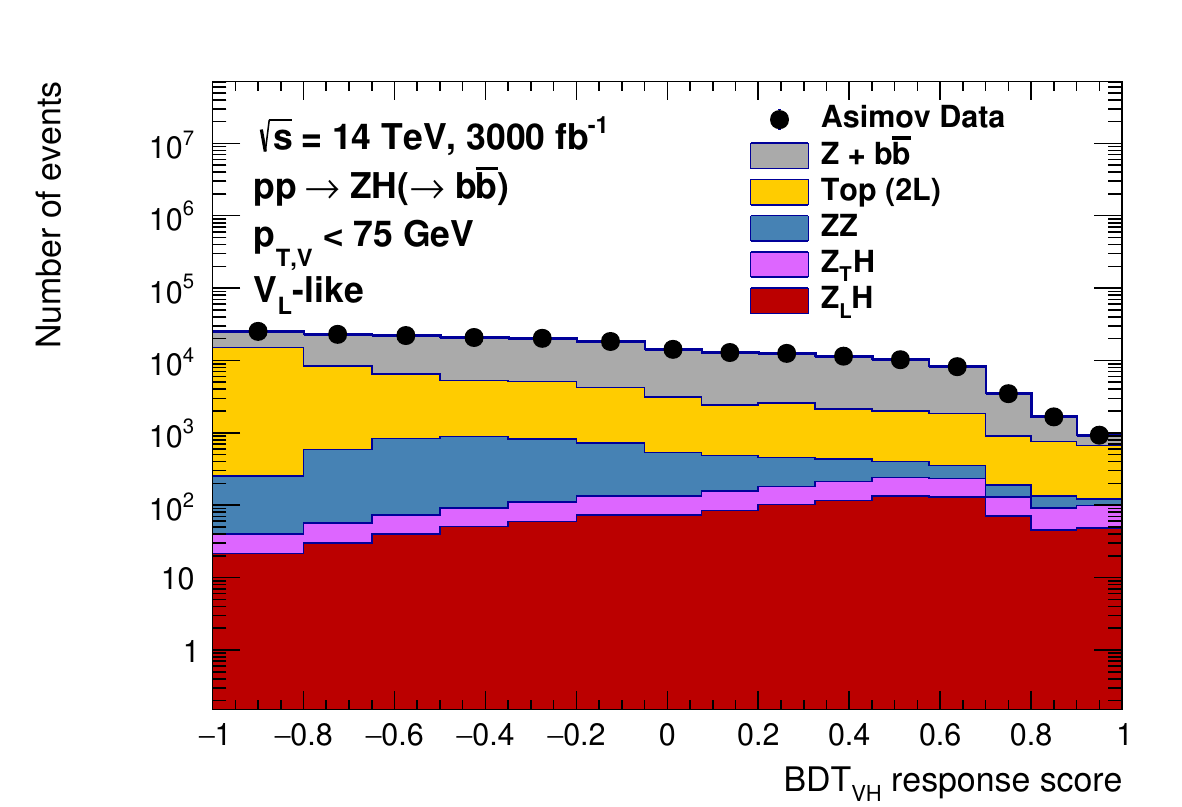}}
\subfloat[]{\includegraphics[width=0.405\textwidth]{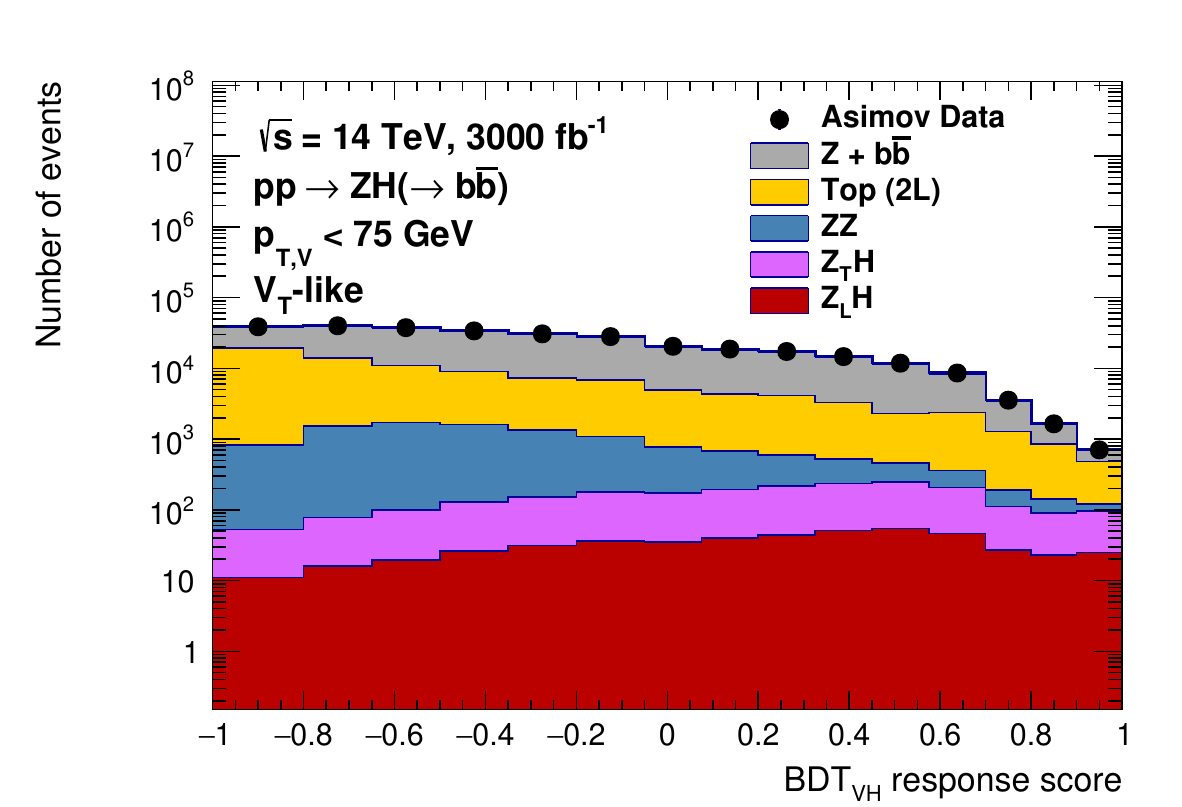}} \\
\subfloat[]{\includegraphics[width=0.405\textwidth]{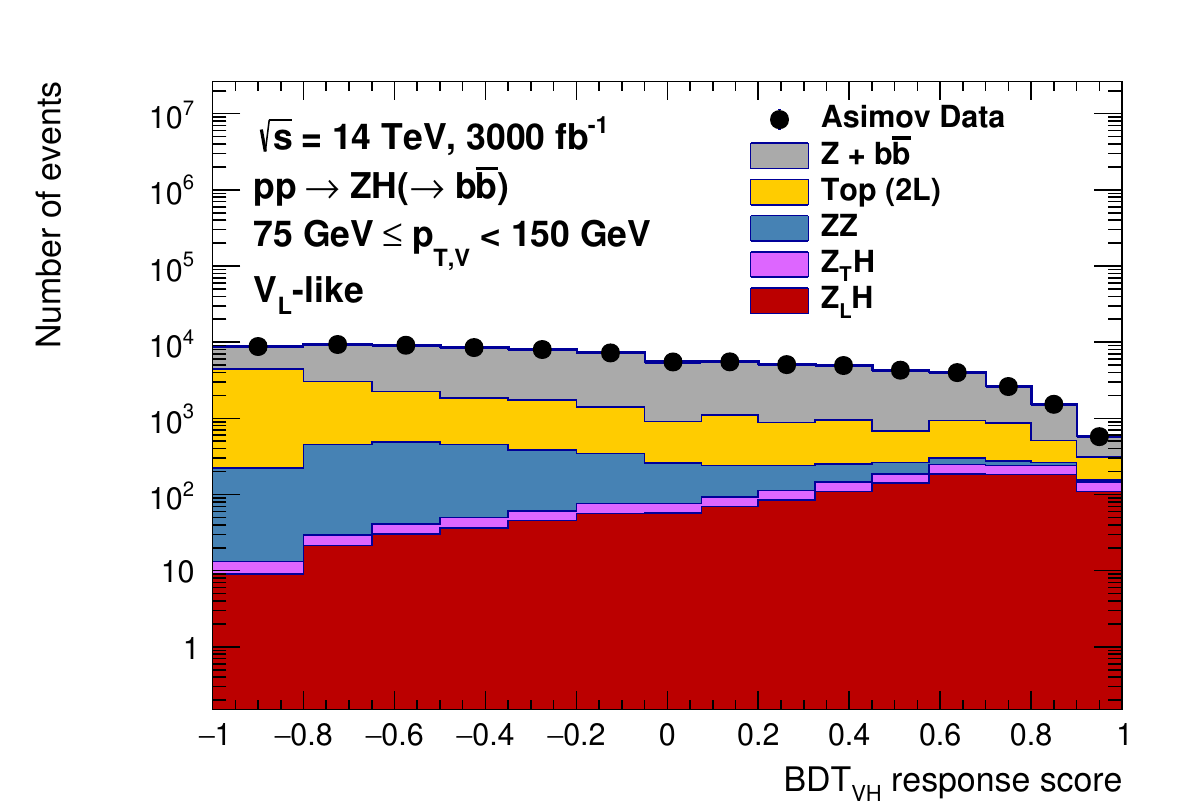}}
\subfloat[]{\includegraphics[width=0.405\textwidth]{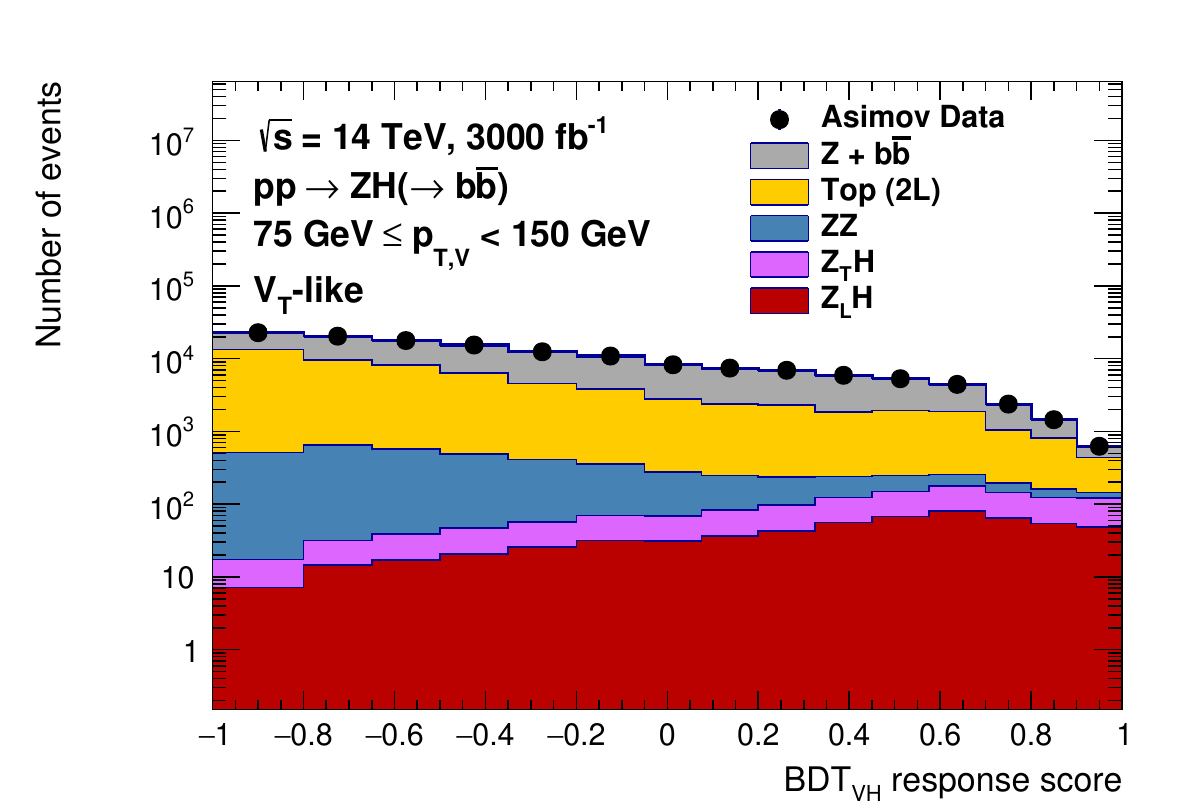}} \\
\subfloat[]{\includegraphics[width=0.405\textwidth]{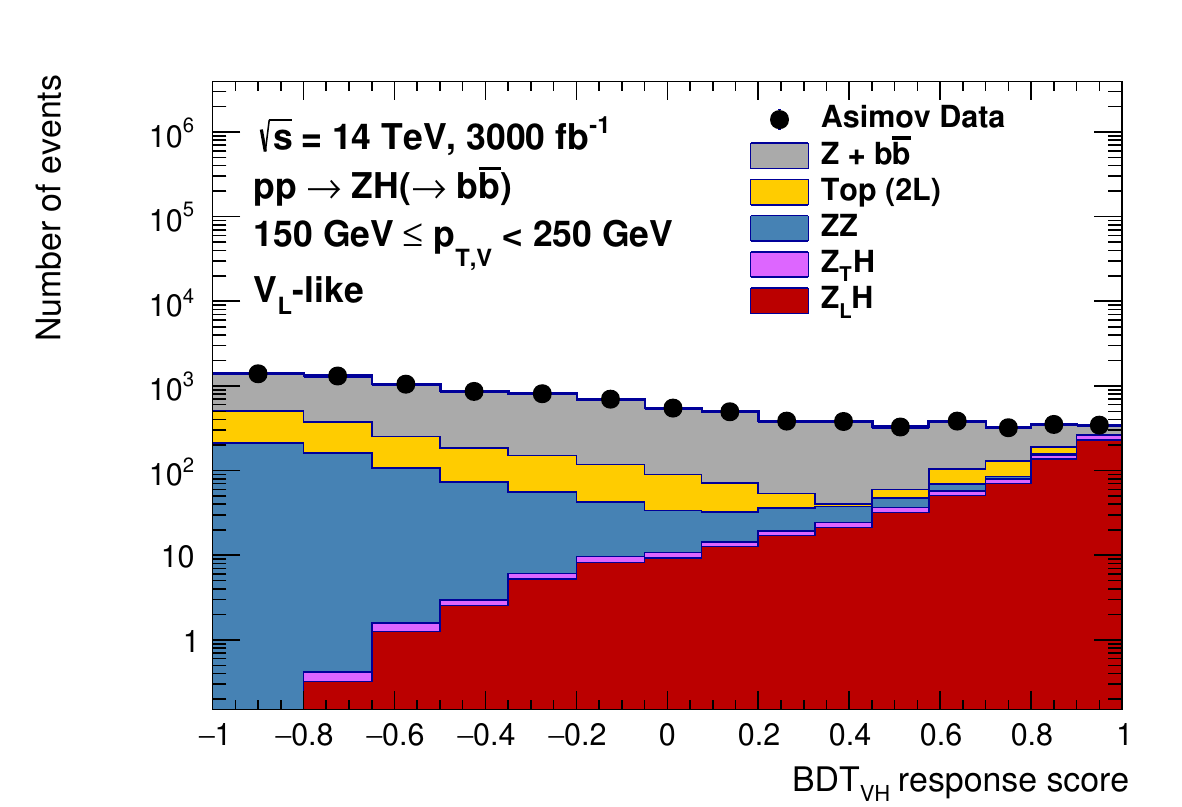}}
\subfloat[]{\includegraphics[width=0.405\textwidth]{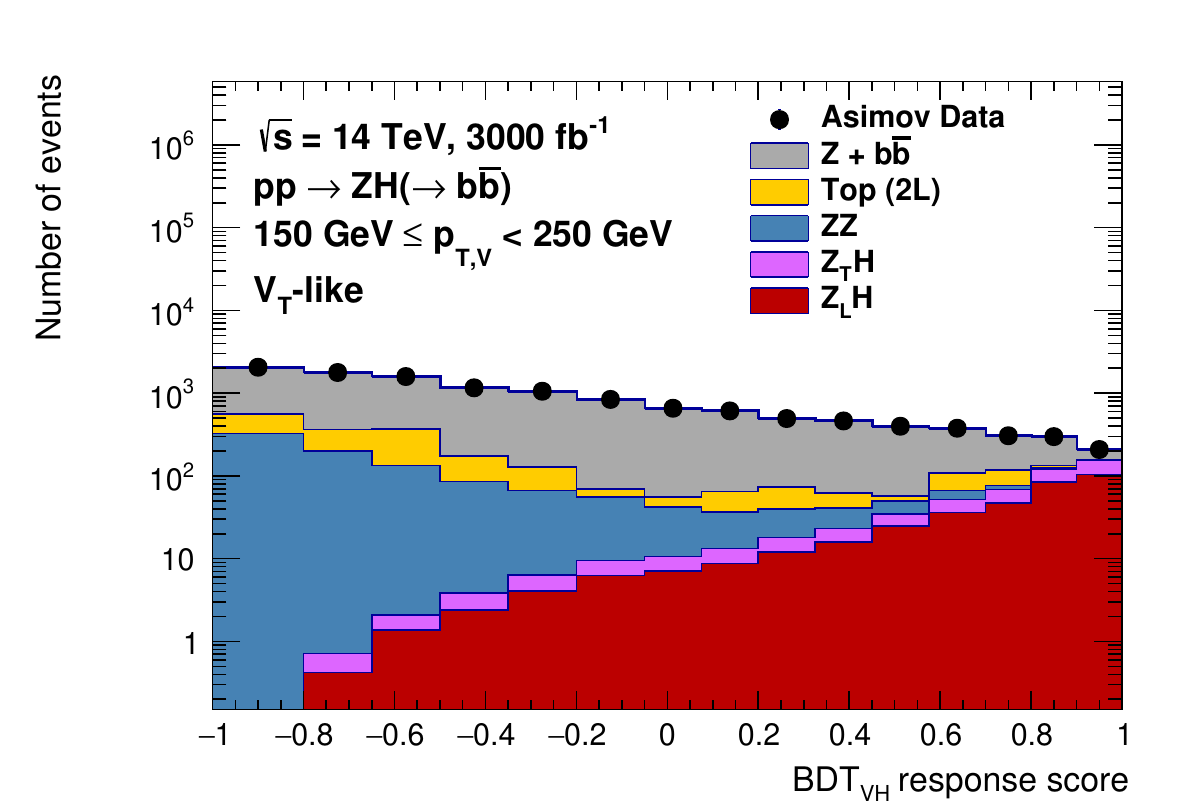}} \\
\subfloat[]{\includegraphics[width=0.405\textwidth]{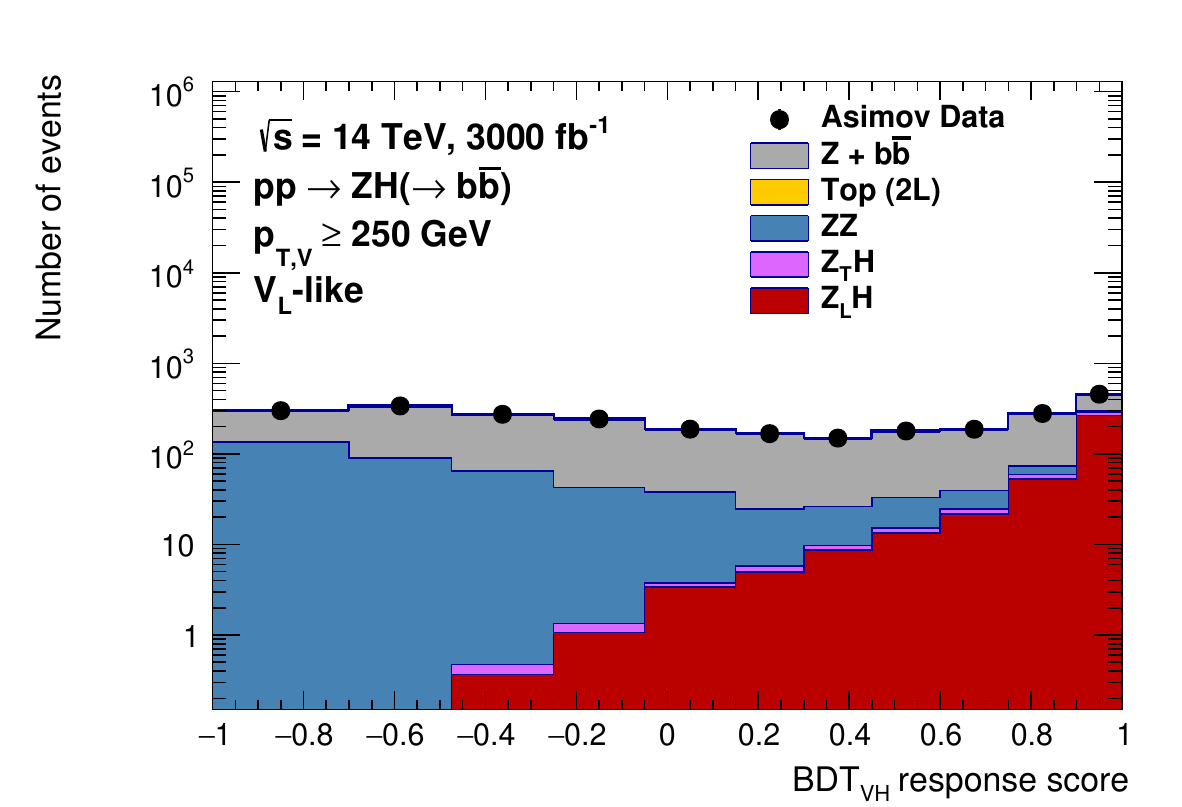}}
\caption{Distributions of the $\mathrm{BDT}_{VH}$ response scores in the $ZH(\rightarrow b\bar{b})$ analysis channel 
presented separately in the four $p_{\mathrm{T},V}$ intervals and in the $V_\mathrm{L}$-like and $V_\mathrm{T}$-like sub-regions.
The predicted signal and background contributions are shown assuming an integrated luminosity of 3000\,fb$^{-1}$.}
\label{fig:FitInputDistributions_ZHbb}
\end{figure}

\begin{figure}[!h]
\centering
\subfloat[]{\includegraphics[width=0.405\textwidth]{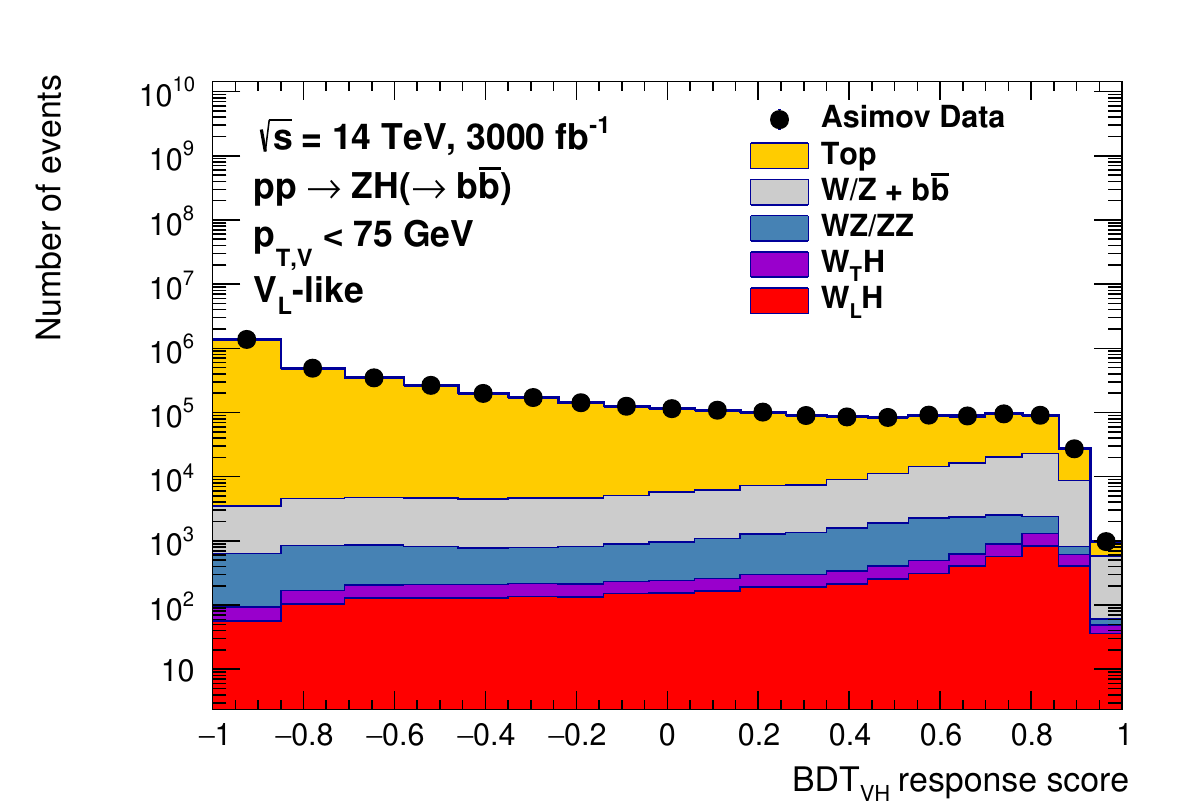}}
\subfloat[]{\includegraphics[width=0.405\textwidth]{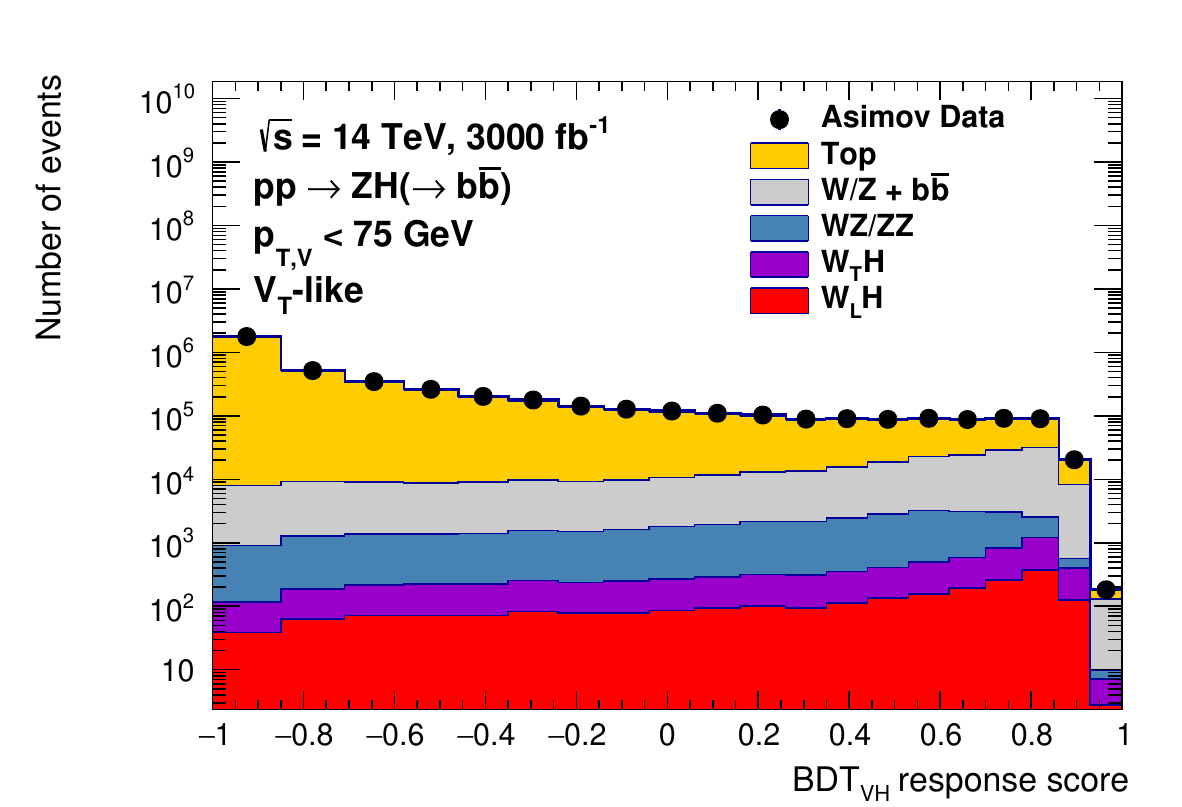}} \\
\subfloat[]{\includegraphics[width=0.405\textwidth]{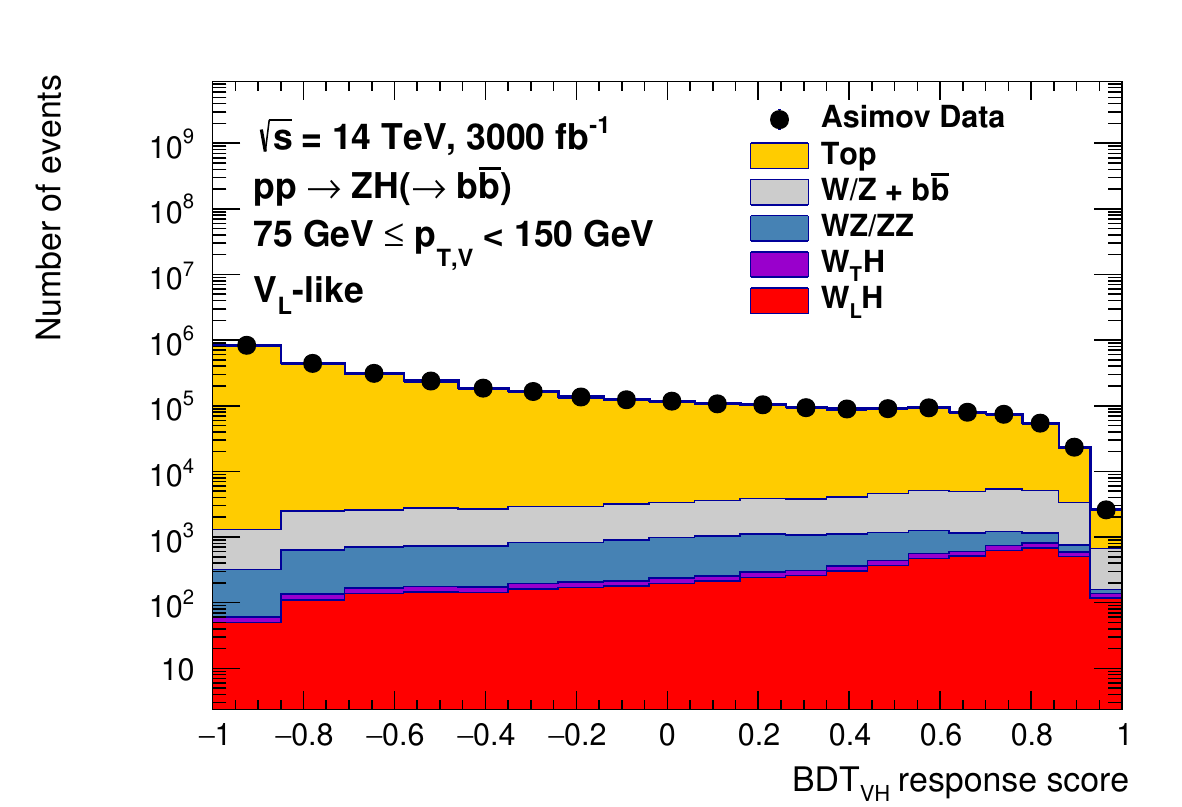}}
\subfloat[]{\includegraphics[width=0.405\textwidth]{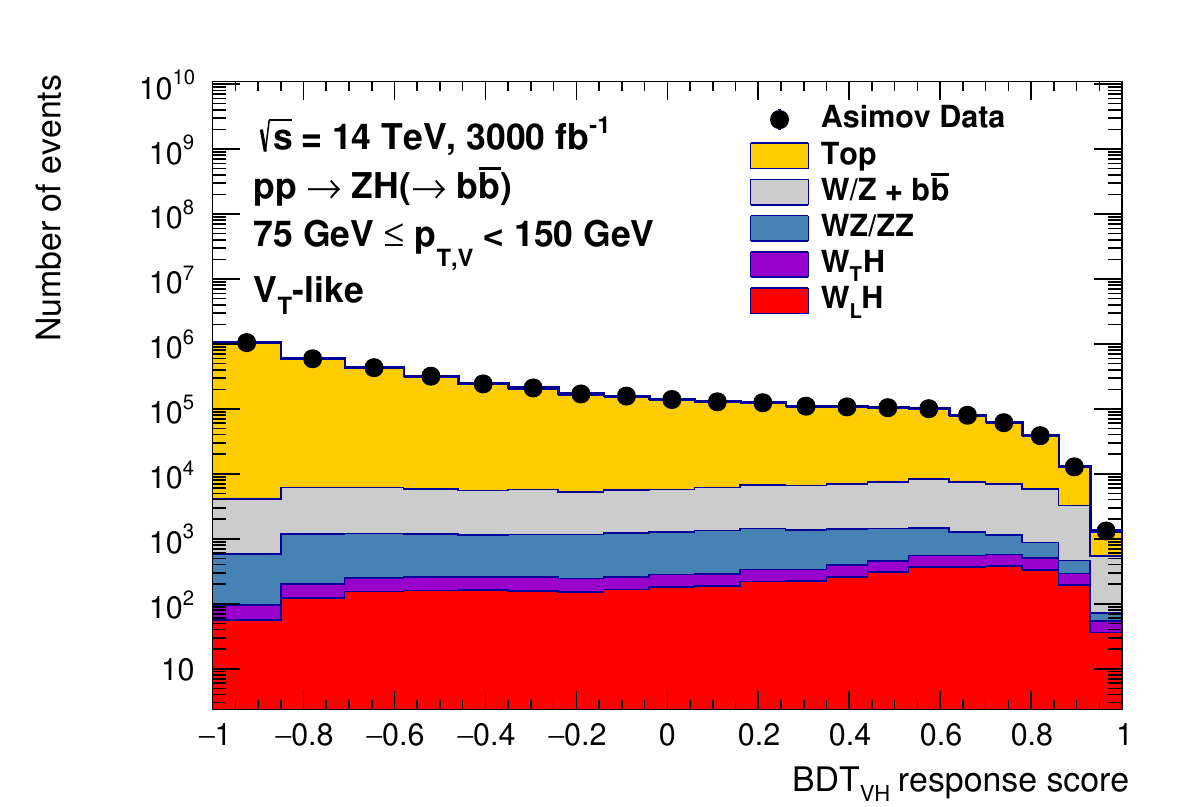}} \\
\subfloat[]{\includegraphics[width=0.405\textwidth]{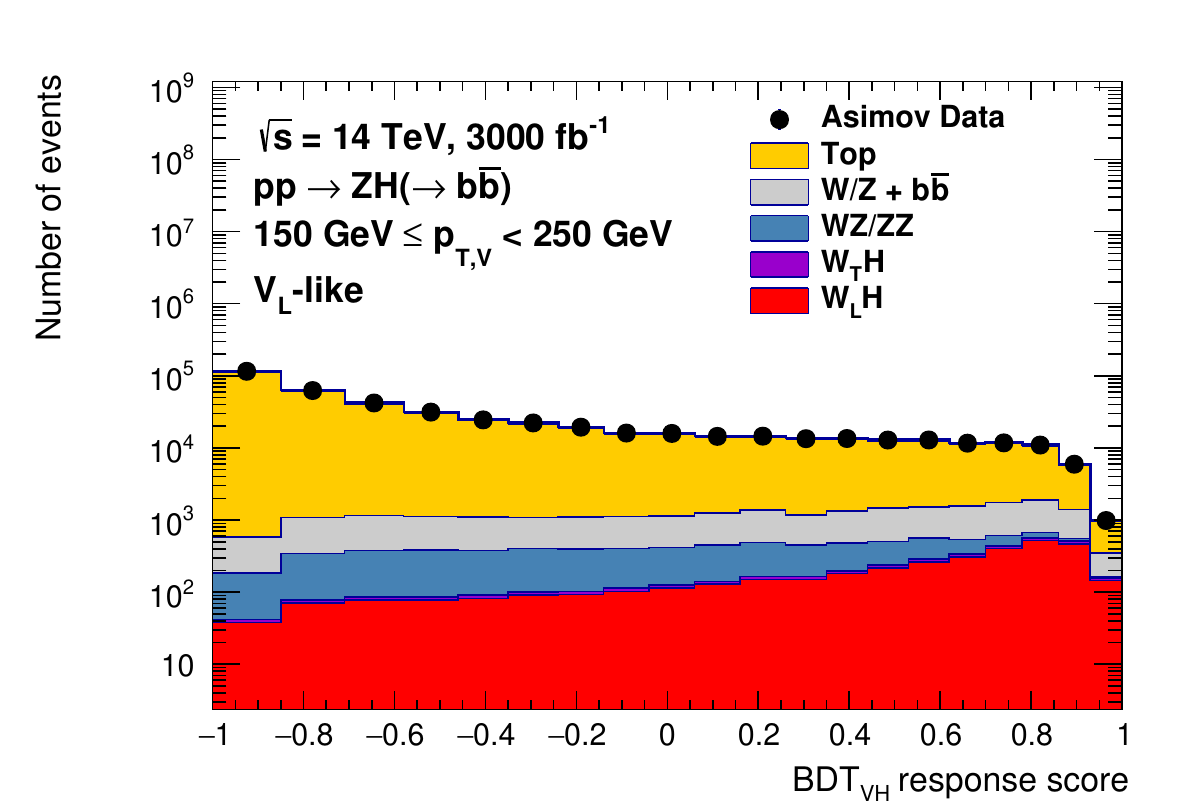}}
\subfloat[]{\includegraphics[width=0.405\textwidth]{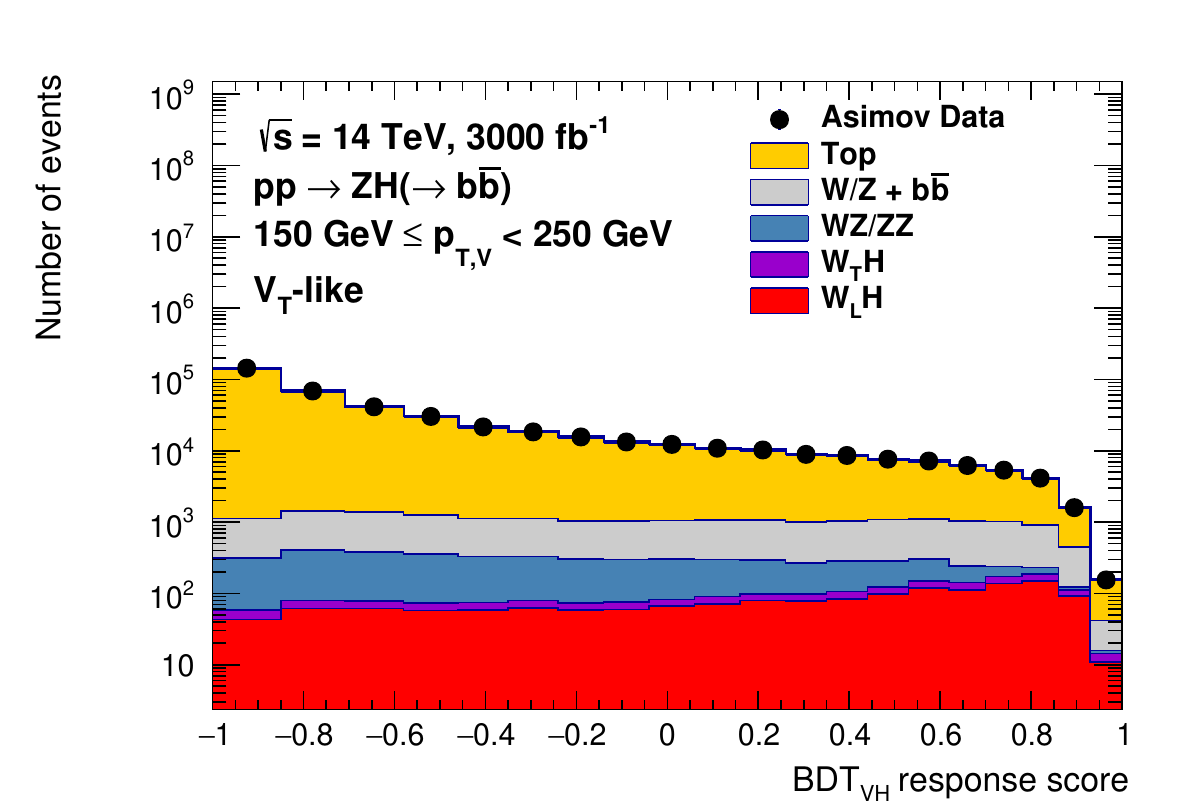}} \\
\subfloat[]{\includegraphics[width=0.405\textwidth]{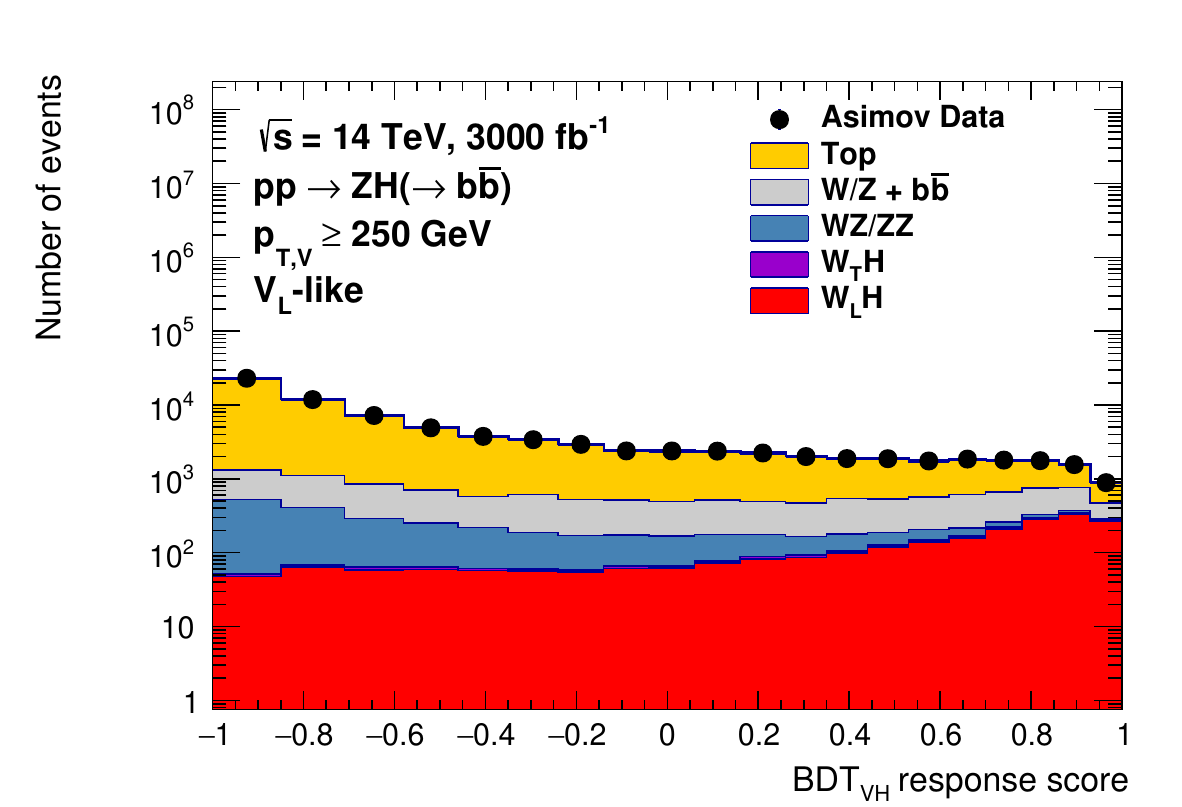}}
\caption{Distributions of the $\mathrm{BDT}_{VH}$ response scores in the $W^\pm H(\rightarrow b\bar{b})$ analysis channel
presented separately in the four $p_{\mathrm{T},V}$ intervals and in the $V_\mathrm{L}$-like and $V_\mathrm{T}$-like sub-regions.
The predicted signal and background contributions are shown assuming an integrated luminosity of 3000\,fb$^{-1}$.}
\label{fig:FitInputDistributions_WHbb}
\end{figure}

Fig.~\ref{fig:FitInputDistributions} shows the expected $\mathrm{BDT}_\mathrm{Pol}$ response score distributions for the 
$W^\pm H$ and $ZH$ analysis channels in $H \rightarrow \gamma\gamma$ decays, comprising 10 equidistant bins of width 
0.2. For the $H \rightarrow b\bar{b}$ decay channel, the corresponding $\mathrm{BDT}_\mathrm{VH}$ score distributions across 
$p_{\mathrm{T},V}$ intervals and $V_\mathrm{L}$-like/$V_\mathrm{T}$-like sub-regions are shown in 
Figs.~\ref{fig:FitInputDistributions_ZHbb} and \ref{fig:FitInputDistributions_WHbb}, where non-equidistant binning is used to 
optimise sensitivity. All predicted signal and background yields are normalised to an integrated luminosity of 
3000\,fb$^{-1}$. \par

\clearpage 

\begin{figure}[!h]
\centering
\subfloat[]{\includegraphics[width=0.405\textwidth]{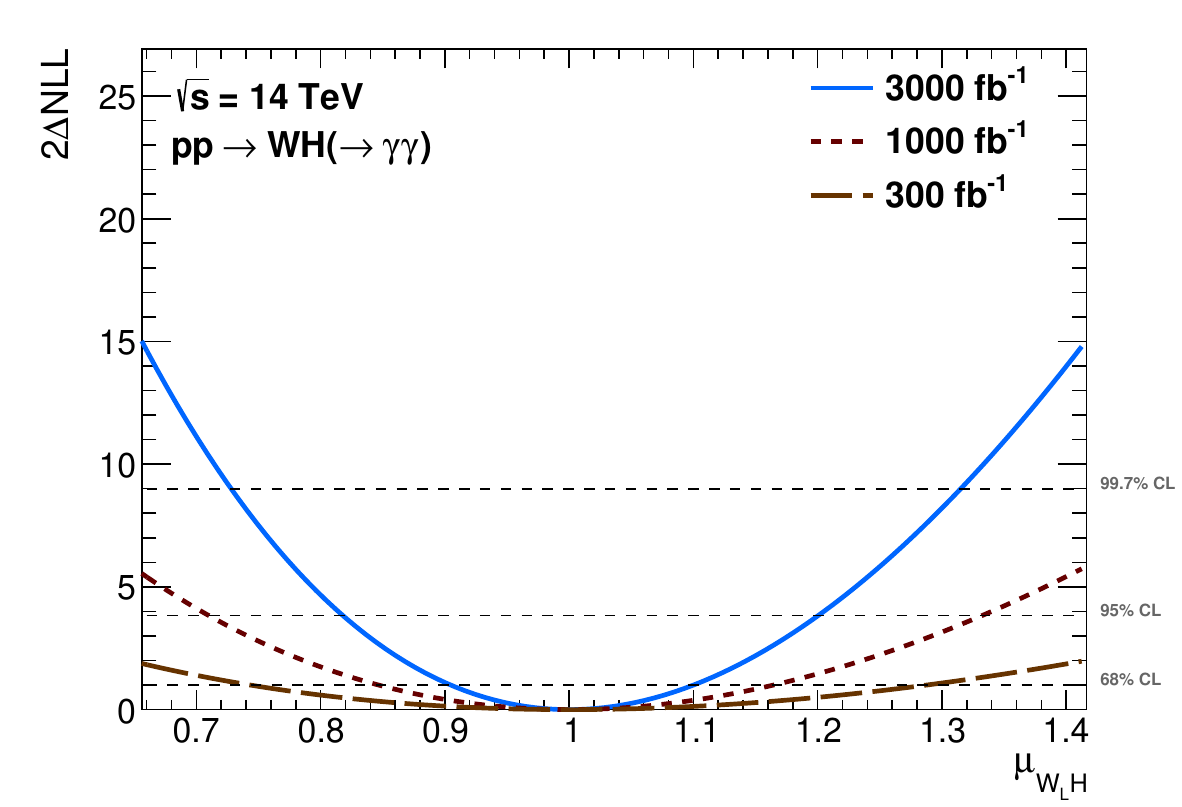}}
\subfloat[]{\includegraphics[width=0.405\textwidth]{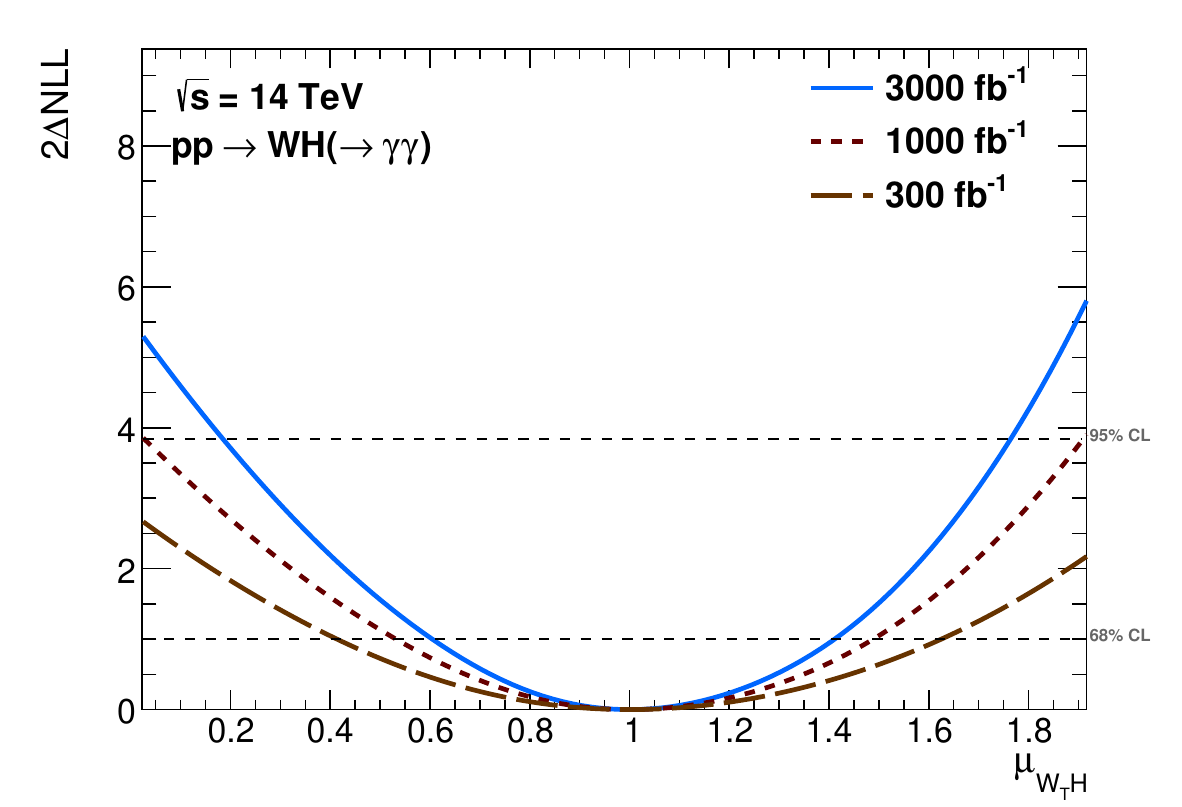}}\\
\subfloat[]{\includegraphics[width=0.405\textwidth]{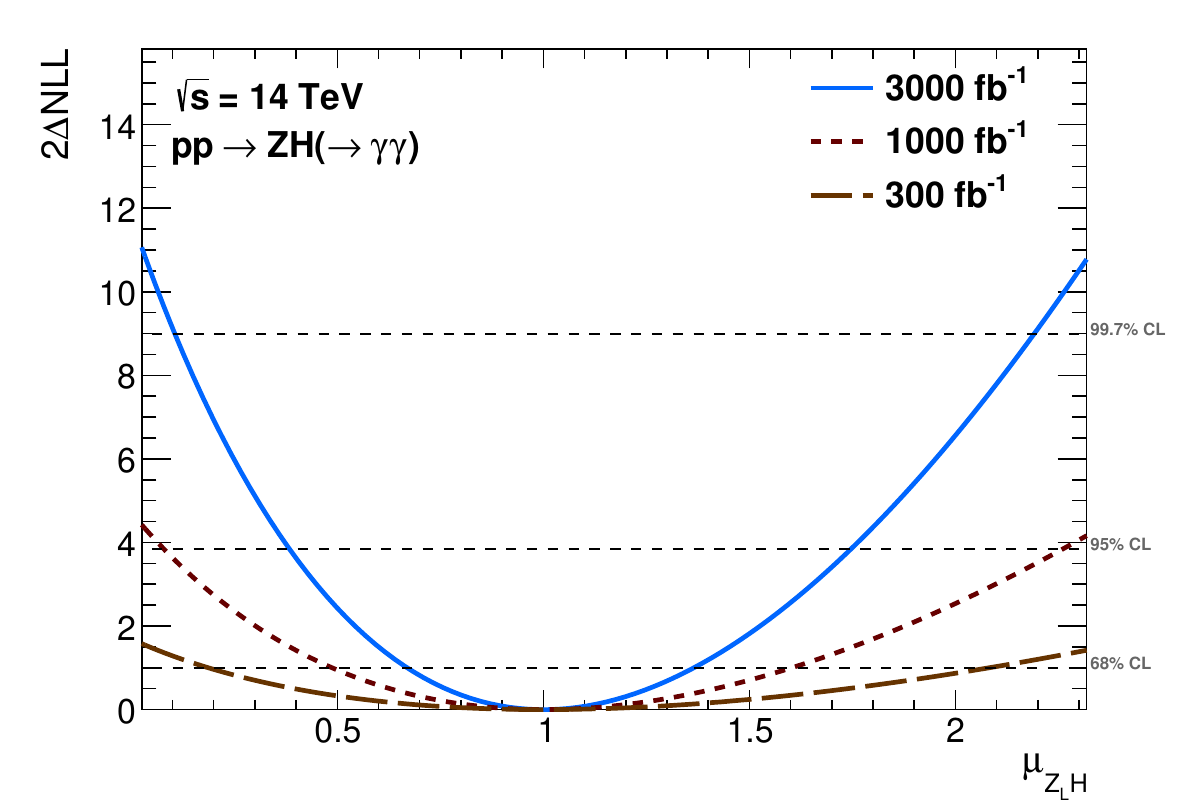}}
\caption{Expected $\Delta\mathrm{NLL}$ curves extracted from the analyses in the $H \rightarrow \gamma\gamma$ decay channel as a 
function of (a) $\mu_{W_\mathrm{L}H}$, (b) $\mu_{W_\mathrm{T}H}$, and (c) $\mu_{Z_\mathrm{L}H}$ for integrated luminosities of 
$300\,\mathrm{fb}^{-1}$, $1000\,\mathrm{fb}^{-1}$, and $3000\,\mathrm{fb}^{-1}$ at $\sqrt{s} = 14\,\mathrm{TeV}$. 
Each parameter is scanned with the other signal strength modifier fixed to its SM value. The values 
of $\mu_{W_\mathrm{L}H}$, $\mu_{W_\mathrm{T}H}$, and $\mu_{Z_\mathrm{L}H}$ where the $\Delta\mathrm{NLL}$ curves intersect the 
dashed black lines correspond to the $68\%$, $95\%$, and $99.7\%$ confidence intervals, respectively.}
\label{fig:NLLCurves_VHyy}
\end{figure}

\begin{figure}[!h]
\centering
\subfloat[]{\includegraphics[width=0.395\textwidth]{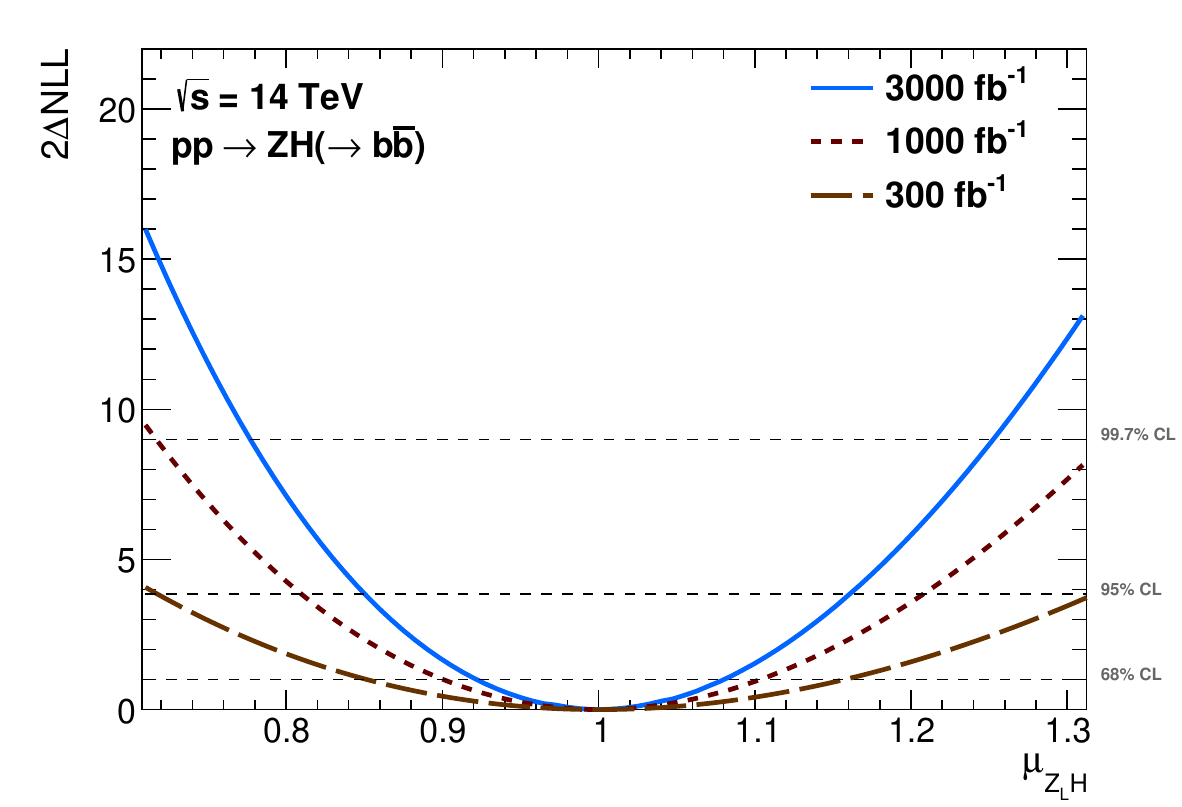}}
\subfloat[]{\includegraphics[width=0.395\textwidth]{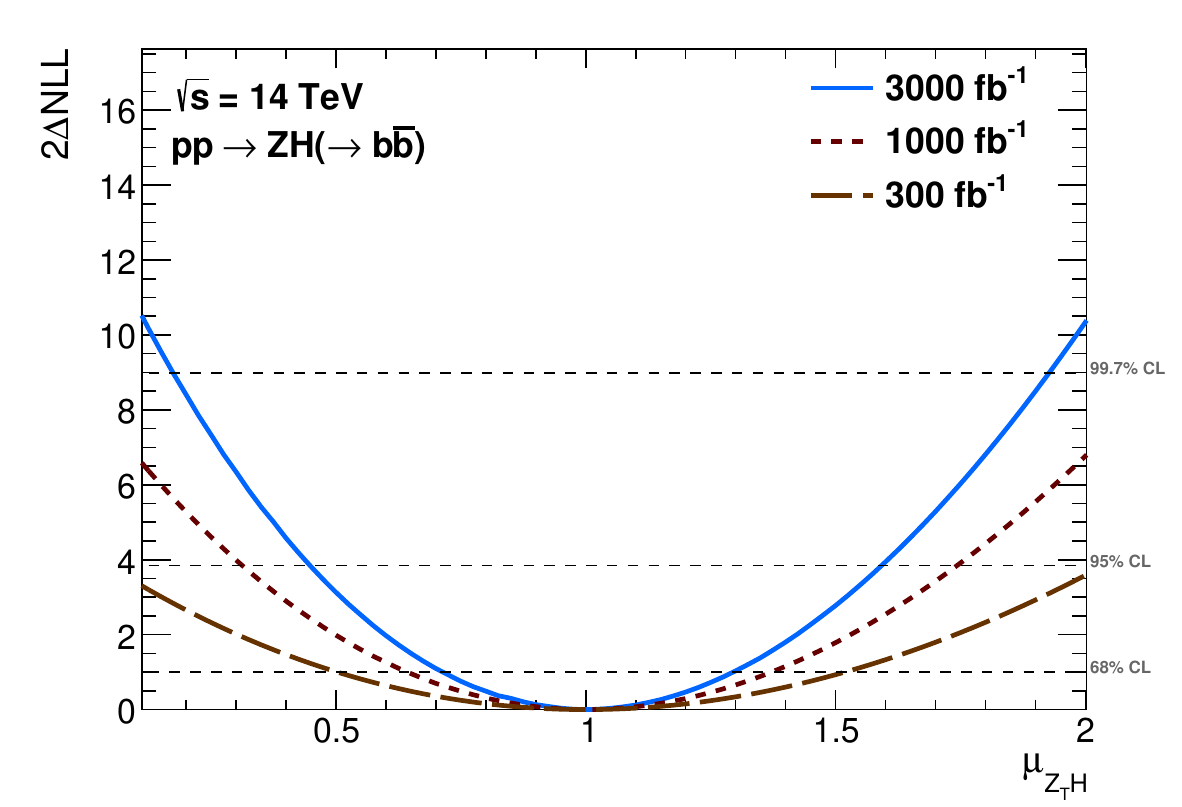}}\\
\subfloat[]{\includegraphics[width=0.395\textwidth]{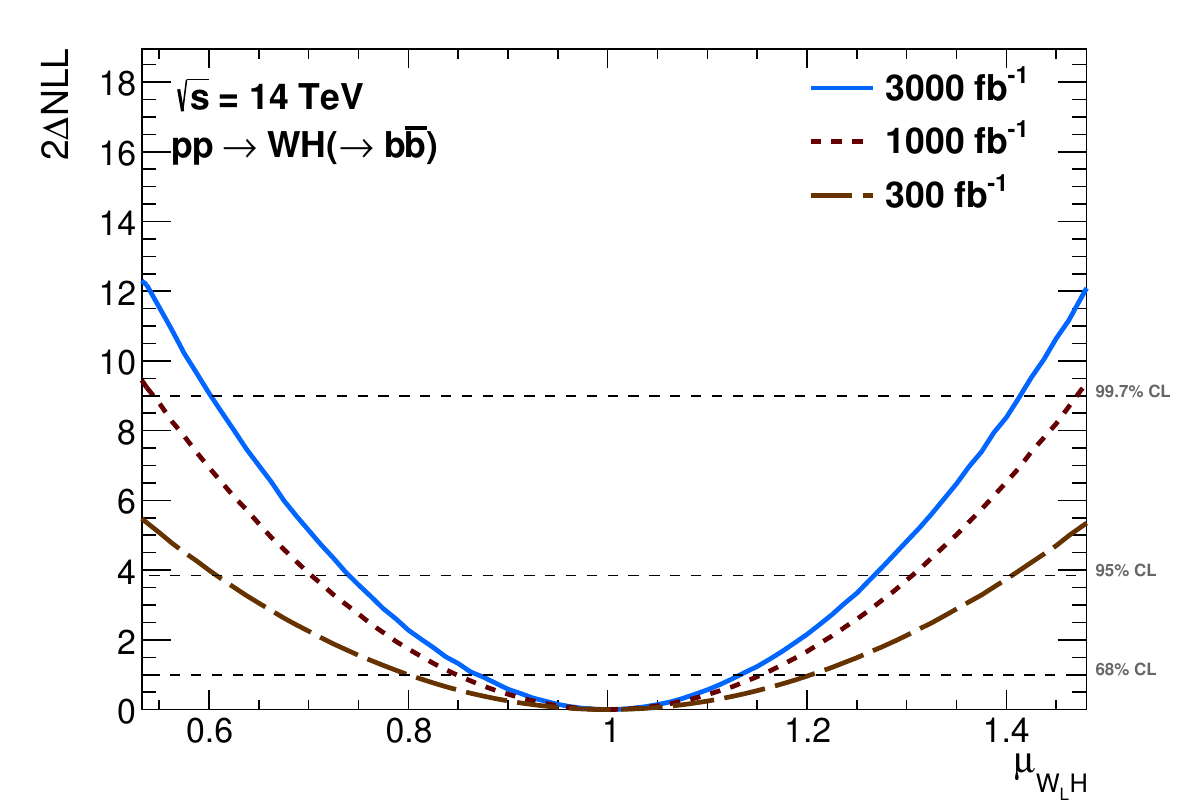}}
\caption{Expected $\Delta\mathrm{NLL}$ curves extracted from the analyses in the $H \rightarrow b\bar{b}$ decay channel as a
function of (a) $\mu_{Z_\mathrm{L}H}$, (b) $\mu_{Z_\mathrm{T}H}$, and (c) $\mu_{W_\mathrm{L}H}$ for integrated luminosities of
$300\,\mathrm{fb}^{-1}$, $1000\,\mathrm{fb}^{-1}$, and $3000\,\mathrm{fb}^{-1}$ at $\sqrt{s} = 14\,\mathrm{TeV}$. 
Each parameter is scanned with the other signal strength modifier fixed to its SM value. The values of 
$\mu_{Z_\mathrm{L}H}$, $\mu_{Z_\mathrm{T}H}$, and $\mu_{W_\mathrm{L}H}$ where the $\Delta\mathrm{NLL}$ curves intersect the
dashed black lines correspond to the $68\%$, $95\%$, and $99.7\%$ confidence intervals, respectively.}
\label{fig:NLLCurves_VHbb}
\end{figure}

The expected negative log-likelihood (NLL) profile scans at 300\,fb$^{-1}$, 1000\,fb$^{-1}$, and 3000\,fb$^{-1}$ are presented 
in Fig.~\ref{fig:NLLCurves_VHyy} as a function of $\mu_{W_\mathrm{L}H}$, $\mu_{W_\mathrm{T}H}$, and $\mu_{Z_\mathrm{L}H}$ in 
the $H \rightarrow \gamma\gamma$ channel, and in Fig.~\ref{fig:NLLCurves_VHbb} as a function of $\mu_{Z_\mathrm{L}H}$, $\mu_{Z_\mathrm{T}H}$, and 
$\mu_{W_\mathrm{L}H}$ in the $H\rightarrow b\bar{b}$ channel. The scans over $\mu_{Z_\mathrm{T}H}$ in $H \rightarrow \gamma\gamma$ and 
$\mu_{W_\mathrm{T}H}$ in $H \rightarrow b\bar{b}$ are omitted due to lack of sensitivity. The NLL curves are sampled with step 
sizes of $\Delta(\mu_{W_\mathrm{L}H}) = 0.01$, $\Delta(\mu_{W_\mathrm{T}H}) = 0.02$, and $\Delta(\mu_{Z_\mathrm{L}H}) = 0.023$ 
for $H \rightarrow \gamma\gamma$, and $\Delta(\mu_{Z_\mathrm{L}H}) = 0.0075$, $\Delta(\mu_{Z_\mathrm{T}H}) = 0.025$, and 
$\Delta(\mu_{W_\mathrm{L}H}) = 0.0125$ for $H \rightarrow b\bar{b}$, with minima at zero corresponding to the Standard Model 
expectation. The resulting expected 68\% and 95\% confidence intervals for all signal strength parameters are summarised in 
Tab.~\ref{tab:ConfidenceIntervals}. \par
\begin{figure}[!h]
\centering
\subfloat[]{\includegraphics[width=0.395\textwidth]{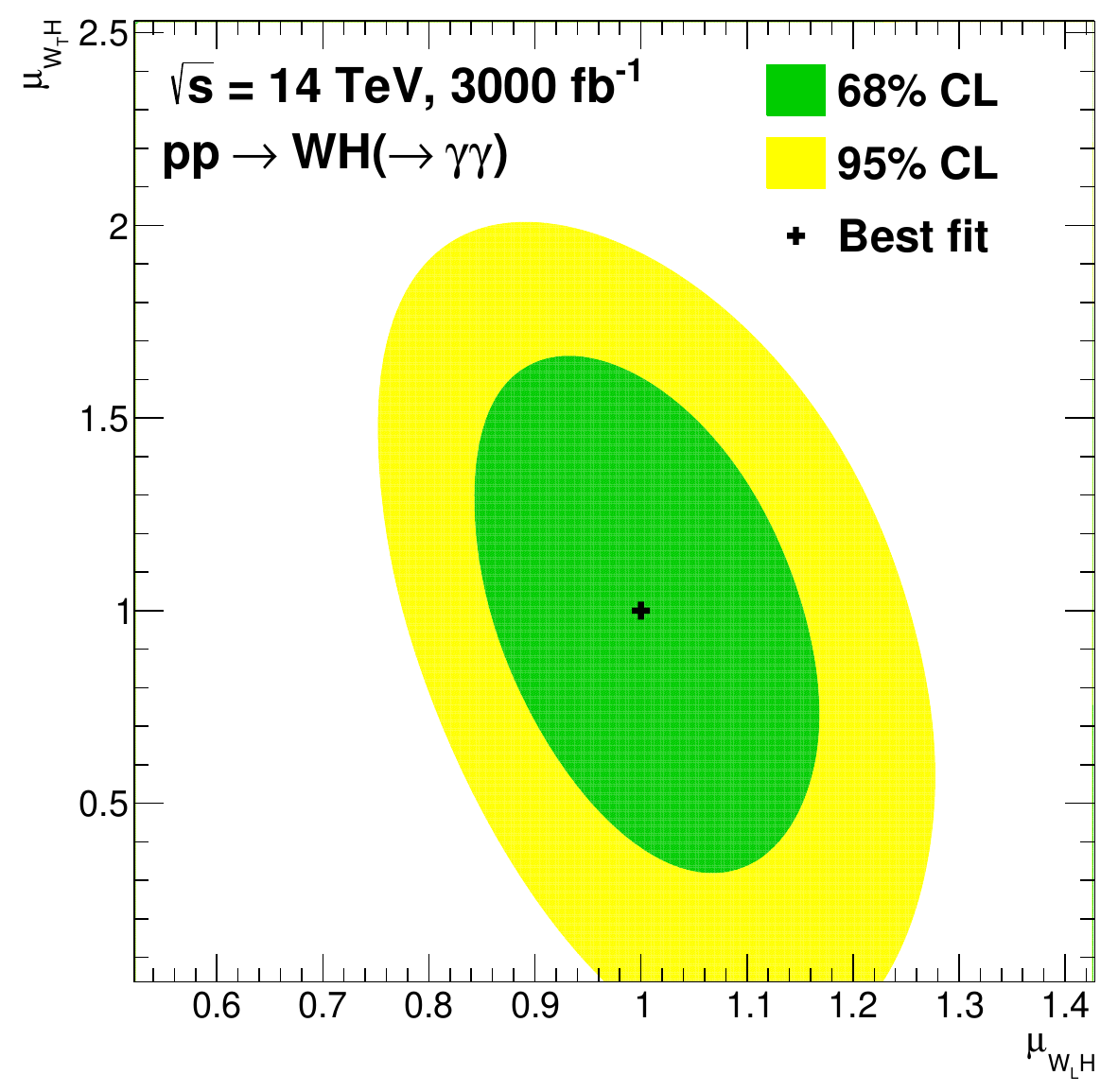}}
\subfloat[]{\includegraphics[width=0.395\textwidth]{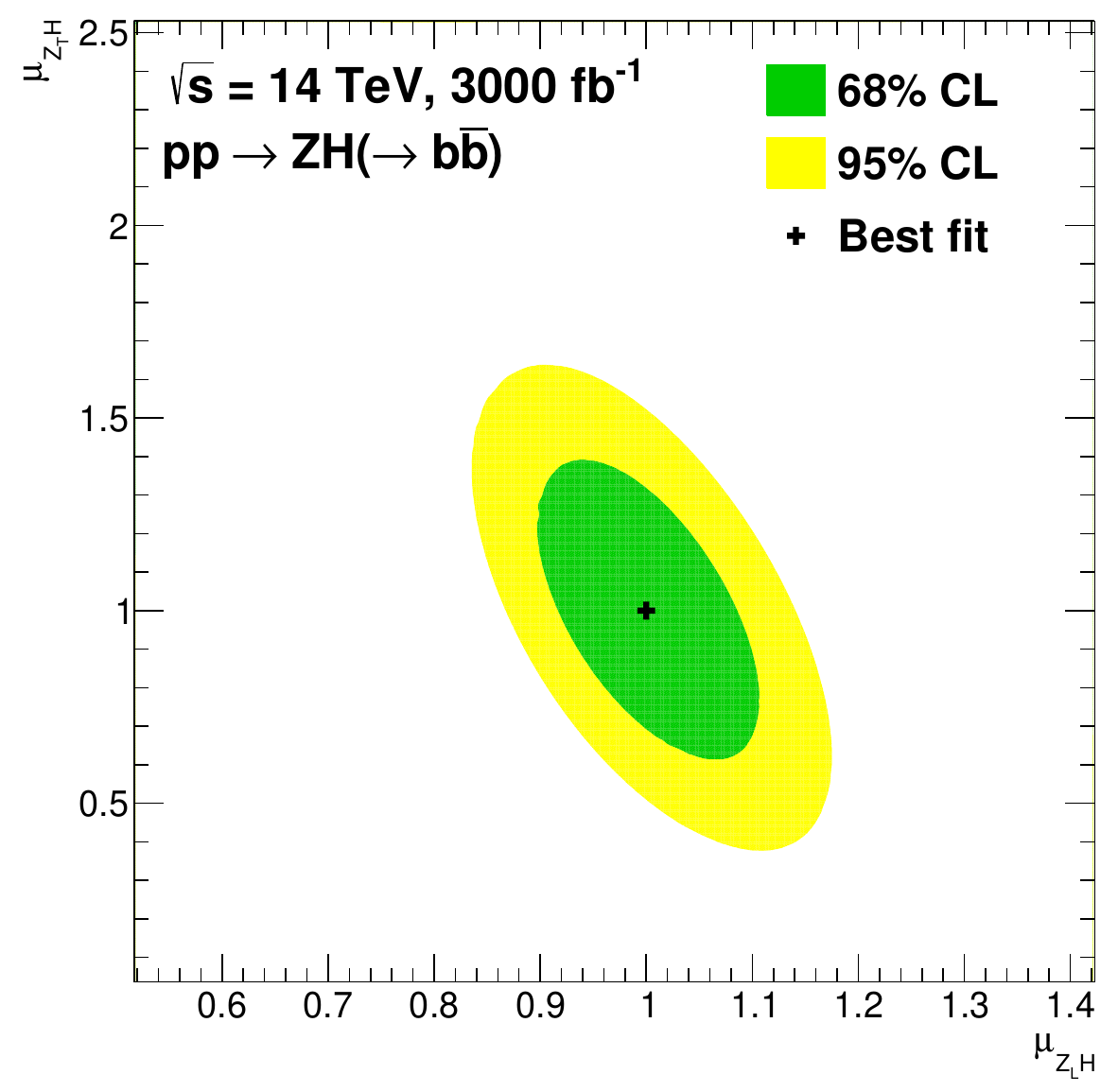}}
\caption{Expected two-dimensional likelihood scans for (a) $\mu_{W_\mathrm{T}H}$ versus $\mu_{W_\mathrm{L}H}$ 
in the $WH(\rightarrow \gamma\gamma)$ channel and (b) $\mu_{Z_\mathrm{T}H}$ versus $\mu_{Z_\mathrm{L}H}$ in the 
$ZH(\rightarrow b\bar{b})$ channel. The $2\Delta\text{NLL}$ values are calculated relative to the SM 
expectation. The best-fit point (cross) as well as the $68\%$ (green) and $95\%$ (yellow) CL contours, 
corresponding to $2\Delta\text{NLL}$ values of $0.0$, $2.30$, and $5.99$, are overlaid.}
\label{fig:2D_NLLContours}
\end{figure}

In addition, two-dimensional likelihood scans for $\mu_{W_\mathrm{T}H}$ versus $\mu_{W_\mathrm{L}H}$ in the $WH(\rightarrow \gamma\gamma)$
channel and $\mu_{Z_\mathrm{T}H}$ versus $\mu_{Z_\mathrm{L}H}$ in the $ZH(\rightarrow b\bar{b})$ channel are shown in
Fig.~\ref{fig:2D_NLLContours}. \par
Based on our simplified fit model, 5.0 standard deviation discovery significances can be achieved with integrated luminosities
of 425\,fb$^{-1}$ for $W^\pm_\mathrm{L}H(\rightarrow \gamma\gamma)$, 210\,fb$^{-1}$ for $Z_\mathrm{L}H(\rightarrow b\bar{b})$,
and 340\,fb$^{-1}$ for $W_\mathrm{L}H(\rightarrow b\bar{b})$. At 3000\,fb$^{-1}$, a significance of 3.4$\sigma$ is reachable
for $Z_\mathrm{L}H(\rightarrow \gamma\gamma)$, whereas $Z_\mathrm{T}H(\rightarrow b\bar{b})$ would require 4000\,fb$^{-1}$ to
reach 3.0$\sigma$. In contrast, $W^\pm_\mathrm{T}H$ in both decay modes and $Z_\mathrm{T}H$ in the diphoton channel remain out
of reach at the HL-LHC. \par

\clearpage

Inclusive measurements of the signal strength parameters for $W^\pm_\mathrm{L}H(\rightarrow \gamma\gamma)$ are projected to
reach precisions of approximately $30\%$, $15\%$, and $10\%$ at 300\,fb$^{-1}$, 1000\,fb$^{-1}$, and 3000\,fb$^{-1}$,
respectively, while $Z_\mathrm{L}H(\rightarrow \gamma\gamma)$ can be measured to within $35\%$ at 3000\,fb$^{-1}$. In the
$H \rightarrow b\bar{b}$ channel, the projected precisions across the same luminosity benchmarks are $15\%$, $10\%$, and $8\%$
for $Z_\mathrm{L}H$, and $20\%$, $15\%$, and $13\%$ for $W^\pm_\mathrm{L}H$. Furthermore, we project the signal strength of
$Z_\mathrm{T}H(\rightarrow b\bar{b})$ to be constrained within $28\%$ at 3000\,fb$^{-1}$. \par
At 3000\,fb$^{-1}$, the sample statistics for $W^\pm H(\rightarrow \gamma\gamma)$, $W^\pm H(\rightarrow b\bar{b})$, and
$ZH(\rightarrow b\bar{b})$ are sufficient to enable differential measurements of $\mu_{W_\mathrm{L}H}$, $\mu_{Z_\mathrm{L}H}$,
and $\mu_{Z_\mathrm{T}H}$ across several simplified template cross-section bins~\cite{Brehmer:2019gmn}, following the framework
discussed in Ref.~\cite{HIGG-2020-16}. Such extensions would significantly enhance sensitivity to BSM effects but are left for
future work.

\begin{table}[h!]
\caption{Expected $68\%$ and $95\%$ CL confidence intervals for the parameters of interest (POIs), corresponding to the 
signal strength scale factors of the $W^\pm_{\mathrm{L},\mathrm{T}}H$ and $Z_{\mathrm{L},\mathrm{T}}H$ production modes. 
Results are shown for integrated luminosities of $300\,\mathrm{fb}^{-1}$, $1000\,\mathrm{fb}^{-1}$, and $3000\,\mathrm{fb}^{-1}$ 
at $\sqrt{s} = 14\,\mathrm{TeV}$ for the $H \rightarrow \gamma\gamma$ (left) and $H \rightarrow b\bar{b}$ (right) decay channels.}
\begin{minipage}{0.48\linewidth}
\begin{center}
\renewcommand{\arraystretch}{1.25}
\begin{tabular}{c | c | c}
\hline
\hline
PoI                               &    68\%  &  95\%   \\
\hline
\multicolumn{3}{c}{For 300\,fb$^{-1}$}     \\
\hline
$\mu_{W_\mathrm{L}H}$             &  [0.74, 1.28] & [0.52, 1.59] \\
$\mu_{W_\mathrm{T}H}$             &  [0.41, 1.63] &      --      \\
$\mu_{Z_\mathrm{L}H}$             &  [0.16, 2.07] &      --      \\
$\mu_{Z_\mathrm{T}H}$             &       --      &      --      \\
\hline
\multicolumn{3}{c}{For 1000\,fb$^{-1}$}    \\
\hline
$\mu_{W_\mathrm{L}H}$             &  [0.84, 1.15] & [0.71, 1.32] \\
$\mu_{W_\mathrm{T}H}$             &  [0.53, 1.49] & [0.03, 1.91] \\
$\mu_{Z_\mathrm{L}H}$             &  [0.47, 1.58] & [0.07, 2.26] \\
$\mu_{Z_\mathrm{T}H}$             &  [0.02, 2.06] &     --       \\
\hline
\multicolumn{3}{c}{For 3000\,fb$^{-1}$}    \\
\hline
$\mu_{W_\mathrm{L}H}$             &  [0.90, 1.09] & [0.82, 1.19] \\
$\mu_{W_\mathrm{T}H}$             &  [0.59, 1.41] & [0.17, 1.75] \\
$\mu_{Z_\mathrm{L}H}$             &  [0.65, 1.35] & [0.37, 1.74] \\
$\mu_{Z_\mathrm{T}H}$             &  [0.21, 1.83] &     --       \\
\hline
\hline
\end{tabular}
\end{center}
\end{minipage}
\hfill
\begin{minipage}{0.48\linewidth}
\begin{center}
\renewcommand{\arraystretch}{1.25}
\begin{tabular}{c | c | c}
\hline
\hline
PoI                               &    68\%  &  95\%   \\
\hline
\multicolumn{3}{c}{For 300\,fb$^{-1}$}   \\
\hline
$\mu_{W_\mathrm{L}H}$             &   [0.79, 1.20]   &  [0.60, 1.40]  \\
$\mu_{W_\mathrm{T}H}$             &      --          &       --       \\
$\mu_{Z_\mathrm{L}H}$             &   [0.85, 1.15]   &  [0.72  1.31]  \\
$\mu_{Z_\mathrm{T}H}$             &   [0.50, 1.50]   &  [0.03, 2.04]  \\
\hline
\multicolumn{3}{c}{For 1000\,fb$^{-1}$}  \\
\hline
$\mu_{W_\mathrm{L}H}$             &  [0.84, 1.15] &  [0.70, 1.30]  \\
$\mu_{W_\mathrm{T}H}$             &      --       &        --      \\
$\mu_{Z_\mathrm{L}H}$             &  [0.90, 1.10] &  [0.81, 1.21]  \\
$\mu_{Z_\mathrm{T}H}$             &  [0.63, 1.35] &  [0.30, 1.73]  \\
\hline
\multicolumn{3}{c}{For 3000\,fb$^{-1}$}  \\
\hline
$\mu_{W_\mathrm{L}H}$             &  [0.86, 1.13] & [0.74, 1.26]  \\
$\mu_{W_\mathrm{T}H}$             &      --       &      --       \\
$\mu_{Z_\mathrm{L}H}$             &  [0.92, 1.08] & [0.85, 1.16]  \\
$\mu_{Z_\mathrm{T}H}$             &  [0.70, 1.28] & [0.43, 1.58]  \\
\hline
\hline
\end{tabular}
\end{center}
\end{minipage}
\label{tab:ConfidenceIntervals}
\end{table}
 \section{Conclusion}\label{sec:Conclusion}

With this work, we demonstrate how measurements of the signal-strength parameters for longitudinally and
transversely polarised $W^\pm$ and $Z$ bosons produced in association with a Higgs boson
can be performed in the $\ell^\pm\nu \gamma\gamma$, $\ell^+\ell^-\gamma\gamma$, $\ell^\pm\nu b\bar{b}$, and
$\ell^+\ell^- b\bar{b}$ final states at the LHC. Unlike previous phenomenology studies that treat
vector-boson polarisation primarily as an indirect probe to constrain dimension-6 Effective Field Theory (EFT)
operators~\cite{Banerjee:2018bio, Banerjee:2019pks}, this work establishes a direct experimental roadmap to 
measure the $W_\mathrm{L}^\pm H$, $W_\mathrm{T}^\pm H$, $Z_\mathrm{L}H$, and $Z_\mathrm{T}H$ production 
cross-sections as fundamental observables. Measuring these cross-sections directly provides the necessary 
baseline to test electroweak unitarity and symmetry restoration, complementing the considerations in 
Ref.~\cite{Huang:2020iya}. Given that this approach does not assume specific BSM coupling structures, 
it establishes a clean, model-independent benchmark for future reinterpretations. \par

To enhance the experimental reach of our approach, we exploit the complementary strengths of the two Higgs boson
decay modes under study. The analysis channels targeting $H \rightarrow \gamma\gamma$ decays are characterised by
relatively low signal statistics, but benefit from an exceptionally clean signature in the detector with low
background contamination. Conversely, the analysis channels targeting $H \rightarrow b\bar{b}$ decays provide high
signal yield, but face substantially larger background rates and a more complex background composition, requiring
more elaborate analysis techniques. \par
The key ingredients for the proposed measurements are polarisation-dependent observables. We have identified
several such observables and trained boosted decision trees using them as input features. The use of these multivariate 
classifiers enhances the separation between longitudinal and transverse states significantly, proving to be a substantial 
improvement over studying individual polarisation-dependent observables or angular asymmetries alone as proposed in 
previous literature~\cite{Hernandez-Juarez:2023dor, Panico:2017frx, Grossi:2024jae, Banerjee:2018bio, Banerjee:2019pks}.
The general measurement strategy resembles the approach employed in measurements of longitudinally polarised $Z$-boson pair 
production in four-lepton final states~\cite{ATLAS:2023zrv, ATLAS:2026fhf}.
To bridge the gap to LHC implementation, our analysis moves beyond parton-level calculations by incorporating
the estimation of dominant backgrounds, effects from parton showering and hadronisation, object-level
detector simulation via \textsc{Delphes}, and the application of realistic reconstruction techniques.
Projections on the signal-strength parameters were obtained separately for $H \rightarrow b\bar{b}$ and
$H \rightarrow \gamma\gamma$ events via maximum likelihood fits using the \texttt{pyhf} framework, incorporating an 
appropriate treatment of statistical and systematic uncertainties. In the $H \rightarrow \gamma\gamma$
channels, the fits exploit the $\mathrm{BDT}_{\mathrm{Pol}}$ response score distributions, whereas in the
$H \rightarrow b\bar{b}$ channels the fits use the $\mathrm{BDT}_{\mathrm{VH}}$ response score distributions
binned across four $p_{\mathrm{T},V}$ intervals, which are further categorised into $V_{\mathrm{L}}$-like
and $V_{\mathrm{T}}$-like sub-regions based on the $\mathrm{BDT}_{\mathrm{Pol}}$ score. Based on these fits,
we report sensitivity projections for three assumed integrated luminosities of 300\,fb$^{-1}$, 1000\,fb$^{-1}$,
and 3000\,fb$^{-1}$.\par

Our studies project that $5.0$ standard deviation discovery significances for $Z_\mathrm{L}H$ and
$W^\pm_\mathrm{L}H$ production are achievable at the LHC, requiring integrated luminosities of 210\,fb$^{-1}$
for $Z_\mathrm{L}H$ in $H \rightarrow b\bar{b}$ and 340\,fb$^{-1}$ (425\,fb$^{-1}$) for $W^\pm_\mathrm{L}H$ in
$H \rightarrow b\bar{b}$ ($H \rightarrow \gamma\gamma$). With the full High-Luminosity LHC dataset of 3000\,fb$^{-1}$,
we project that the signal-strength parameters $\mu_{W_\mathrm{L}H}$ and $\mu_{Z_\mathrm{L}H}$ can be measured
with precisions of $13\%$ and $8\%$ in $H \rightarrow b\bar{b}$ final states, and $10\%$ and $35\%$ in
$H \rightarrow \gamma\gamma$ final states, respectively. These findings demonstrate that isolating longitudinally
polarised vector bosons produced in association with a Higgs boson is well within the experimental reach of future
LHC analyses. Remarkably, these results highlight that $WH$ production in the $H \rightarrow \gamma\gamma$ channel
(unexplored in previous literature) is highly competitive with $H \rightarrow b\bar{b}$ at 3000\,fb$^{-1}$, as
its exceptionally low background contamination enables high signal purity despite the smaller branching ratio. \par

Furthermore, we project 68\% CL constraints on $W^\pm_\mathrm{T}H$ production with a precision of $41\%$ in
$H \rightarrow \gamma\gamma$ and on $Z_\mathrm{T}H$ production with a precision of $28\%$ in $H \rightarrow b\bar{b}$
final states. For studies of $W^\pm_\mathrm{T}H$ production, final states with $H \rightarrow \gamma\gamma$ decays are
particularly crucial, given that analyses targeting $WH(\rightarrow b\bar{b})$ decays lack sensitivity in the
low-$p_{\mathrm{T},V}$ regime dominated by the $W^\pm_\mathrm{T}H$ process (as further discussed in App.~\ref{sec:AppI}).
While we expect the performance for transversely polarised states could be further enhanced through dedicated optimisations,
we leave such dedicated studies for future work. \par

We note that the fit model employed by the maximum likelihood procedure in our analyses is simplified in several respects. 
For the results presented in this article, we have not taken interference effects between $V_\mathrm{L}H$ and $V_\mathrm{T}H$ 
production into account directly in the fit. These interference effects appear to be negligible at tree-level within the 
expected precision of the signal-strength measurements for $V_\mathrm{L}H$ and $V_\mathrm{T}H$ at the LHC, as indicated by our 
dedicated studies presented in App.~\ref{App:Interference}. Nevertheless, such studies will need to be refined in future work 
incorporating higher-order electroweak and QCD corrections into the cross-section calculations for $V_\mathrm{L}H$ and 
$V_\mathrm{T}H$ production~\cite{Denner:2021csi, Poncelet:2021jmj, Le:2022lrp}. At present, we are confident that our proposed 
measurement strategies are robust, given that contributions from interference effects between longitudinal and transversely 
polarised vector bosons~\cite{Panico:2017frx, Denner:2021csi} play only a sub-dominant role in cross-section measurements of 
$V_\mathrm{L}V_\mathrm{L}$ production~\cite{ATLAS:2023zrv,ATLAS:2026fhf}. The likelihood procedure also neglects contributions 
from backgrounds containing misidentified photons and leptons, as well as most experimental and several theoretical systematic 
uncertainties. Given that statistical uncertainties will dominate in the $H \rightarrow \gamma\gamma$ channels and that 
background estimates can be derived using data-driven methods, thereby strongly constraining both their normalisation and 
shape, we argue that our results nevertheless provide a reliable estimate of the achievable sensitivity. The situation in the 
$H \rightarrow b\bar{b}$ channel differs, as the results are not statistically limited. For this study, we have assumed 
conservative uncertainties following those obtained for dominant backgrounds in recent $VH(\rightarrow b\bar{b})$ analyses by 
the ATLAS collaboration. Therefore, we consider these projections to be realistic, with the expectation that the ultimate 
precision achieved by the experiments may be even superior as systematic uncertainties will improve over the lifespan of the 
HL-LHC. \par

Based on these results, we recommend pursuing such studies using LHC proton-proton collision data. We also suggest extending 
these investigations to other Higgs boson decay modes, in particular $H \rightarrow \tau^+\tau^-$, which features a 
significantly higher branching ratio than $H \rightarrow \gamma\gamma$ (though lower than $H \rightarrow b\bar{b}$) and a 
cleaner background composition than $H \rightarrow b\bar{b}$ (though less clean than $H \rightarrow \gamma\gamma$). However, 
these decay modes will require substantial effort to understand the impact that additional neutrinos from $\tau$-lepton decays 
will have on the resolution of polarisation-dependent observables.\par

Further improvements on the measurements discussed in this work can be obtained by extending our approach to differential
$p_{\mathrm{T},V}$ bins within the Simplified Template Cross Sections (STXS) framework~\cite{Brehmer:2019gmn}, which
will directly map back to constraints on high-energy primary parameters and dimension-6 EFT operators proposed in previous
studies~\cite{Franceschini:2017xkh, Banerjee:2018bio, Banerjee:2019pks}. \par
 \section*{Acknowledgements}

D.D. is supported by the UK Science and Technology Facilities Council (STFC) and wishes to thank Xiaoli Chen for the design of 
the schematic representation of polarisation-dependent observables depicted in Fig.~\ref{fig:VectorBosonDecayPlanes}. Furthermore, 
we would like to thank Giovanni Pelliccioli for useful discussions.
 \begin{appendices}
\section{Polarisation fraction dependence on the vector boson transverse momentum}\label{sec:AppI}

\begin{figure}[!h]
\centering
\subfloat[]{\includegraphics[width=0.5\textwidth]{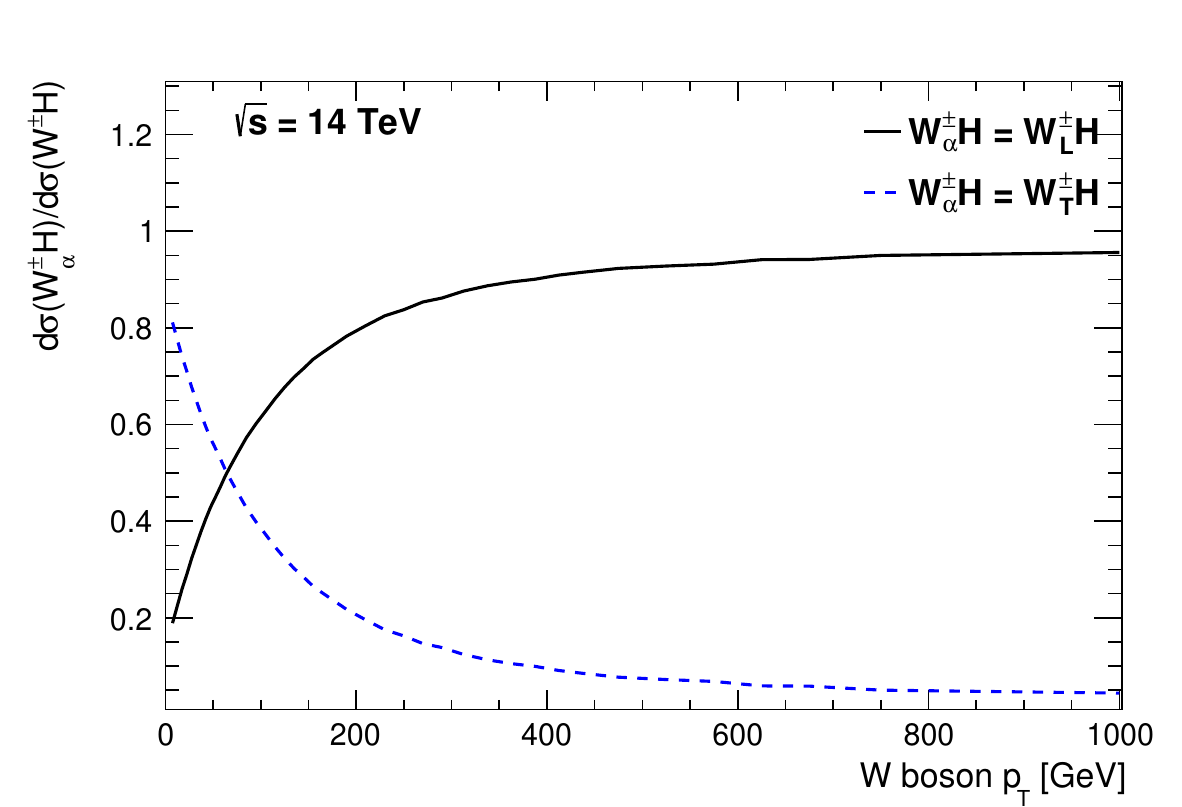}}
\subfloat[]{\includegraphics[width=0.5\textwidth]{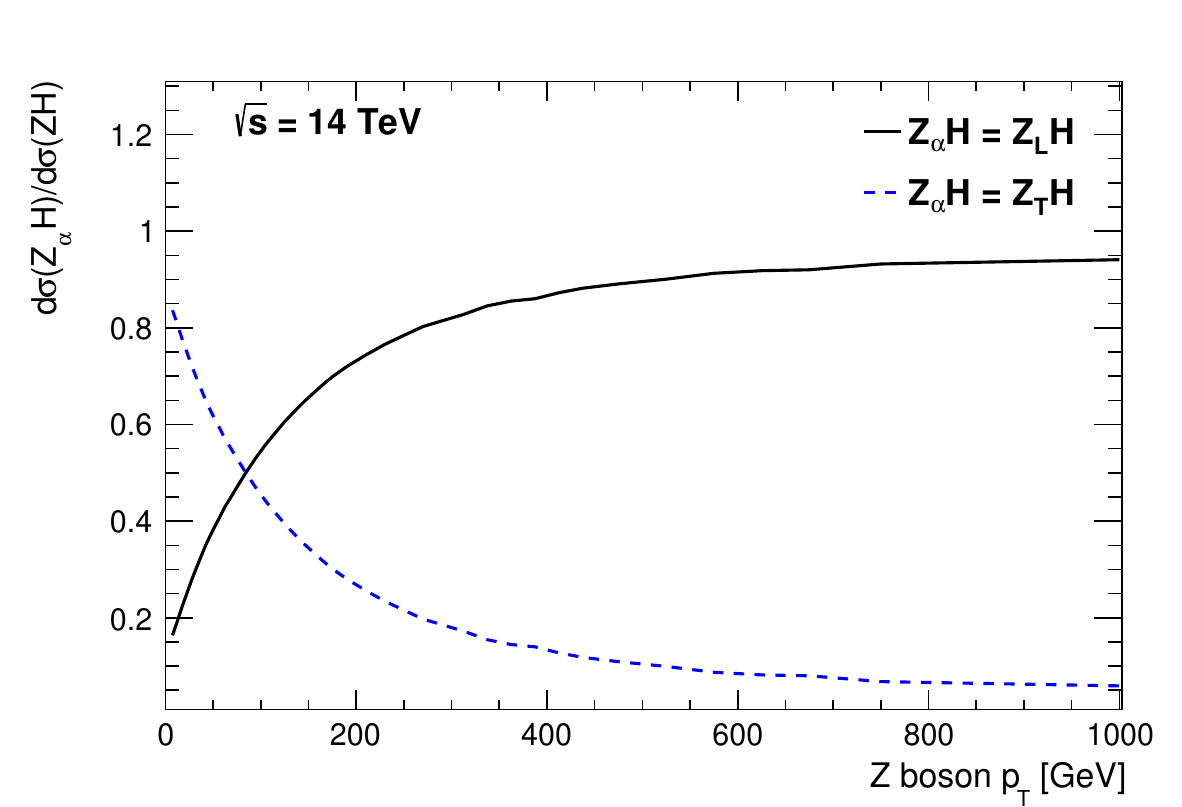}}
\caption{Fraction of longitudinally ($V_\mathrm{L}$) and transversely ($V_\mathrm{T}$) polarised vector bosons in
Higgs-Strahlung ($VH$) production, shown as a function of the transverse momentum, $p_\mathrm{T}$, of the (a) $W^\pm$
boson and (b) $Z$ boson. The cross-sections for $V_\mathrm{L}H$ and $V_\mathrm{T}H$ production were calculated
using \MGMCatNLOV3.5.1 and \PYTHIA8.1.3, following the details in Section~\ref{sec:EventGeneration}.}
\label{fig:VL_Fractions}
\end{figure}

The polarisation of vector bosons produced in association with a Higgs boson can be either transverse or longitudinal. As shown in 
Fig.~\ref{fig:VL_Fractions}, their relative fractions vary with the vector boson's transverse momentum. To maximise sensitivity to 
BSM effects, future production cross-section measurements should ideally be performed in distinct intervals of the vector boson 
$p_\mathrm{T}$.\par
One good option for this is to use simplified template cross-sections as discussed in Ref.~\cite{Brehmer:2019gmn}. Since the
separation strength of characteristic observables can change significantly with the vector boson's transverse momentum, separate
machine learning techniques might need to be trained for different kinematic regions. We will leave more detailed investigations
into this dependence for future work.\par
Given the dominance of longitudinally polarised vector bosons in $VH$ events at high $p_\mathrm{T}$, this channel is an ideal probe
for testing the Goldstone boson equivalence theorem, as argued in Ref.~\cite{Huang:2020iya}.\par

 \section{Event classification for the $VH(\rightarrow b\bar{b})$ signal extraction}\label{App:BDT_Training}

Sets of boosted decision trees, denoted as $\mathrm{BDT}_{VH}$, are used to further improve the sensitivity to both the $V_{L}H(\rightarrow b\bar{b})$ and $V_{T}H(\rightarrow b\bar{b})$ 
signal processes. The general training strategy for these BDTs resembles that of recent ATLAS analyses targeting $H\rightarrow b\bar{b}$ decays~\cite{ATLAS:2024yzu}, 
with minor modifications applied to remove observables exhibiting significant sensitivity to the vector boson polarisation. The $V_{L}H$ and $V_{T}H$ signal 
processes are combined into a single signal class and trained against the sum of all background processes, with individual background contributions weighted 
according to their cross-section predictions. \par
The discrimination is performed across four intervals of vector boson transverse momentum ($p_{\mathrm{T},V}$): $p_{\mathrm{T},V} < 75\,\mathrm{GeV}$, 
$75\,\mathrm{GeV} \le p_{\mathrm{T},V} < 150\,\mathrm{GeV}$, $150\,\mathrm{GeV} \le p_{\mathrm{T},V} < 250\,\mathrm{GeV}$, and 
$p_{\mathrm{T},V} \ge 250\,\mathrm{GeV}$. Dedicated BDTs are trained for each of the lower two regions, whereas a single BDT is trained across the 
upper two intervals ($p_{\mathrm{T},V} \ge 150\,\mathrm{GeV}$). Similar to the training of the $\mathrm{BDT}_\mathrm{Pol}$ (see Sec.~\ref{sec:BDT_Training}), 
a two-fold cross-validation scheme is employed for the $\mathrm{BDT}_{VH}$ training, in which events within each $p_{\mathrm{T},V}$ interval are randomly split 
into two equal-sized subsamples.\par
The BDTs are implemented using the TMVA package~\cite{TMVA}, where each ensemble consists of 600 decision trees trained with the Gradient 
Boost algorithm~\cite{GradientBoost}, using a learning rate of $0.1$ and a maximum tree depth of five. The input features used 
across the various BDTs are listed in Table~\ref{tab:MVA_Training_Inputs} and described below. Dedicated feature sets are used for 
the 1- and 2-lepton channels.\par
The invariant mass of the reconstructed Higgs boson candidate, $m_H$, is the most discriminating observable in both channels across
all $p_{\mathrm{T},V}$ regions, as no background process yields a peaking structure at the Higgs boson mass. In the 1-lepton channel,
the second highest-ranked feature is the reconstructed top-quark mass, $m_{\ell\nu b}$, calculated from the four-momentum sum of the
charged lepton, the neutrino, and the $b$-tagged jet that minimises $|m_{\ell\nu b} - 172.5\,\mathrm{GeV}|$. \par
Other observables common to both channels include the invariant mass of the three-jet system ($m_{j_{1}j_{2}j_{3}}$) constructed from
the two Higgs-candidate jets and the leading additional jet, the transverse momenta of the two $b$-tagged jets ($p_{\mathrm{T},b_{1}}$
and $p_{\mathrm{T},b_{2}}$), the scalar sum of transverse momenta of additional jets with $p_{\mathrm{T}} > 20\,\mathrm{GeV}$ ($\sum_{i} p_{\mathrm{T},j_{i}}$),
the transverse momenta of the reconstructed Higgs boson ($p_{\mathrm{T},H}$) and vector boson ($p_{\mathrm{T},V}$) candidates,
the pseudorapidity difference between the Higgs and vector boson candidates ($\Delta\eta_{VH}$), and the angular separation between
the two $b$-tagged jets ($\Delta R_{b\bar{b}}$).\par

\begin{table}[!h]
\centering
\renewcommand{\arraystretch}{1.15}
\caption{Observables used as input features for the $\mathrm{BDT}_{VH}$ training in the
$VH(H\rightarrow b\bar{b})$ channels. The $\checkmark$ symbol indicates that an observable is included in the
corresponding training.}
\begin{tabular}{c | c c}
\hline
\hline
Observable                                       & 1-lepton channel & 2-lepton channel \\
\hline
$m_{H}$                                          &    $\checkmark$  &  $\checkmark$  \\
$m_{j_{1}j_{2}j_{3}}$                            &    $\checkmark$  &  $\checkmark$  \\
$p_{\mathrm{T},b_{1}}$                           &    $\checkmark$  &  $\checkmark$  \\
$p_{\mathrm{T},b_{2}}$                           &    $\checkmark$  &  $\checkmark$  \\
$\sum_{i} p_{\mathrm{T},L_{i}}$                  &    $\checkmark$  &  $\checkmark$  \\
$p_{\mathrm{T},V}$                               &    $\checkmark$  &  $\checkmark$  \\
$p_{\mathrm{T},H}$                               &    $\checkmark$  &  $\checkmark$  \\
$\eta_{VH}$                                      &    $\checkmark$  &  $\checkmark$  \\
$\Delta R_{b\bar{b}}$                            &    $\checkmark$  &  $\checkmark$  \\
$E^\mathrm{miss}_\mathrm{T}$                     &    $\checkmark$  &                \\
$m_{\mathrm{T},W}$                               &    $\checkmark$  &                \\
$m_\mathrm{\ell\nu b}$                           &    $\checkmark$  &                \\
$\min\left[\Delta\Phi(\ell, j_{i}) \right]$      &    $\checkmark$  &                \\
$m_{\ell\ell}$                                   &                  &  $\checkmark$  \\
$E^\mathrm{miss}_\mathrm{T}/\sqrt{H_\mathrm{T}}$ &                  &  $\checkmark$  \\
\hline
\hline
\end{tabular}
\label{tab:MVA_Training_Inputs}
\end{table}
\clearpage

The 1-lepton channel additionally incorporates the missing transverse momentum ($E_\mathrm{T}^\mathrm{miss}$), the transverse mass of 
the $W$ boson candidate ($m_{\mathrm{T},W}$), and the minimum azimuthal separation between either Higgs-candidate jet and any 
additional jet ($\min\left[\Delta\Phi(\ell, j_{i}) \right]$). The 2-lepton channel further includes the dilepton invariant 
mass ($m_{\ell\ell}$) and the missing transverse momentum significance ($E_\mathrm{T}^\mathrm{miss}/\sqrt{H_\mathrm{T}}$), where 
$H_\mathrm{T}$ is defined as the scalar sum of the $p_\mathrm{T}$ of all selected leptons and jets.\par

\begin{figure}[h!]
\centering
\subfloat[]{\includegraphics[width=0.245\textwidth]{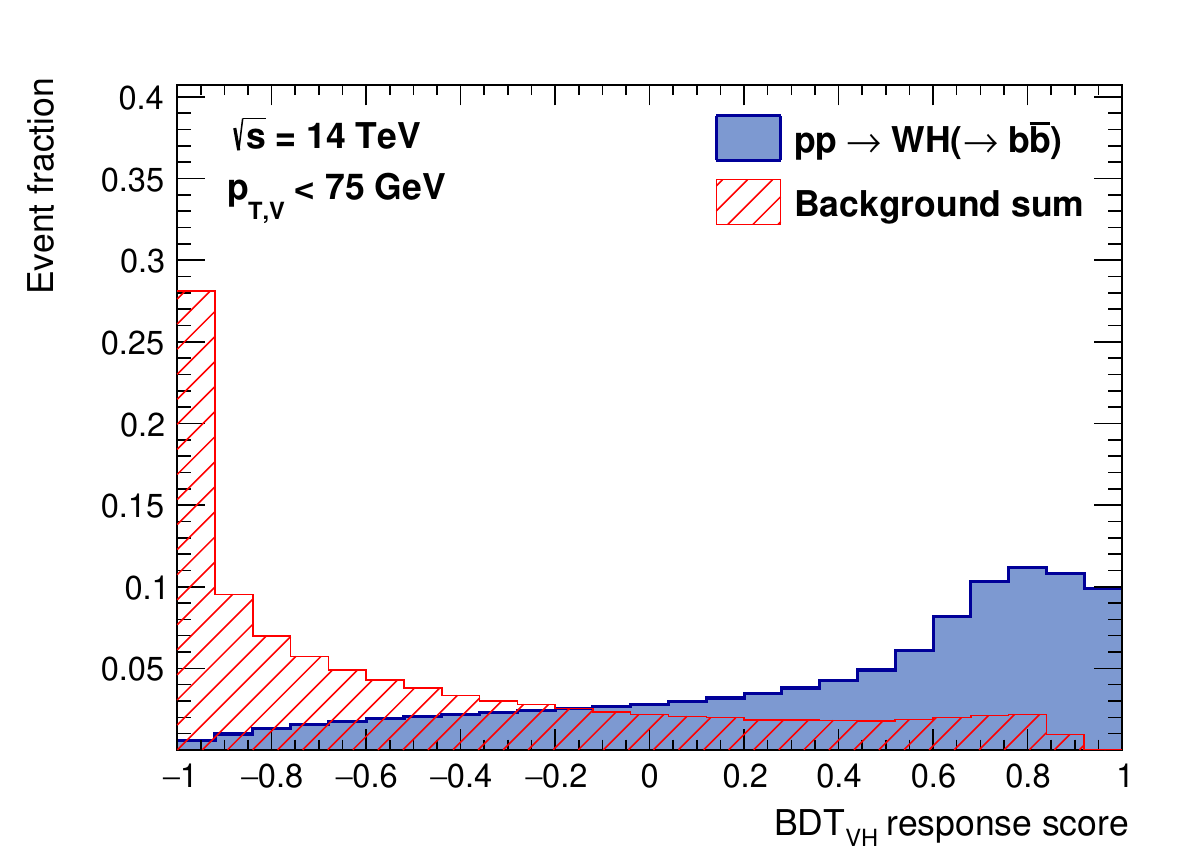}}
\subfloat[]{\includegraphics[width=0.245\textwidth]{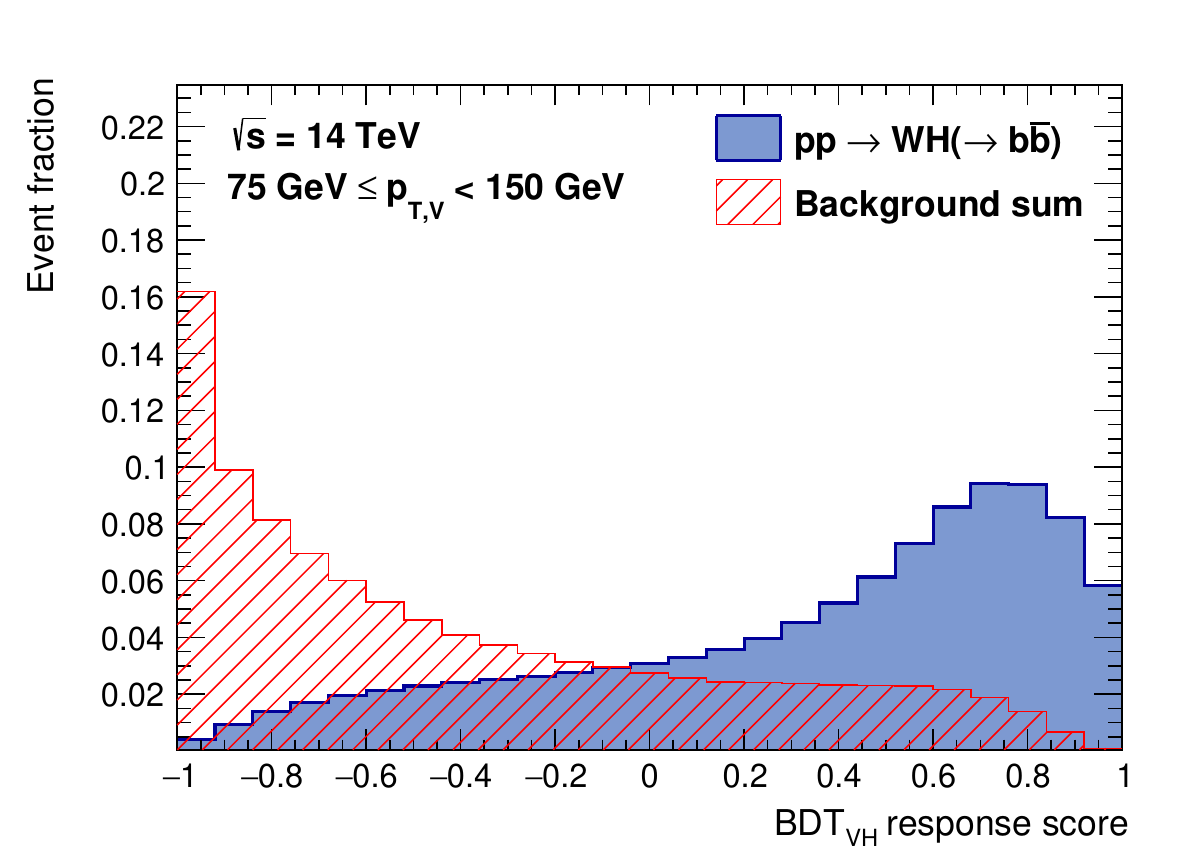}}
\subfloat[]{\includegraphics[width=0.245\textwidth]{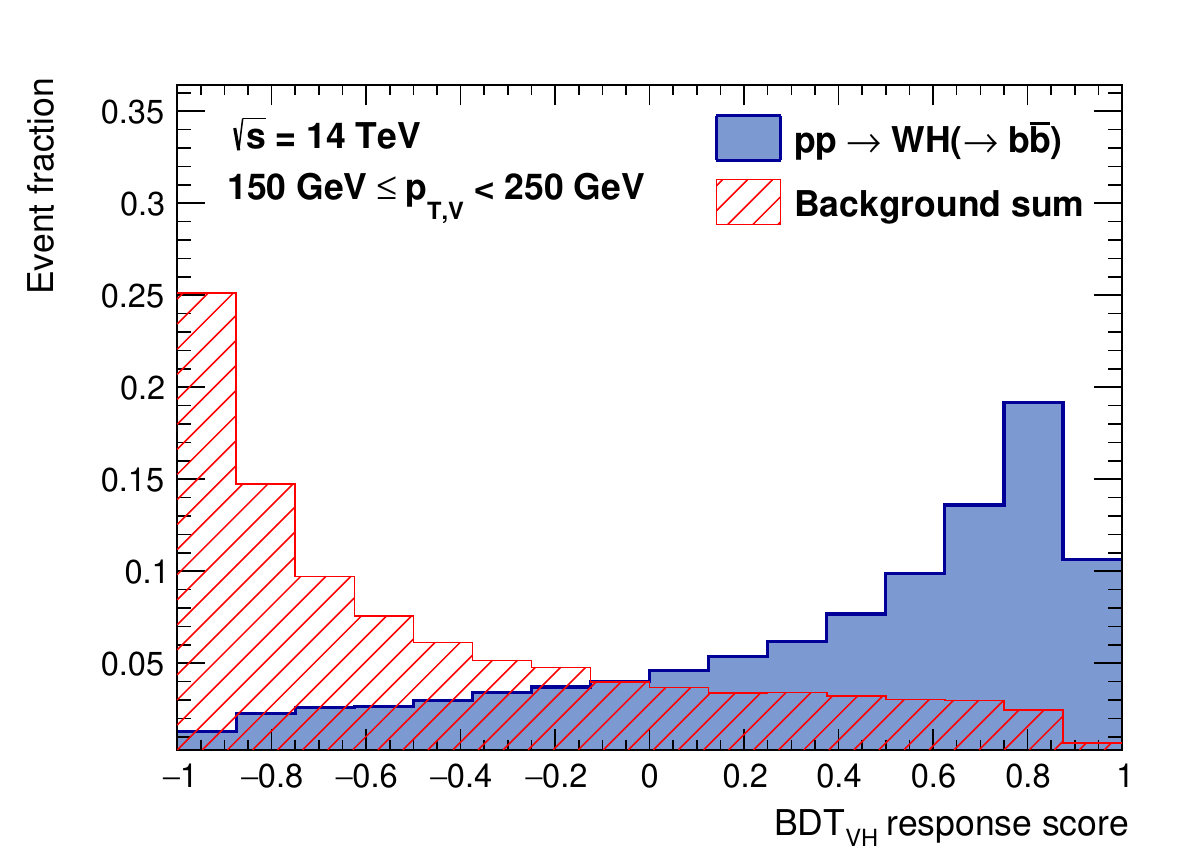}}
\subfloat[]{\includegraphics[width=0.245\textwidth]{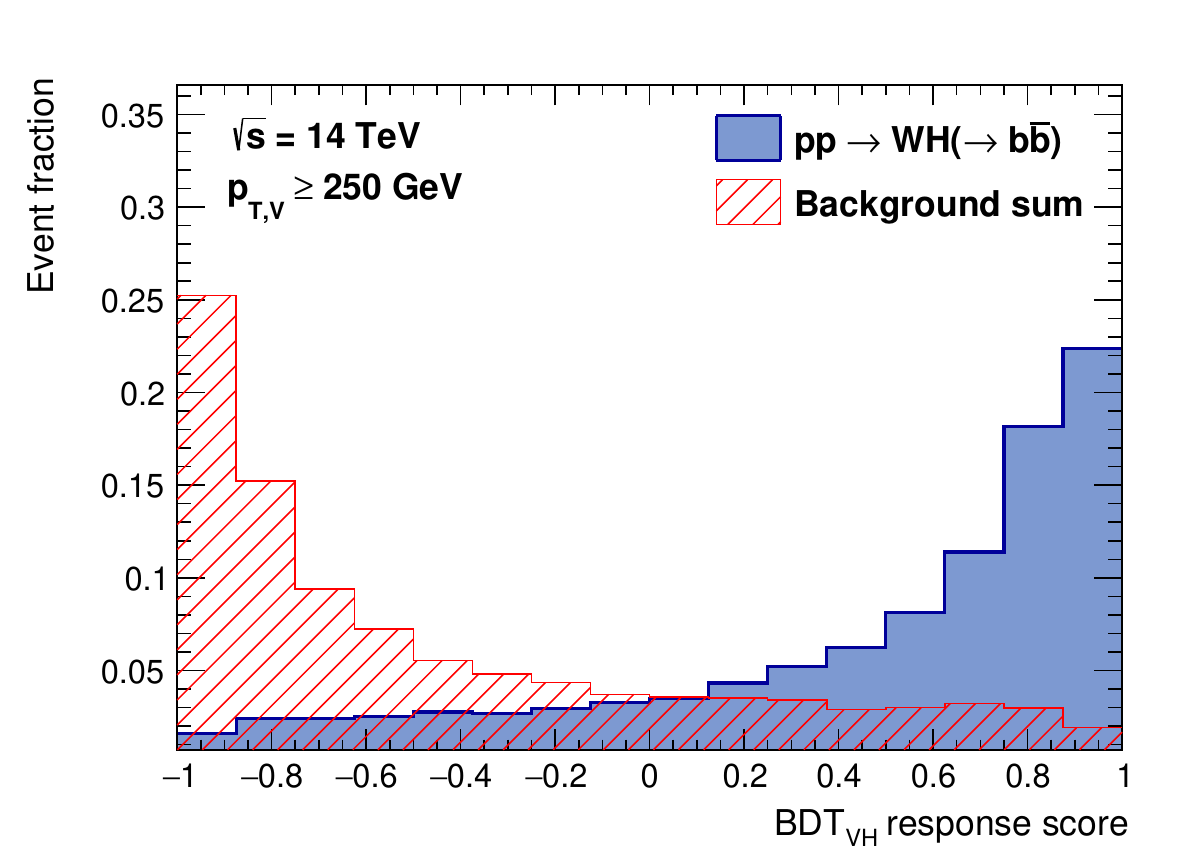}}\\
\subfloat[]{\includegraphics[width=0.245\textwidth]{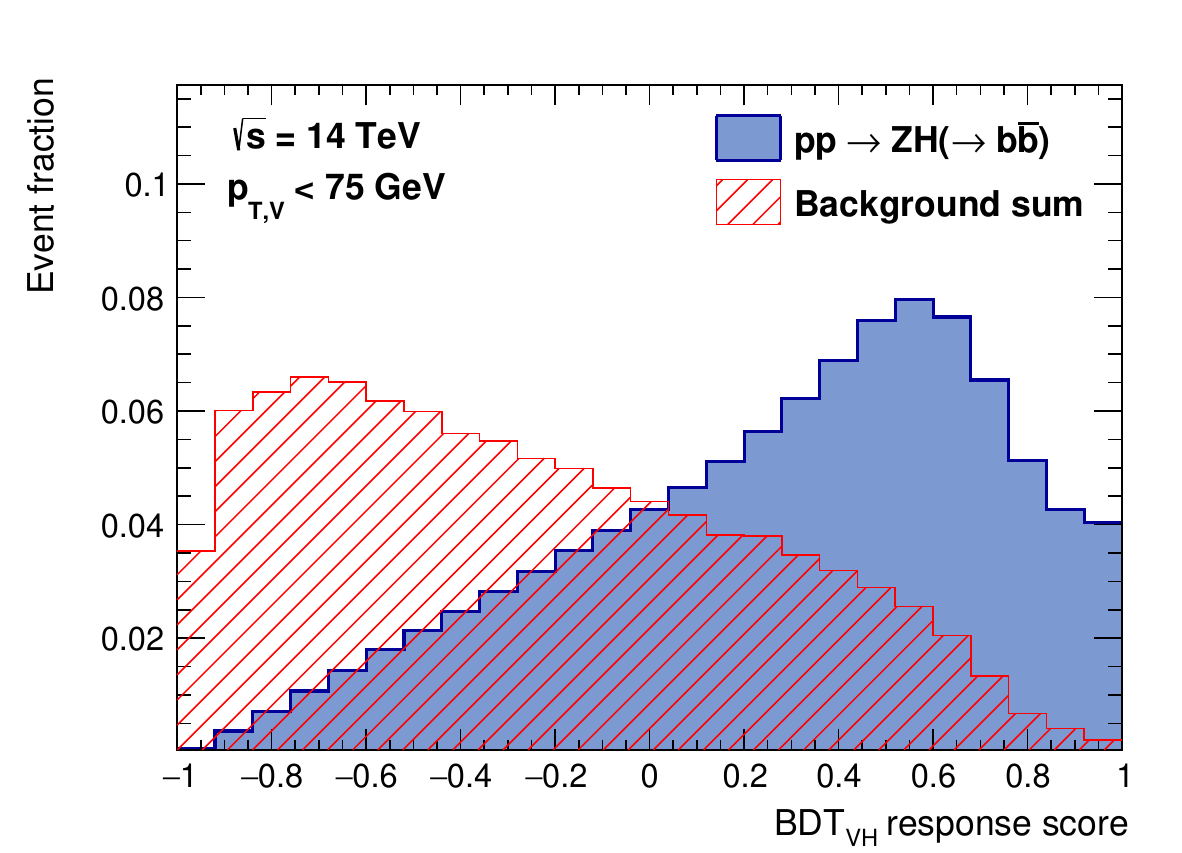}}
\subfloat[]{\includegraphics[width=0.245\textwidth]{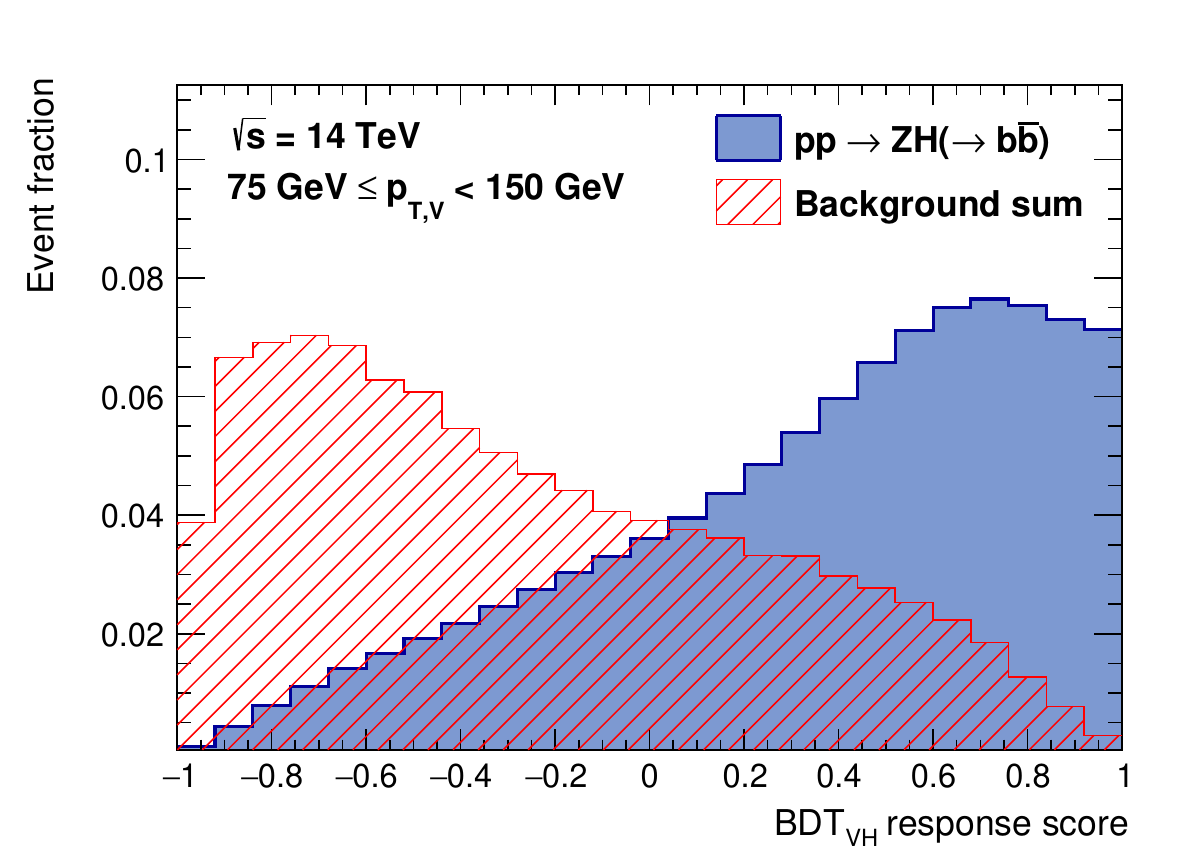}}
\subfloat[]{\includegraphics[width=0.245\textwidth]{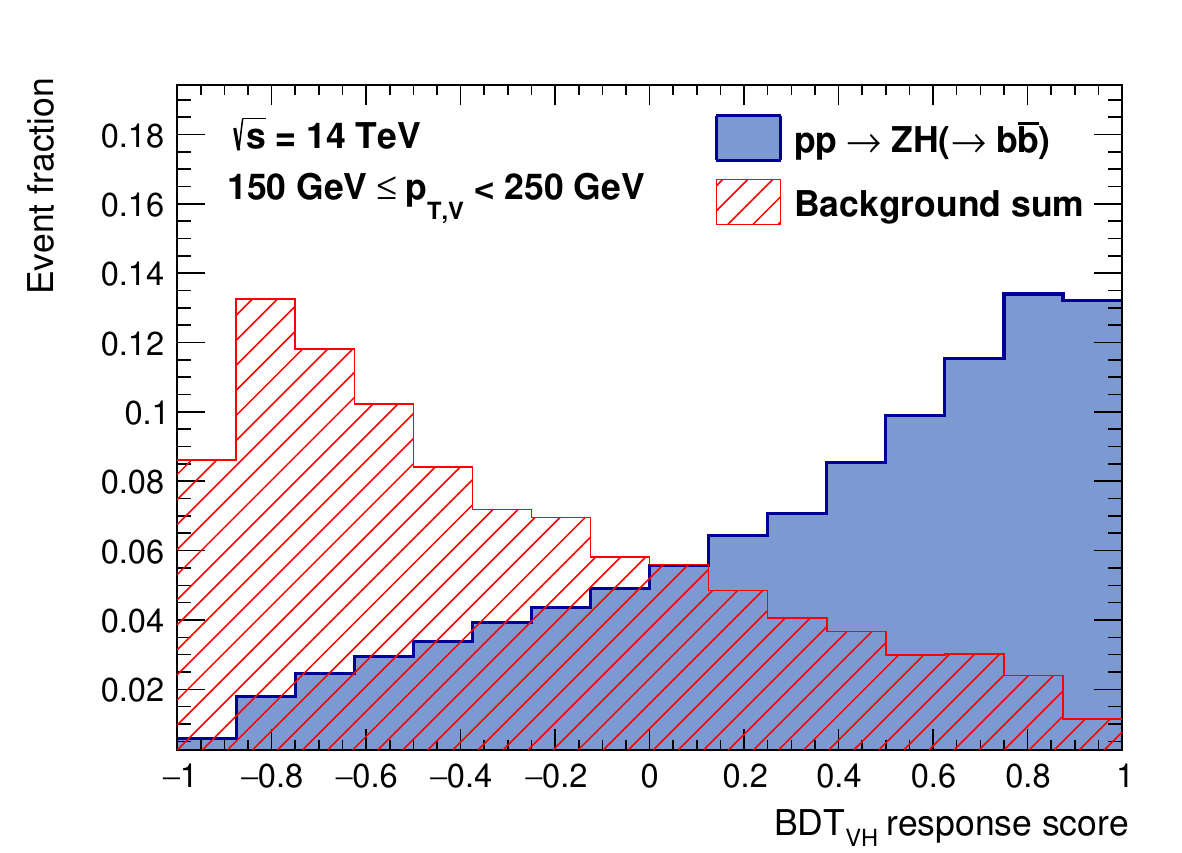}}
\subfloat[]{\includegraphics[width=0.245\textwidth]{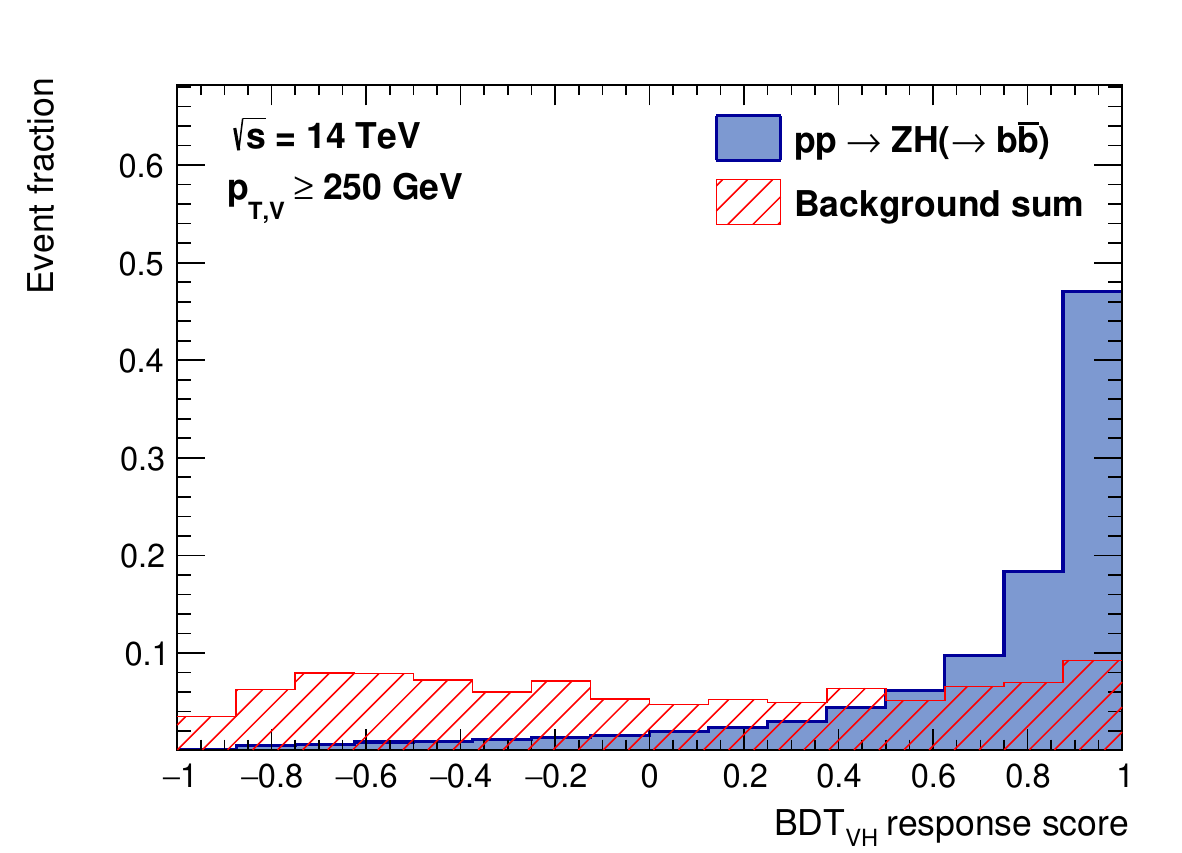}}
\caption{Distributions of the $\mathrm{BDT}_{\mathrm{VH}}$ response score for $W^\pm H(\rightarrow b\bar{b})$ (top row) and 
$ZH(\rightarrow b\bar{b})$ (bottom row) events as well as the corresponding backgrounds, presented separately in the four 
$p_{\mathrm{T},V}$ intervals. All distributions are normalised to unit area. }
\label{fig:BDTVH_Training}
\end{figure}
 \section{Interference Effects}\label{App:Interference}

The simulated polarised samples described in Section~\ref{sec:EventGeneration} do not consider 
the interference effects between the $V_\mathrm{L}H$ and $V_\mathrm{T}H$ production modes. 
This appendix is dedicated to quantifying the scale of these neglected interference effects.\par
The study is performed specifically using $ZH$ production as a representative benchmark, 
exploiting the fully reconstructible decay kinematics of the $Z$ boson to cleanly isolate 
the polarisation features. This is done by comparing simulated $Z_\mathrm{L}H$ and $Z_\mathrm{T}H$ 
samples to an inclusive sample of unpolarised $ZH$ production, which implicitly contains 
the full interference contributions. This unpolarised sample is produced with the exact same 
configuration of the \MGMCatNLOV{} generator as the two polarised samples.\par
Interference effects are evaluated across all input features used when training the 
polarisation-separating BDT (described in Section~\ref{sec:PolarisationObservables}), as well as on the 
resulting BDT response distribution.\footnote{For the purpose of this study, the BDT was re-trained 
using parton-level kinematic features.} All comparisons are evaluated at parton-level in order to isolate 
the intrinsic quantum-mechanical interference at the matrix-element level without detector resolution 
and acceptance effects obscuring the kinematic distributions.\par

\begin{figure}[h!]
\centering
\subfloat[]{\includegraphics[width=0.575\textwidth]{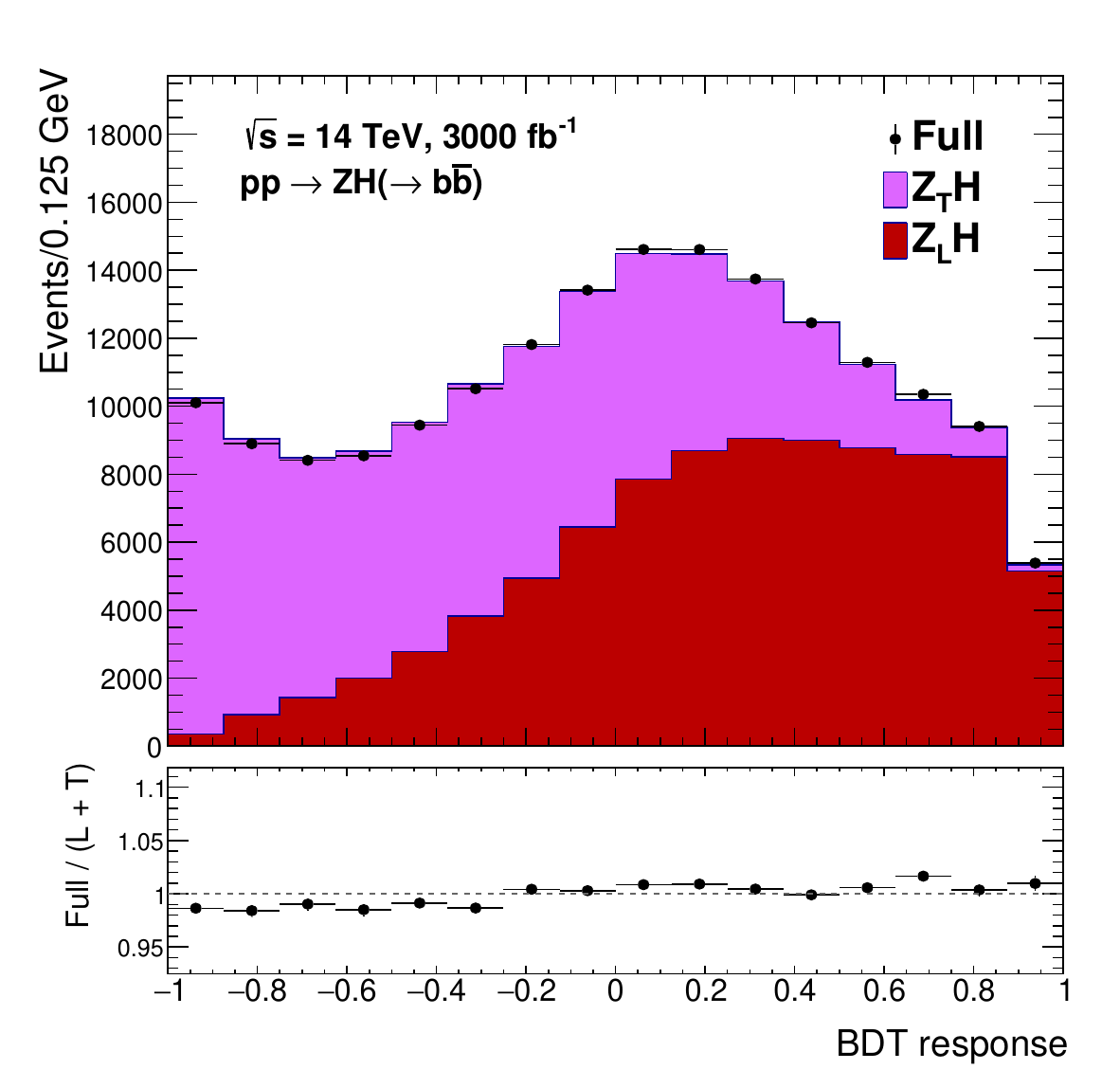}}
\caption{Distribution of the $\mathrm{BDT}_\mathrm{Pol}$ response score for simulated
$ZH \rightarrow \ell^+\ell^- b\bar{b}$ events. The stacked histograms represent the sum of longitudinally ($Z_\text{L}H$)
and transversely ($Z_\text{T}H$) polarised components compared to an inclusive sample of unpolarised $ZH$ production. All
distributions are normalised to their expected cross sections. The lower panel displays the ratio of the unpolarised sample
to the sum of the polarised samples.}
\label{fig:PolarisationBDTInterference}
\end{figure}

\begin{figure}[h!]
\centering
\subfloat[]{\includegraphics[width=0.2675\textwidth]{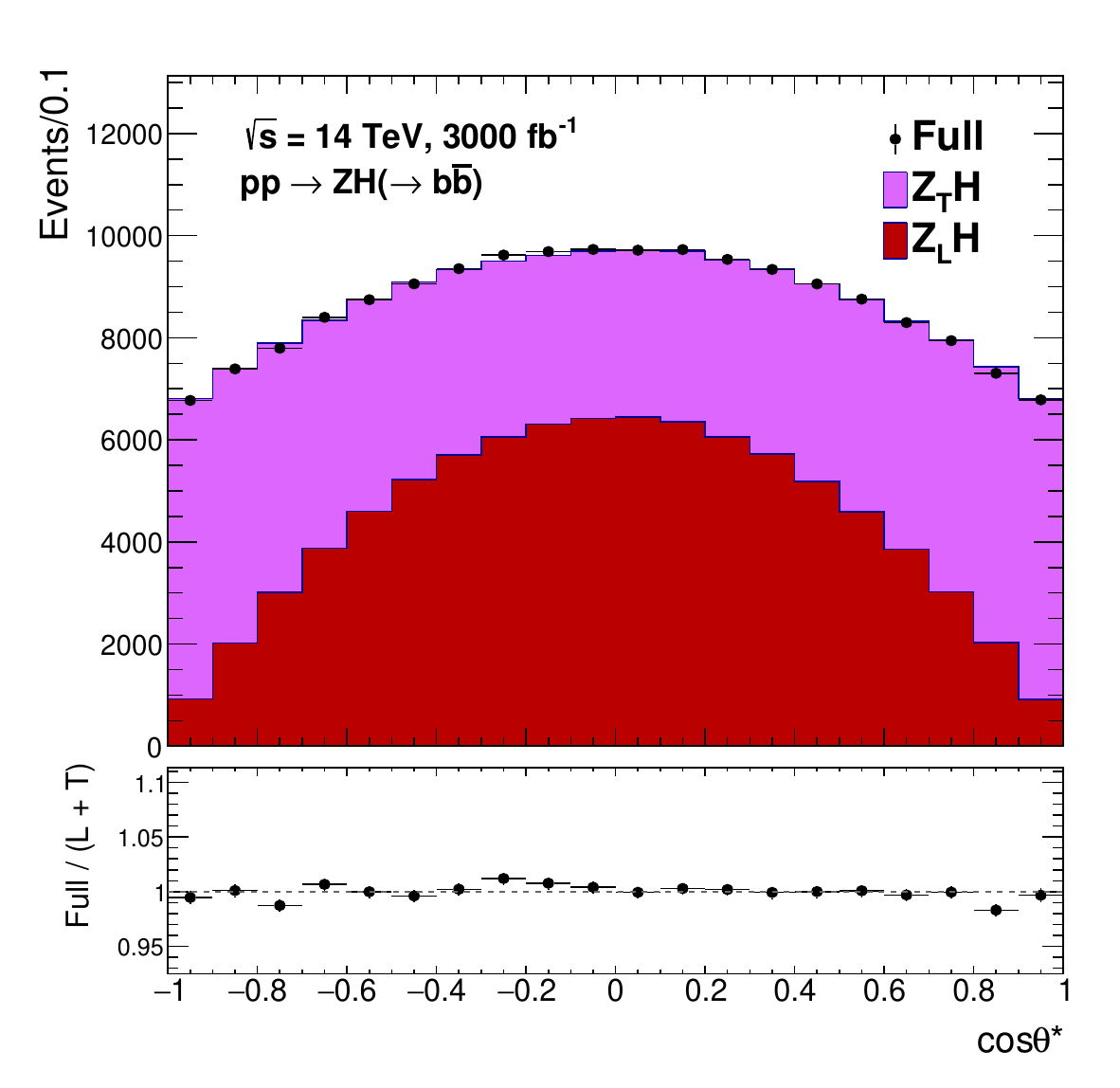}}
\subfloat[]{\includegraphics[width=0.2675\textwidth]{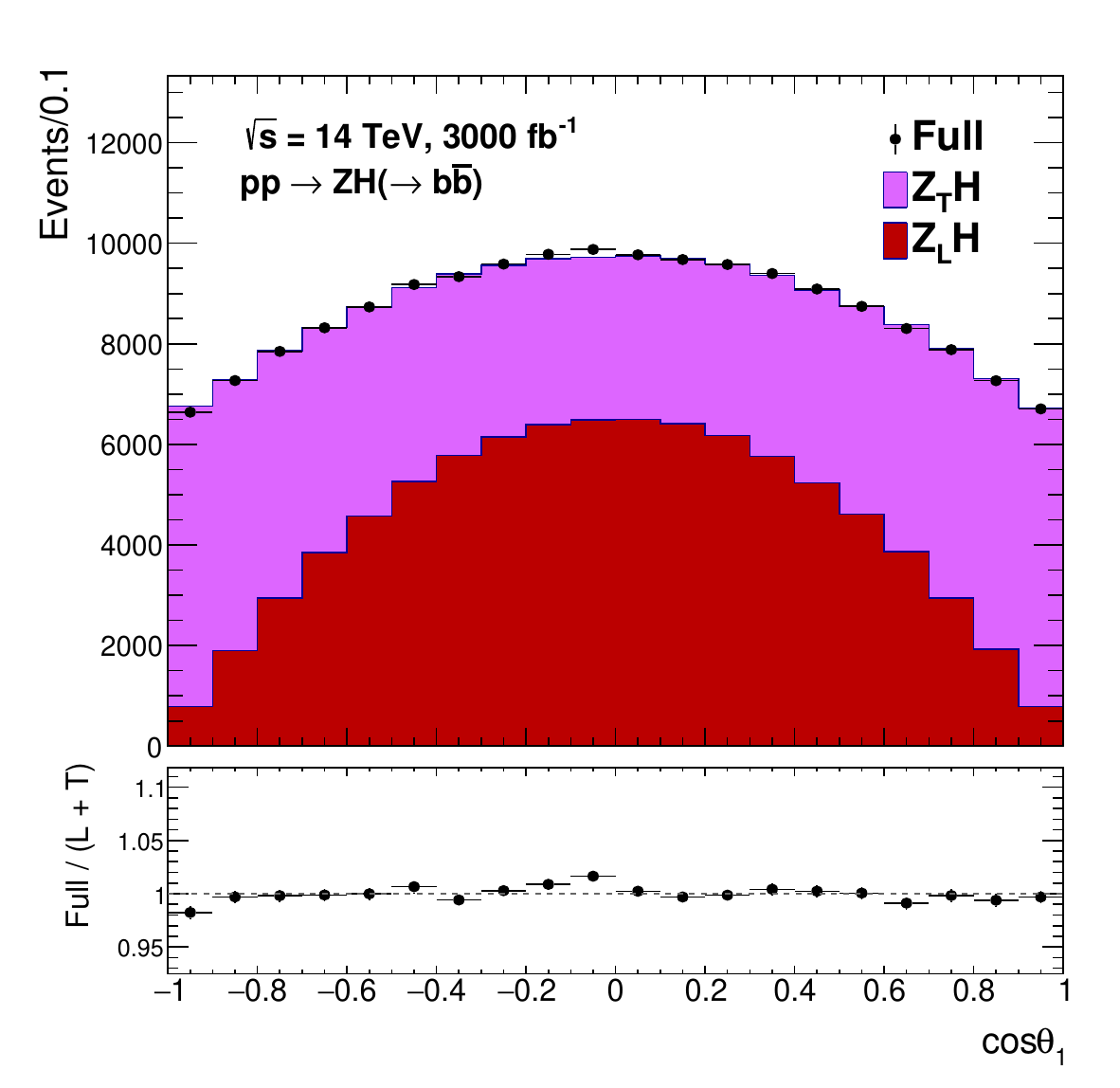}}
\subfloat[]{\includegraphics[width=0.2675\textwidth]{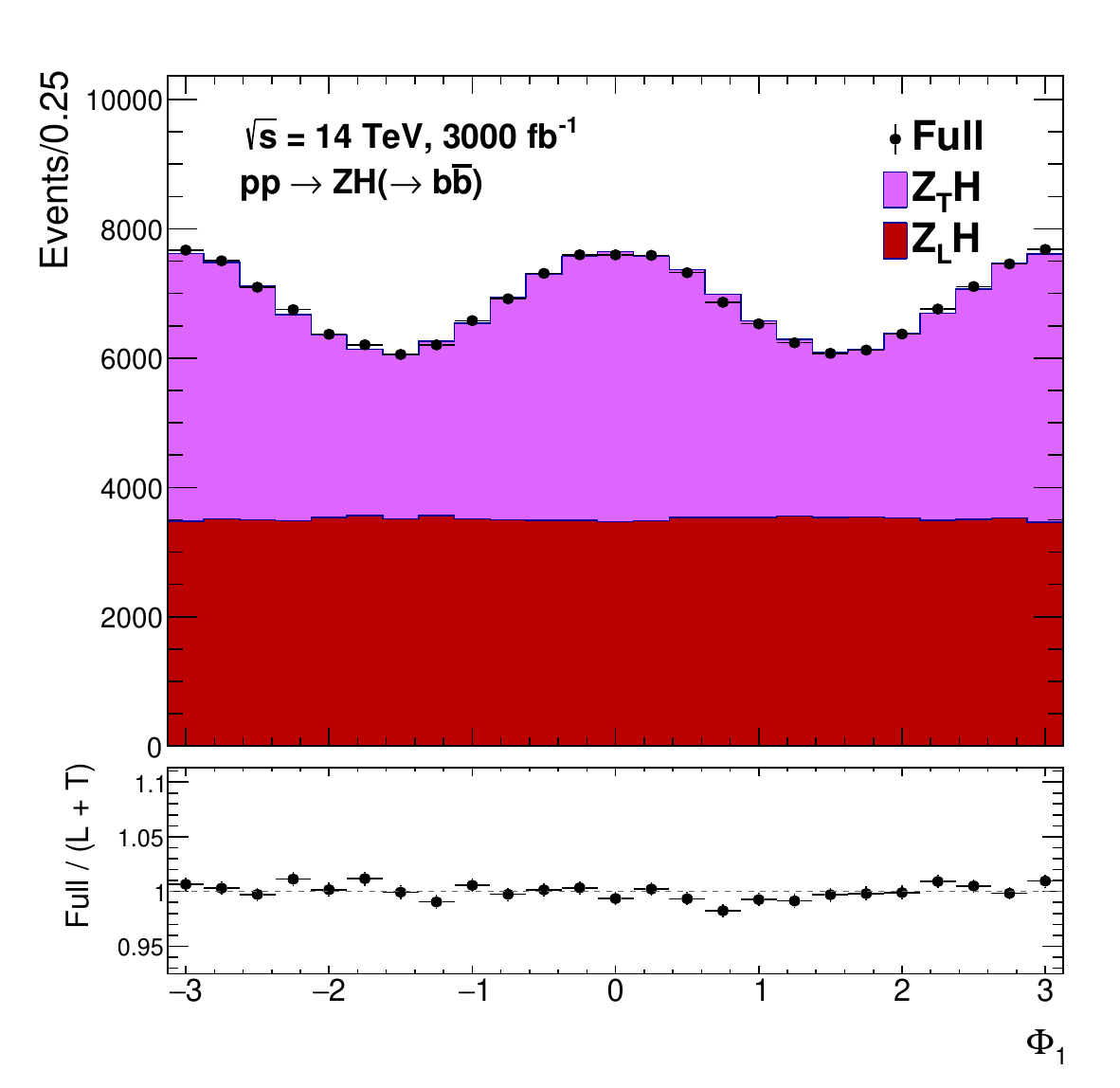}}\\
\subfloat[]{\includegraphics[width=0.2675\textwidth]{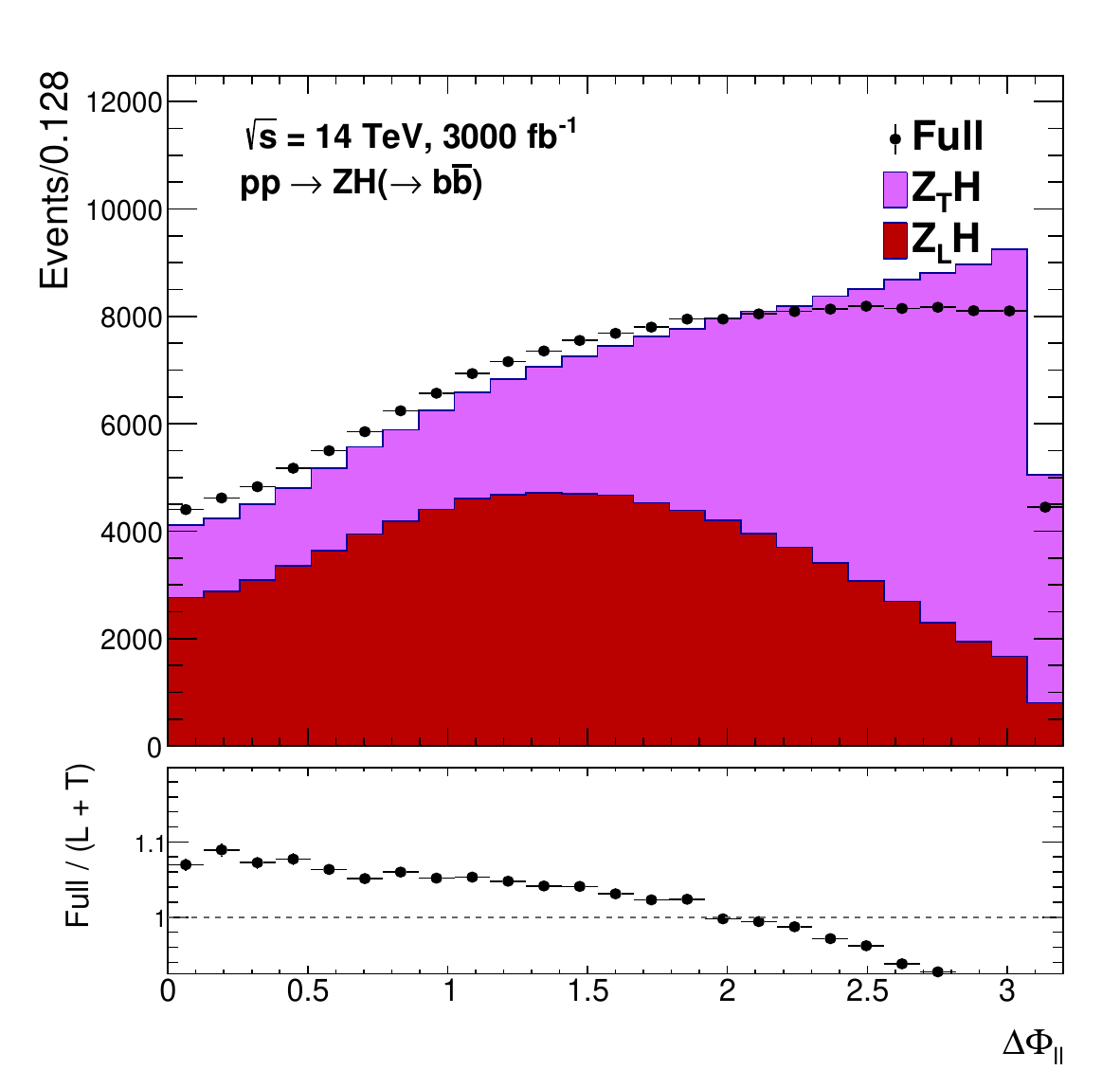}}
\subfloat[]{\includegraphics[width=0.2675\textwidth]{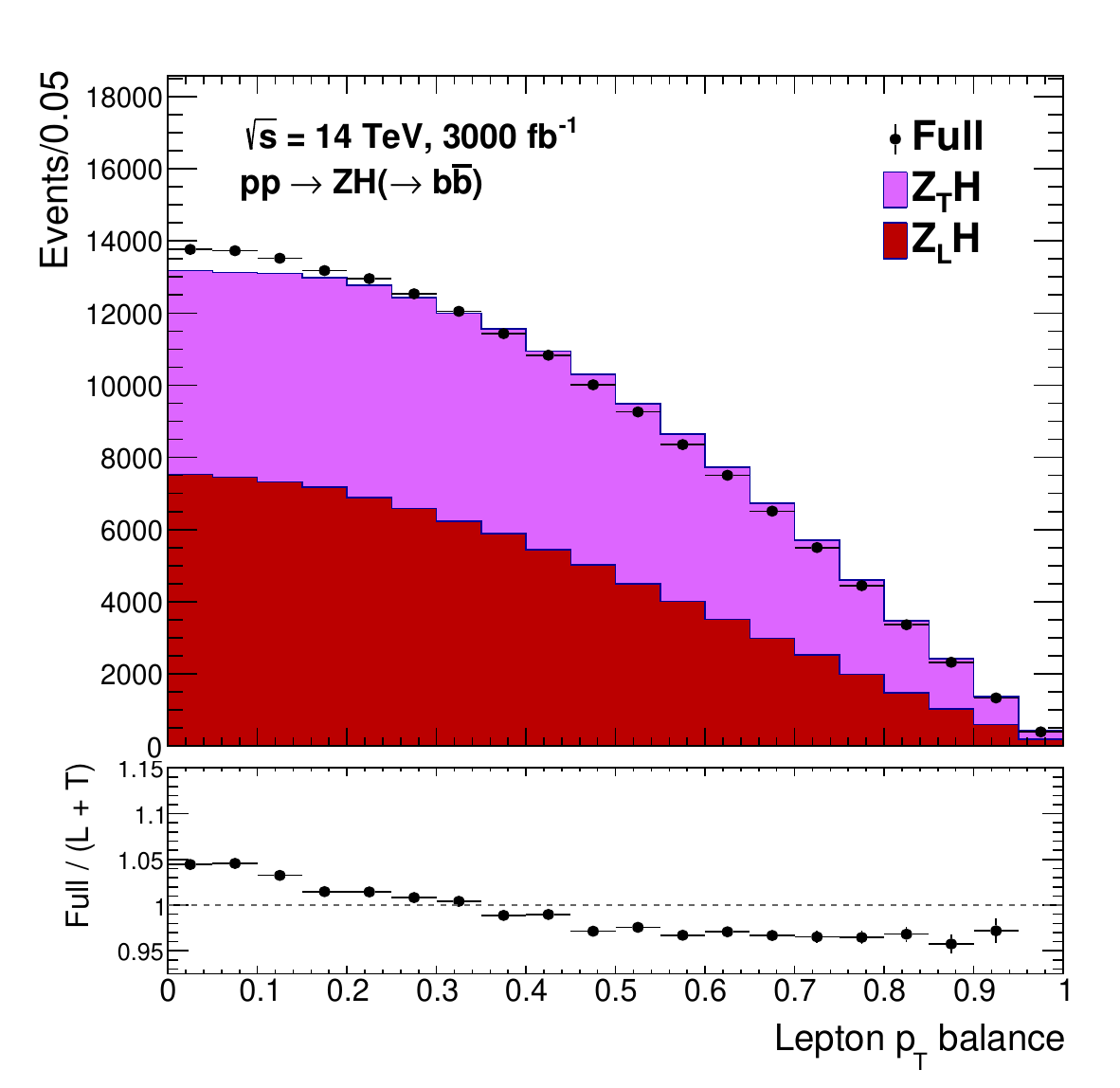}}\\
\subfloat[]{\includegraphics[width=0.2675\textwidth]{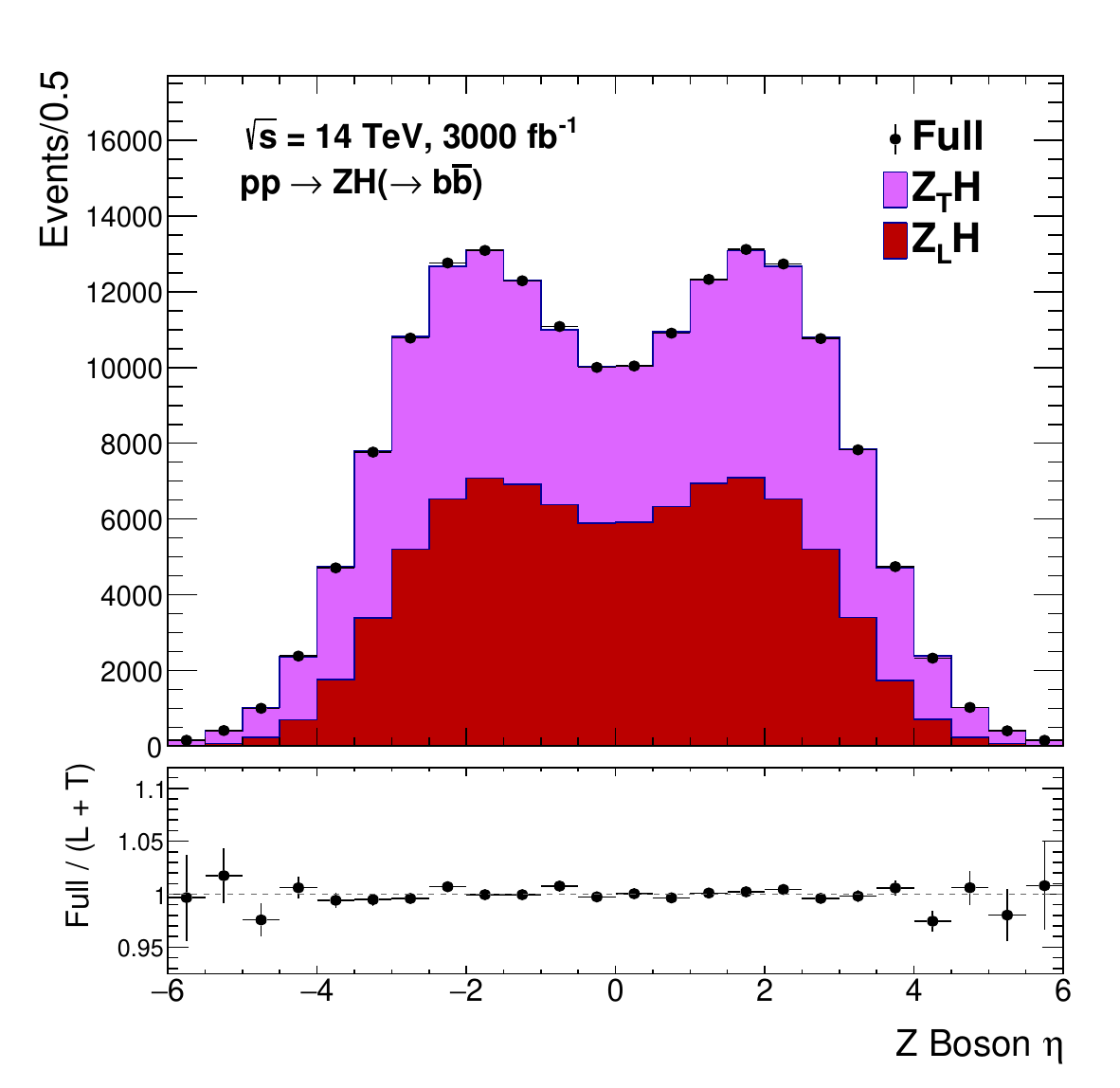}}
\subfloat[]{\includegraphics[width=0.2675\textwidth]{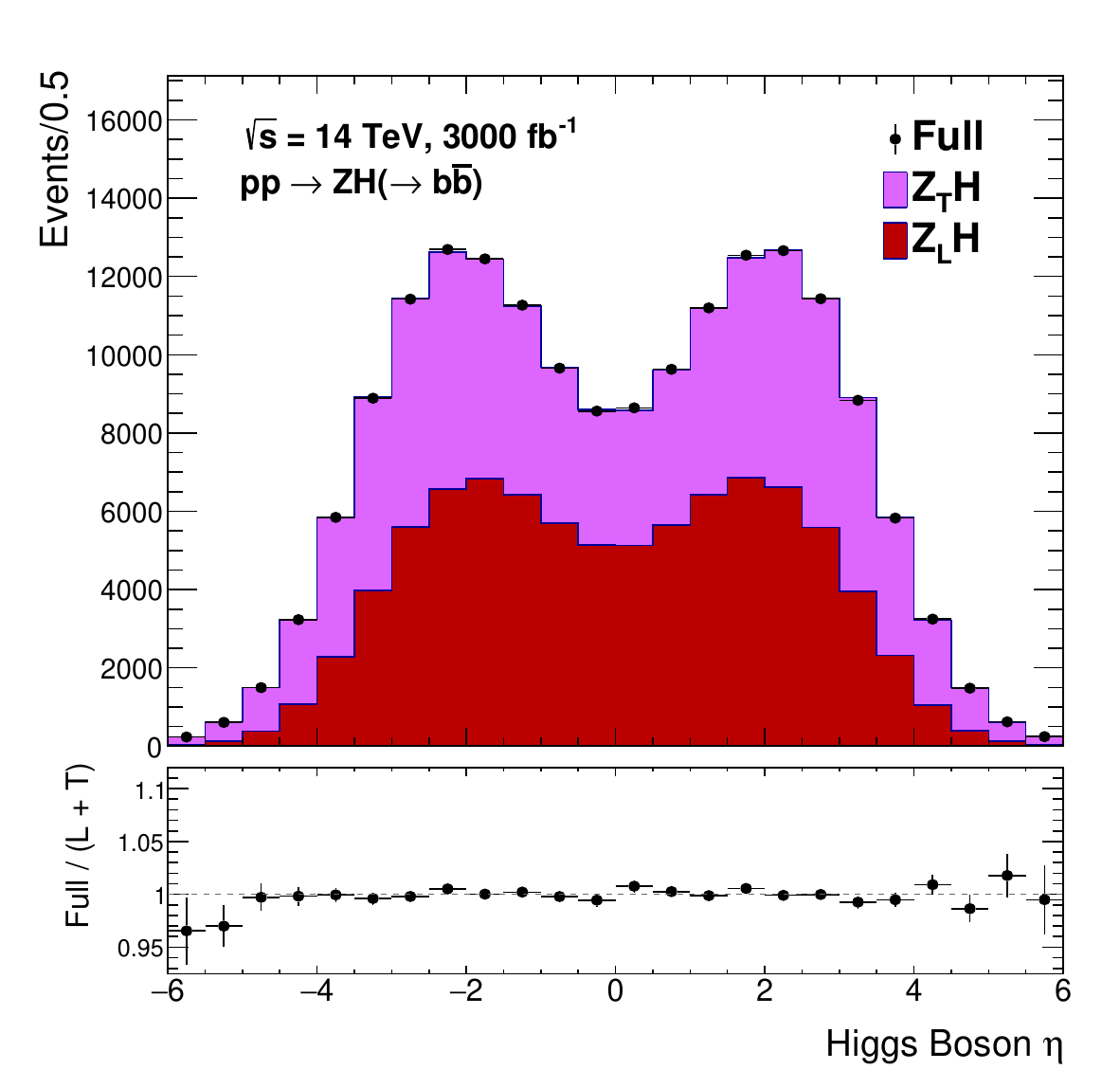}}\\
\subfloat[]{\includegraphics[width=0.2675\textwidth]{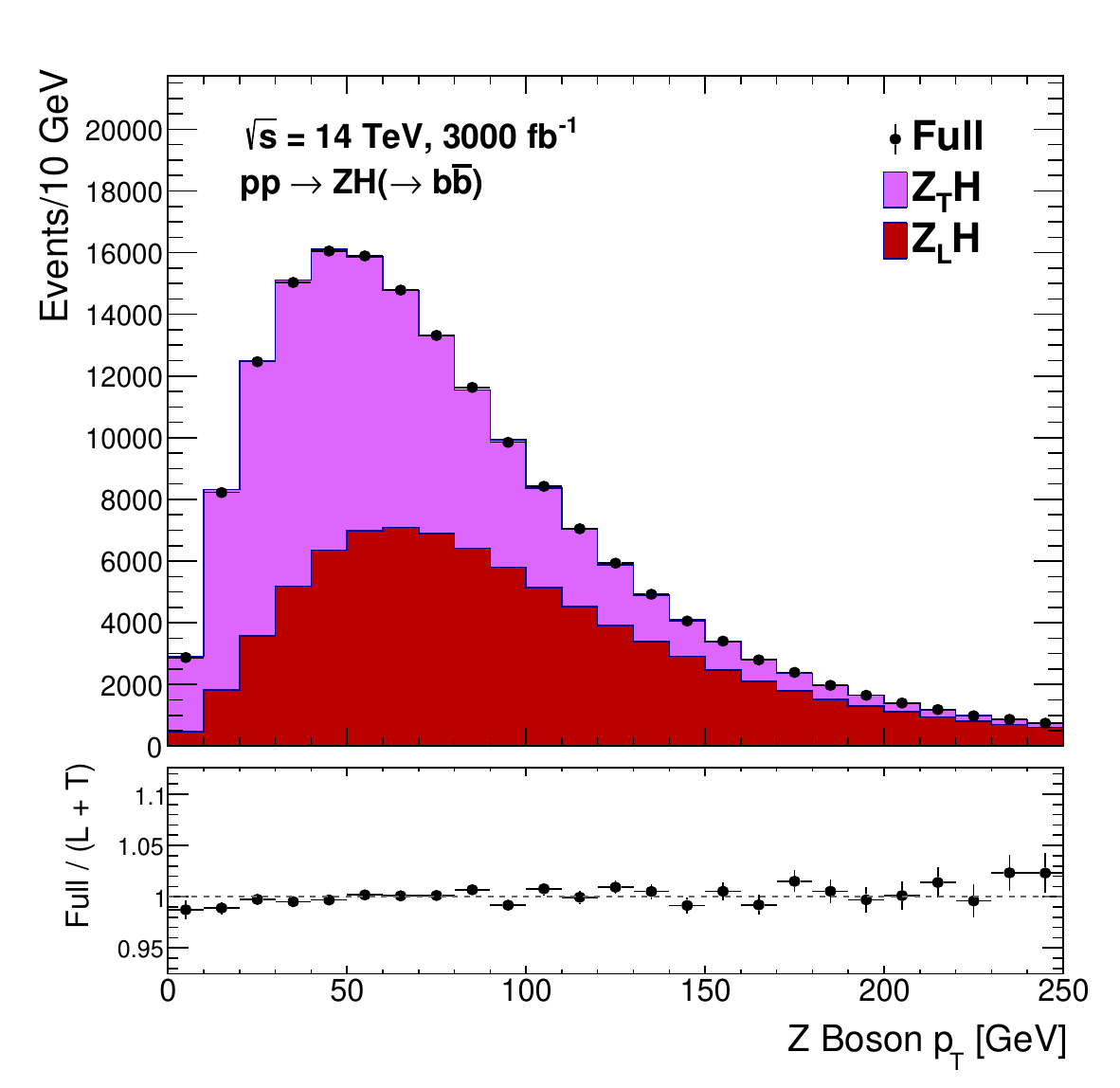}}
\subfloat[]{\includegraphics[width=0.2675\textwidth]{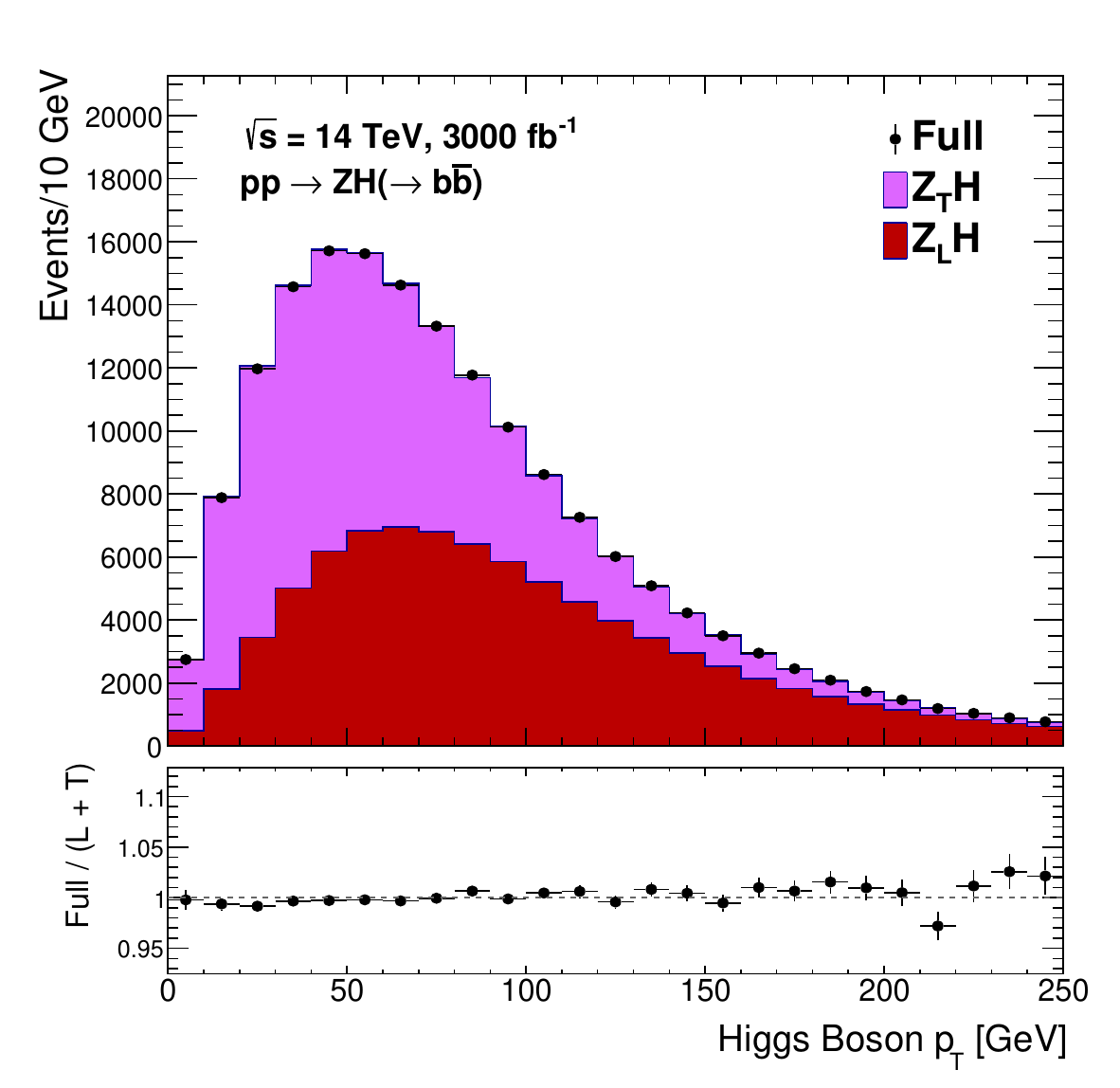}}
\subfloat[]{\includegraphics[width=0.2675\textwidth]{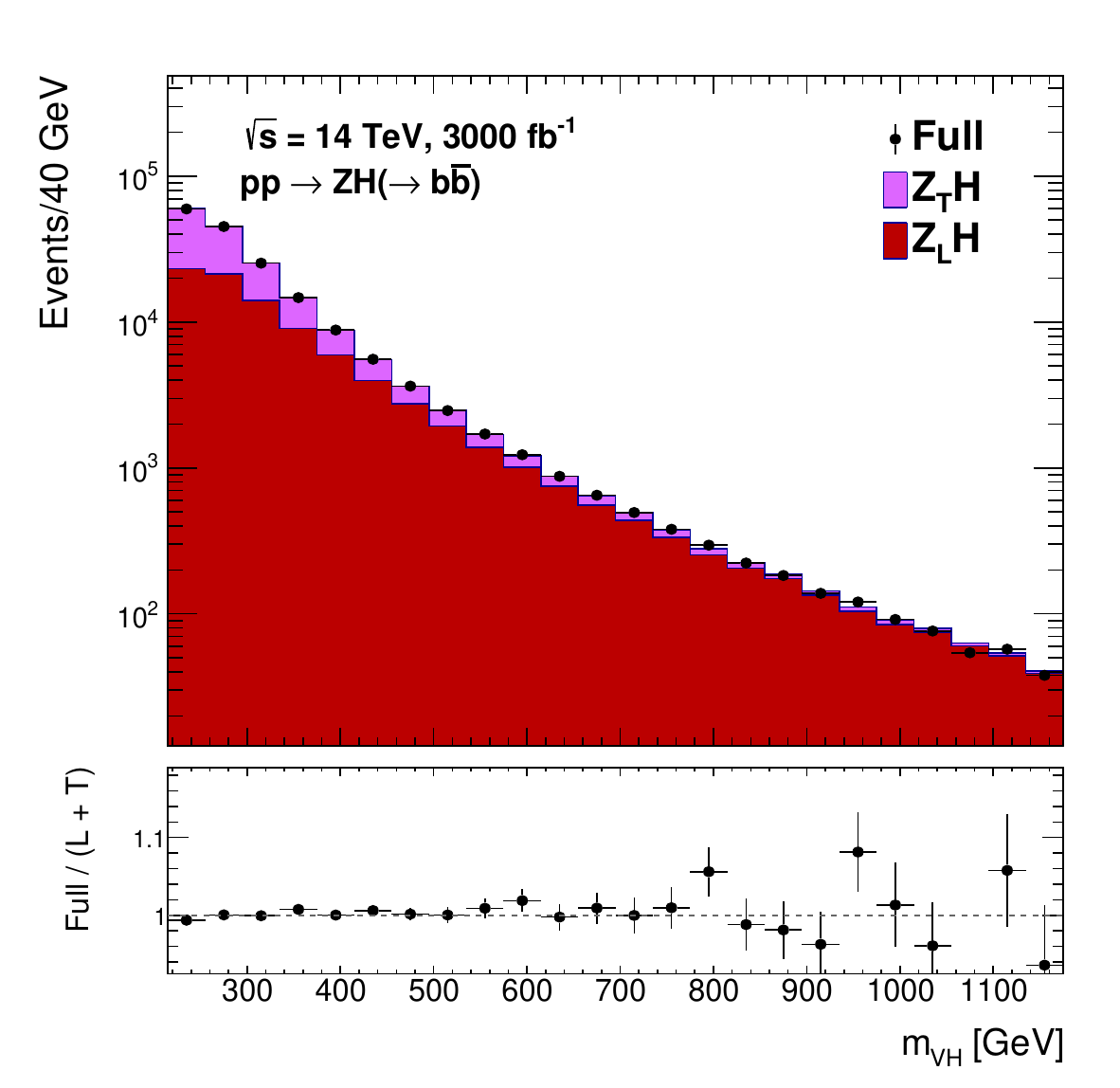}}
\caption{ Distributions of the (a) $\cos\theta^*$, (b) $\cos\theta_1$, (c) $\Phi_1$, (d) $\Delta\Phi(L, \bar{L})$, 
(e) lepton $p_\text{T}$ balance, (f) $Z$ boson $\eta$, (g) Higgs boson $\eta$, (h) $Z$ boson $p_\text{T}$, 
(i) Higgs boson $p_\text{T}$, and (j) invariant mass of the $ZH$ system ($m_{ZH}$) for simulated 
$ZH \rightarrow \ell^+\ell^- b\bar{b}$ events. The stacked histograms represent the sum of longitudinally ($Z_\text{L}H$) 
and transversely ($Z_\text{T}H$) polarised components compared to an inclusive sample of unpolarised $ZH$ production. All 
distributions are normalised to their expected cross sections. The lower panel displays the ratio of the unpolarised sample 
to the sum of the polarised samples.}
\label{fig:PolarisationObservablesInterference}
\end{figure}

The BDT response score distribution (shown in Fig.~\ref{fig:PolarisationBDTInterference}) appears to be only mildly
impacted by the missing interference effects, as the deviations between the inclusive sample and the sum of the
polarised samples extend to about 1\% at most. This demonstrates that neglecting the
$Z_\mathrm{L}H\text{--}Z_\mathrm{T}H$ interference term does not introduce any significant bias to the results
discussed in Section~\ref{sec:Results}. This robustness is also reflected at the level of individual input features.
Across all considered input observables (shown in Fig.~\ref{fig:PolarisationObservablesInterference}), the sum of the
two polarised samples ($Z_\mathrm{L}H + Z_\mathrm{T}H$) reproduces the distribution of the inclusive sample of $ZH$
production to a high degree of precision for most features, with lower panels displaying ratios generally
consistent with unity within statistical uncertainties. Significant differences in individual features are only
visible for the $\Delta\phi_{\ell\ell}$ and lepton $p_\mathrm{T}$ balance observables, with deviations ranging up
to 10\% and 5\%, respectively. \par

\clearpage
 \end{appendices}
\printbibliography[heading=bibintoc,title={Bibliography}]
\end{document}